\documentclass[a4paper,12pt]{article}
\pdfoutput=1
\usepackage{graphicx,subfigure,amsmath,amssymb,multirow}
\usepackage[titletoc,title]{appendix}
\usepackage{cite}
\usepackage{longtable}
\usepackage{graphicx}
\usepackage{color}
\usepackage{booktabs}
\usepackage[colorlinks]{hyperref}

\newlength{\dinwidth}
\newlength{\dinmargin}
\allowdisplaybreaks[4]
\def\qb{{ \bf q}_\bot}

\def\kb{{ \bf k}_\bot}

\begin{document}
\title{\bf  Semileptonic $\bar{B}^*_{c} \to (P, V) \ell^- \bar{\nu}_{\ell}$ decays within the self-consistent light-front quark model}

\author{Qin Chang$^{a,b}$, Jianying Qian$^{a}$, Liting Wang$^{a}$\thanks{wanglt@htu.edu.cn}, Xiaolin Wang$^{a}$, Shuai Xu$^{c}$, Jie Zhu$^{d}$\\
{ $^a$\small Centre for Theoretical Physics,}
        {\small Henan Normal University, Xinxiang 453007, China}\\
{ $^b$\small Center for High Energy Physics, Henan Academy of Sciences, Zhengzhou 455004, China}\\
{ $^c$\small School of Physics and Telecommunications Engineering,}\\
        {\small Zhoukou Normal University, Zhoukou 466001, China}\\
{ $^d$\small School of Physics and Electrical Engineering,}
        {\small Anyang Normal University, Anyang 455000, China}}
\date{}
 \maketitle

\begin{abstract}
 Motivated by the recent observation of the $B_c^*$ meson by the ATLAS Collaboration and its promising discovery potential at future high-luminosity colliders, we investigate the semileptonic decays $\bar{B}^*_{c} \to (P, V) \ell^- \bar{\nu}_{\ell}$ ($P=D$, $B_{d,s}$, $\eta_c(1S,\,2S,\,3S)$; $V=D^*$, $B^*_{d,s}$, $\psi(1S,\,2S,\,3S)$) in the Standard Model. Using the relevant form factors obtained in the self-consistent covariant light-front quark model, we present theoretical predictions for several physical observables, including the branching fraction, the ratios of branching fractions $R_{(L)}$, the longitudinal polarization fraction of the daughter vector meson, the $\tau$ lepton spin asymmetry and the forward-backward asymmetry. It is found that the branching fractions for the Cabibbo-favored $\bar{B}^*_{c}$ $\to$ $\bar{B}^*_{s} e^- \bar{\nu}_e$ and $\bar{B}^*_{s} \mu^- \bar{\nu}_\mu$ decays reach the order of ${\cal O}(10^{-6})$, which might be firstly observed at the future high-luminosity Large Hadron Collider.
\end{abstract}

\newpage
\section{Introduction}
The semileptonic decays of $B$ mesons induced by $b \to (c,u) \ell^- \bar{\nu}_\ell$ and $b \to s \ell^+ \ell^-$ transitions provide critical channels for precision measurements of Cabibbo-Kobayashi-Maskawa (CKM) matrix elements and serve as sensitive probes of lepton flavor universality (LFU). Recent experimental observations about the ratios of the branching fractions in the semileptonic $B$ meson decays, $R_{D}$, $R_{D^{*}}$ and $R_{J/\psi}$ deviate from the SM predictions by $1.9\sigma$\cite{Bigi:2016mdz,Bordone:2019vic,Martinelli:2021onb,Bernlochner:2022ywh,Ray:2023xjn,FlavourLatticeAveragingGroupFLAG:2024oxs,BaBar:2012obs,BaBar:2013mob,Belle:2015qfa,Belle:2019rba,LHCb:2023zxo,LHCb:2024jll,Belle-II:2025yjp,HFLAVcollaboration}, $2.7\sigma$\cite{Bordone:2019vic ,Bernlochner:2022ywh,Ray:2023xjn ,BaBar:2019vpl,Gambino:2019sif,Martinelli:2023fwm,BaBar:2012obs ,BaBar:2013mob ,Belle:2015qfa,Belle:2016dyj,Belle:2017ilt,Belle:2019rba,LHCb:2023zxo,LHCb:2023uiv,Belle-II:2024ami,LHCb:2024jll,Belle-II:2025yjp,HFLAVcollaboration}, and $1.8\sigma$\cite{Lytle:2016ixw,Colquhoun:2016osw,Watanabe:2017mip,Tran:2018kuv,Issadykov:2018myx,Leljak:2019eyw,Hu:2019qcn,Azizi:2019aaf,LHCb:2017vlu}, respectively.
These anomalies have motivated extensive searches for physics beyond the SM in semileptonic heavy-flavor decays, and various new physics (NP) scenarios have been proposed to explain them~\cite{Chang:2018sud,Sheng:2022peg,Bauer:2015knc,Iguro:2018vqb,Crivellin:2019dwb,Deshpand:2016cpw,Hu:2020yvs,Altmannshofer:2020axr,Tanaka:2012nw,Crivellin:2012ye,Celis:2016azn,Asadi:2018wea,Babu:2018vrl,1801.00917}.

The $B_c^*$ meson, with quantum numbers $J^P=1^-$, lies below the $BD$ pair threshold, meaning that its strong decays are kinematically forbidden. Its mass difference from the ground state is predicted by HPQCD to be $m_{B_c^*}-m_{B_c}=54\pm3$ MeV~\cite{Dowdall:2012ab}. Although the electromagnetic transition $B_c^* \rightarrow B_c \gamma$ dominates the total width, which results in a very short lifetime $\tau_{B_c^*} \sim \mathcal{O}(10^{-18})~\text{s}$~\cite{Simonis:2016pnh} despite being phase-space suppressed, its weak decay channels remain highly important. In particular, the semileptonic decays offer a complementary test of the LFU anomalies observed in $B$ meson channels. Recently, the ATLAS Collaboration reported the first observation of the $B_c^*$ meson in the decay $B_c^{*+}\to B_c^+\gamma$~\cite{ATLAS:2026ubk}, making a detailed theoretical study of its decay properties timely.

Although no weak decay of the $B_c^*$ meson has been observed so far, the upcoming high-luminosity runs of the LHC and future facilities, such as the Circular Electron Positron Collider and the Future Circular Collider, are expected to produce large samples of $B_c^*$ events~\cite{Chen:2018obq,LHCb:2018roe,Chang:2005wd,CEPCStudyGroup:2018ghi,FCC:2018byv}, rendering experimental studies of these decays feasible.
Over the past years, various theoretical approaches have already been applied to $B_c^*$ decays, including the QCD sum rules\cite{Wang:2012hu,Zeynali:2014wya,Bashiry:2014qia}, the QCD factorization approach\cite{Sun:2017lup,Chang:2018mva}, Non-Relativistic QCD factorization approach\cite{Chang:2025rna,Tao:2025kqf}, the perturbative QCD approach\cite{Sun:2017lla}, the relativistic independent quark model\cite{Patnaik:2017cbl}, the Bauer-Stech-Wirbel model\cite{Chang:2016cdi,R:2019uyb}, the Bethe-Salpeter (BS) method\cite{Wang:2018ryc} and the covariant light-front quark model (CLFQM)~\cite{Chang:2020xvu,Yang:2022jqu,Wang:2024cyi}. While these studies have provided valuable estimates of form factors and branching fractions for the ground-state transitions such as $B_c^* \rightarrow \eta_c$ and $B_c^* \rightarrow J/\psi$, the radially excited states have received considerably less attention. Consequently, the corresponding observables, including the LFU ratios and angular asymmetries, have not yet been examined in detail. These observables are of particular interest, as they are sensitive to possible NP and can provide complementary tests of the SM. A systematic investigation of these quantities is therefore desirable in light of the growing experimental interest in $B_c^*$ decays.

The CLFQM, originally developed by Jaus~\cite{Jaus:1999zv}, provides a simple and clear framework for computing hadronic matrix elements, with proper treatment of zero-mode contributions. However, when applied to mesons with higher quantum numbers, the traditional version suffers from self-consistency issues~\cite{Jaus:2002sv,Choi:2013mda,Chang:2018zjq}.
In this work, we adopt the self-consistent CLFQM, in which spurious contributions are explicitly eliminated, ensuring that different but theoretically equivalent expressions for a given form factor yield consistent results. This approach has been successfully applied to $P\to T$ transitions in our previous work~\cite{Chen:2021ywv} and is extended here to investigate the $B_c^*\to(P,V)$ transitions.

This work presents a systematic study of the semileptonic decays $\bar{B}^*_c \to (P,V) \ell^-\bar{\nu}_{\ell}$, with $P=D$, $B_{d,s}$, $\eta_c(1S,2S,3S)$, $V=D^*$, $B^*_{d,s}$, $\psi(1S,2S,3S)$ and $\ell=e,\mu,\tau$. Using the self-consistent CLFQM, we calculate the transition form factors and parametrize them with the Bourrely-Caprini-Lellouch (BCL) $z$-expansion approach. We then extend the phenomenological analysis to include branching fractions, lepton-universality ratios, polarization fractions, and angular asymmetries. {A complete error budget is provided by propagating all input parameter uncertainties, which has not been fully addressed in some earlier studies. These observables, especially for the radially excited states, are calculated for the first time,} Our results are expected to be valuable for future experimental searches at LHCb/HL-LHC and other high-luminosity colliders.

This paper is organized as follows. Section 2 is devoted to the theoretical framework. The helicity amplitudes and differential decay rates for the $\bar{B}^*_{c} \to (P, V) \ell^- \bar{\nu}_{\ell}$ decays are explicitly derived. In Section 3, we present the numerical results and discussion for these decays. Finally, we give our conclusions in Section 4. The explicit expressions for the form factors are given in Appendix~\ref{sec:AppendixA}.

\section{Theoretical Framework}
\subsection{Effective Hamiltonian and Amplitude}
Within the SM, the semileptonic $\bar{B}^*_{c} \to (P, V) \ell^- \bar{\nu}_{\ell}$ ($P=D$, $B_{d,s}$, $\eta_c(1S,\,2S,\,3S)$; $V=D^*$, $B^*_{d,s}$, $\psi(1S,\,2S,\,3S)$) decays are induced by $b \to q_1 \ell^- \bar{\nu}_\ell$ ($q_1=u,\,c$) or $\bar{c} \to \bar{q}_2 \ell^- \bar{\nu}_\ell$ ($q_2=d,\,s$) transition at quark level via W-exchange. As an example, the $b \to u \ell^- \bar{\nu}_\ell$ transition can be described by the effective low-scale $\mathcal{O}(m_b)$ Hamiltonian as
	\begin{eqnarray}
		\mathcal{H}_{\rm eff}(b \to u \ell^- \bar{\nu}_\ell)&=& \frac{G_F}{\sqrt{2}}V_{ub}
    [\bar{u}\gamma_{\mu}(1-\gamma_5)b][\bar{\ell}\gamma^{\mu}(1-\gamma_5)\nu_{\ell}]\label{eq:btou}\,,
	\end{eqnarray}
where $G_F$ is Fermi coupling constant, and $V_{ub}$ denotes the CKM matrix element.

Using Eq.~\eqref{eq:btou}, the square amplitude for $\bar{B}^*_c \to \bar{D}^{(*)} \ell^- \bar{\nu}_\ell$ decays induced by $b \to u \ell^- \bar{\nu}_\ell$ transition can be expressed as
\begin{eqnarray}\label{eq:M2}
	|{\cal M}(\bar{B}^*_{c} \to \bar{D}^{(*)} \ell^- \bar{\nu}_\ell)|^2&=&\frac{G^2_F\,|V_{ub}|^2}{2}
	|\langle \bar{D}^{(*)}|\bar{u}\gamma_{\mu}(1-\gamma_5)b|\bar{B}^*_{c}\rangle\,
	\bar{\ell}\gamma^{\mu}(1-\gamma_5)\nu_{\ell}|^2 \nonumber\\
	&\equiv&\frac{G^2_F\,|V_{ub}|^2}{2} L_{\mu\nu}H^{\mu\nu}\,,
\end{eqnarray}
in which, the leptonic ($L_{\mu\nu}$) and hadronic ($H^{\mu\nu}$) tensors are built from the respective products of the lepton and hadron currents.

The contraction of leptonic and hadronic tensor $L_{\mu\nu}H^{\mu\nu}$ could be rewritten as
\begin{eqnarray}\label{eq:M2LI}
L_{\mu\nu}H^{\mu\nu}=\sum_{m,m^{\prime},n,n^{\prime}} L(m,n)H(m^{\prime},n^{\prime})g_{mm^{\prime}}g_{nn^{\prime}}\,,
\end{eqnarray}
by inserting the completeness relation for the polarization four-vectors
\begin{eqnarray}
\sum_{m,n}\bar{\epsilon}_{\mu}(m) \bar	{\epsilon}_{\nu}^*(n)g_{mn}=g_{\mu\nu}\,,
\end{eqnarray}
in which, the quantities $L(m,n)\equiv L_{\mu\nu}\bar{\epsilon}^{\mu}(m)\bar{\epsilon}^{*\nu}(n)$ and $H(m,n)\equiv H^{\mu\nu}\bar{\epsilon}^*_{\mu}(m)\bar{\epsilon}_{\nu}(n)$ in Eq.~\eqref{eq:M2LI} are Lorentz invariant and therefore could be evaluated in different reference frames. For convenience, $H(m,n)$ and $L(m,n)$ will be evaluated in the $B^*_c$-meson rest frame and the $\ell^--\bar{\nu}_\ell$ center-of-mass frame, respectively.

\subsection{Kinematics for $\bar{B}^*_c \to \bar{D}^{(*)} \ell^- \bar{\nu}_\ell$ Decays}
  In the rest frame of $B^*_c$-meson with daughter $D^{(*)}$-meson moving in the positive $z$-direction, the momenta of particles $B^*_c$, $D^{(*)}$ and virtual $W^*$ are
 \begin{eqnarray}
 p_{B^*_c}^{\mu}=(m_{B^*_c},0,0,0)\,,\quad  p_{D^{(*)}}^{\mu}=(E_{D^{(*)}},0,0,|\vec{p}|)\,,\quad q^{\mu}=(q^0,0,0,-|\vec{p}|)\,,
\end{eqnarray}
where $q^0=m_{B^*_c}-E_{D^{(*)}}=(m_{B^*_c}^2-m_{D^{(*)}}^2+q^2)/2m_{B^*_c}$ and $|\vec{p}|=\lambda^{1/2}(m_{B^*_c}^2,m_{D^{(*)}}^2,q^2)/2m_{B^*_c}$ with $\lambda(a,b,c)=a^2+b^2+c^2-2(ab+bc+ca)$ are the energy and momentum of virtual $W^*$. For the four polarization vectors of virtual $W^*$, $\bar{\epsilon}^{\mu}(\lambda_{W^*}=t,0,\pm)$, one can conveniently choose~\cite{Korner:1987kd,Korner:1989qb}
 \begin{eqnarray}\label{eq:polW}
\bar{\epsilon}^{\mu}(t)=\frac{1}{\sqrt{q^2}}(q^0,0,0,-|\vec{p}|)\,,\quad \bar{\epsilon}^{\mu}(0)=\frac{1}{\sqrt{q^2}}(|\vec{p}|,0,0,-q^0)\,,\quad  \bar{\epsilon}^{\mu}(\pm)=\frac{1}{\sqrt{2}}(0,\pm1,-i,0)\,,
\end{eqnarray}
in which, $\lambda_{W^*}=t$ has to be understood as $\lambda_{W^*}=0$ with $J=0$.

Meanwhile, the polarization vectors of initial $B^*_c$-meson and daughter $D^*$-meson, $\epsilon_1^{\mu}(0,\pm)$ and $\epsilon_2^{\mu}(0,\pm)$, can be written as
 \begin{eqnarray}\label{eq:polBc}
\epsilon_1^{\mu}(0)=(0,0,0,1)\,,\quad  \epsilon_1^{\mu}(\pm)=\frac{1}{\sqrt{2}}(0,\mp1,-i,0)\,;\\
\epsilon_2^{\mu}(0)=\frac{1}{m_{D^*}}(|\vec{p}|,0,0,E_{D^*})\,,\quad  \epsilon_2^{\mu}(\pm)=\frac{1}{\sqrt{2}}(0,\mp1,-i,0)\,.
\end{eqnarray}
Turning to the $\ell^--\bar{\nu}_\ell$ center-of-mass frame, the four momenta of lepton and antineutrino are given as
 \begin{eqnarray}
 p_\ell^{\mu}=(E_{\ell}, |\vec{p}_{\ell}|\sin\theta,0,|\vec{p}_{\ell}|\cos\theta)\,,\quad  p_{\nu_\ell}^{\mu}=(|\vec{p}_{\ell}|, -|\vec{p}_{\ell}|\sin\theta,0,-|\vec{p}_{\ell}|\cos\theta)\,,
 \end{eqnarray}
where $E_{\ell}=(q^2+m_{\ell}^2)/2\sqrt{q^2}$ and $|\vec{p}_{\ell}|=(q^2-m_{\ell}^2)/2\sqrt{q^2}$ are the energy and the magnitude of the three-momentum of the charged lepton, respectively; and $\theta$ is the angle between the $D^{(*)}$-meson and $\ell$ three-momenta. In this reference frame, the polarization vectors of virtual $W^*$ boson have the form
 \begin{eqnarray}
\bar{\epsilon}^{\mu}(t)=(1,0,0,0)\,,\quad \bar{\epsilon}^{\mu}(0)=(0,0,0,1)\,,\quad \bar{\epsilon}^{\mu}(\pm)=\frac{1}{\sqrt{2}}(0,\mp1,-i,0)\,.
\end{eqnarray}

\subsection{Hadronic helicity amplitudes $H_{\lambda_{W^*}\lambda_{B^*_c}\lambda_{D^{(*)}}}$}
For the $\bar{B}^*_{c} \to \bar{D}^{(*)} \ell^- \bar{\nu}_\ell$ decays, the hadronic helicity amplitudes $H_{\lambda_{W^*}\lambda_{B^*_c}\lambda_{D^{(*)}}}$ are defined by the following expressions.
 \begin{eqnarray}\label{eq:hha}
 H_{\lambda_{W^*}\lambda_{B^*_{c}}\lambda_{D^{(*)}}}=
 \left\langle \bar{D}^{(*)}(p_{D^{(*)}},\lambda_{D^{(*)}})|\bar{u}\gamma_{\mu}(1-\gamma_{5})b|\bar{B}^*_{c}(p_{B^*_c},\lambda_{B^*_c})
 \right\rangle\bar{\epsilon}^{*\mu}(\lambda_{W^*})\,,
\end{eqnarray}
which describe the decay of three helicity states of $B^*_c$ meson into the daughter pseudoscalar $D$ or vector $D^{*}$ meson and the four helicity states of virtual $W^*$.

For the $\bar{B}^*_{c} \to \bar{D}^{(*)}$ transitions, the hadronic matrix elements $\left\langle \bar{D}^{(*)}(p_{D^{(*)}},\lambda_{D^{(*)}})|\bar{u}\gamma_{\mu}(1-\gamma_{5})b\right.$ $\left.|\bar{B}^*_{c}(p_{B^*_c},\lambda_{B^*_c})
 \right\rangle$ can be factorized in terms of fourteen form factors $A_i(q^2)$ and $V_i(q^2)$ as
\begin{eqnarray}\label{eq:ff}
	\left\langle \bar{D}\left(p_D\right)\left|\bar{u} \gamma_\mu b\right| \bar{B}^*_{c}\left(\epsilon_1, p_{B^*_{c}}\right)\right\rangle
		&=& -\frac{2 \text{i} V\left(q^2\right)}{m_{B^*_{c}}+m_D} \varepsilon_{\mu \nu \alpha \beta} \epsilon_1^\nu p_D^\alpha p_{B^*_{c}}^\beta\,, \label{eq:ffPV} \\
	\left\langle \bar{D}\left(p_D\right)\left|\bar{u} \gamma_\mu \gamma_5 b\right| \bar{B}^*_{c}\left(\epsilon_1, p_{B^*_{c}}\right)\right\rangle
		&=& 2 m_{B^*_{c}} A_0\left(q^2\right) \frac{\epsilon_1 \cdot q}{q^2} q_\mu\nonumber\\
        &&+\left({m_{B^*_{c}}+m_D}\right) A_1\left(q^2\right)\left(\epsilon_{1\mu}-\frac{\epsilon_1 \cdot q}{q^2} q_\mu\right)\nonumber \\
		&& +A_2\left(q^2\right) \frac{\epsilon_1 \cdot q}{m_{B^*_{c}}+m_D}\left(P_\mu-\frac{m_{B^*_{c}}^2-m_D^2}{q^2} q_\mu\right)\,,\label{eq:ffPA}\\
	\langle \bar{D}^{*}(\epsilon_2, p_{D^{*}})|\bar{u}\gamma_{\mu} b|\bar{B}^*_{c}(\epsilon_1, p_{B^*_{c}})\rangle
		&=&(\epsilon_1 \cdot\epsilon_2^{*})\left[-P_{\mu}\,V_1(q^2)+q_{\mu}\,V_2(q^2)\right]\nonumber\\
		&&+\frac{(\epsilon_1 \cdot q)(\epsilon_2^{*}\cdot q)}{m_{B^*_{c}}^2-m_{D^{*}}^2}\left[P_{\mu}\,V_3(q^2)-q_{\mu}\,V_4(q^2)\right]\nonumber\\
		&&-(\epsilon_1 \cdot q)\,\epsilon_{2\mu}^* \,V_5(q^2)+(\epsilon_2^* \cdot q)\,\epsilon_{1\mu}\,V_6(q^2)\,,\label{eq:ffVV}\\
	\langle \bar{D}^{*}(\epsilon_2, p_{D^{*}})|\bar{u} \gamma_{5} \gamma_{\mu} b|\bar{B}^*_{c}(\epsilon_1, p_{B^*_{c}})\rangle
		&=&-\text{i}\varepsilon_{\mu\nu\alpha\beta}\epsilon_1^{\alpha}\epsilon_2^{*\beta}\left[P^{\nu}\,A_1(q^2)-q^{\nu}\,A_2(q^2)\right]\nonumber\\
		&&-\frac{\text{i}\epsilon_2^{\ast}\cdot q}{m_{B^*_{c}}^2-m_{D^{*}}^2}\varepsilon_{\mu\nu\alpha\beta}\epsilon_1^{\nu}P^{\alpha}q^{\beta}\,A_3(q^2)\nonumber\\
		&&+\frac{\text{i}\epsilon_1\cdot q}{m_{B^*_{c}}^2-m_{D^{*}}^2}\varepsilon_{\mu\nu\alpha\beta}\epsilon_2^{*\nu}P^{\alpha}q^{\beta}A_4(q^2)\,,\label{eq:ffVA}
\end{eqnarray}
with the sign convention $\varepsilon_{0123}=-1$ and the momentum $P^\mu=p_{B^*_{c}}^\mu+p_{D^{(*)}}^\mu$.

Then, by contracting above hadronic matrix elements with the $W^*$ polarization vectors in the $B^*_c$-meson rest frame, we obtain fourteen non-vanishing helicity amplitudes $H_{\lambda_{W^*}\lambda_{B^*_c}\lambda_{D^{(*)}}}$
\begin{eqnarray}
		H_{t0}(q^2) & =&\frac{2 m_{B^*_{c}}|\vec{p}|}{\sqrt{q^2}} A_0\left(q^2\right)\label{eq:Ht0}\,,\\
		H_{00}(q^2) & =&\frac{1}{2 m_{B^*_{c}} \sqrt{q^2}}\Bigg[\left(m_{B^*_{c}}+m_D\right)\left(m_{B^*_{c}}^2-m_D^2{+q^2}\right) A_1\left(q^2\right)\nonumber\\
        &&+\frac{4 m_{B^*_{c}}^2|\vec{p}|^2}{m_{B^*_{c}}+m_D} A_2\left(q^2\right)\Bigg]\,, \\
		H_{\mp \pm }(q^2) & =&-\left(m_{B^*_{c}}+m_D\right) A_1\left(q^2\right) \mp \frac{2 m_{B^*_{c}}|\vec{p}|}{m_{B^*_{c}}+m_D} V\left(q^2\right)\label{eq:Hpm}\,,\\
		H_{0++}(q^2)&=&-\frac{m_{B^*_{c}}^2-m_{D^*}^2} {\sqrt{q^2}}A_1(q^2)+\sqrt{q^2}A_2(q^2)+\frac{2m_{B^*_{c}}|\vec{p}|}{\sqrt{q^2}}V_1(q^2)\label{eq:H0pp}\,,\\
		H_{t++}(q^2)&=&-\frac{2m_{B^*_{c}}|\vec{p}|}{\sqrt{q^2}}A_1(q^2)+\frac{m_{B^*_{c}}^2-m_{D^*}^2}{\sqrt{q^2}}V_1(q^2)-\sqrt{q^2}V_2(q^2)\,,\\
		H_{-+0}(q^2)&=&-\frac{m_{B^*_{c}}^2+3m_{D^*}^2-q^2}{2m_{D^*}}A_1(q^2)+\frac{m_{B^*_{c}}^2-m_{D^*}^2-q^2}{2m_{D^*}}A_2(q^2)\nonumber\\
					&&-\frac{2m_{B^*_{c}}^2|\vec{p}|^2}{m_{D^*}(m_{B^*_{c}}^2-m_{D^*}^2)}A_3(q^2)-\frac{m_{B^*_{c}}|\vec{p}|}{m_{D^*}}V_6(q^2)\,,\\
		H_{0--}(q^2)&=&\frac{m_{B^*_{c}}^2-m_{D^*}^2}{\sqrt{q^2}}A_1(q^2)-\sqrt{q^2}A_2(q^2)+\frac{2m_{B^*_{c}}|\vec{p}|}{\sqrt{q^2}}V_1(q^2)\,,\\
		H_{t--}(q^2) &=&\frac{2m_{B^*_{c}}|\vec{p}|}{\sqrt{q^2}}A_1(q^2)+\frac{m_{B^*_{c}}^2-m_{D^*}^2}{\sqrt{q^2}}V_1(q^2)
		-\sqrt{q^2}V_2(q^2)\,,\\
H_{+-0}(q^2)&=&\frac{m_{B^*_{c}}^2+3m_{D^*}^2-q^2}{2m_{D^*}}A_1(q^2)-\frac{m_{B^*_{c}}^2-m_{D^*}^2-q^2
		}{2m_{D^*}}A_2(q^2)\nonumber\\
		&&+\frac{2m_{B^*_{c}}^2|\vec{p}|^2}{m_{D^*}(m_{B^*_{c}}^2-m_{D^*}^2)}A_3(q^2)
		-\frac{m_{B^*_{c}}|\vec{p}|}{m_{D^*}}V_6(q^2)\,,\\
		H_{+0+}(q^2)&=&\frac{3m_{B^*_{c}}^2+m_{D^*}^2-q^2}{2m_{B^*_{c}}}A_1(q^2)
		-\frac{m_{B^*_{c}}^2-m_{D^*}^2+q^2}{2m_{B^*_{c}}}A_2(q^2)\nonumber\\
		&&+\frac{2m_{B^*_{c}}|\vec{p}|^2}{m_{B^*_{c}}^2-m_{D^*}^2}A_4(q^2)-|\vec{p}|V_5(q^2)\,,\\
		H_{-0-}(q^2)&=&-\frac{3m_{B^*_{c}}^2+m_{D^*}^2-q^2}{2m_{B^*_{c}}}A_1(q^2)
		+\frac{m_{B^*_{c}}^2-m_{D^*}^2+q^2}{2m_{B^*_{c}}}A_2(q^2)\nonumber\\
		&&-\frac{2m_{B^*_{c}}|\vec{p}|^2}{m_{B^*_{c}}^2-m_{D^*}^2}A_4(q^2)-|\vec{p}|V_5(q^2)\,,\\
H_{000}(q^2)&=&\frac{|\vec{p}|(m_{B^*_{c}}^2+m_{D^*}^2-q^2)}{\sqrt{q^2}m_{D^*}}V_1(q^2)
		+\frac{2m_{B^*_{c}}^2|\vec{p}|^3}{\sqrt{q^2}m_{D^*}(m_{B^*_{c}}^2-m_{D^*}^2)}V_3(q^2)\nonumber\\
		&&-\frac{|\vec{p}|(m_{B^*_{c}}^2-m_{D^*}^2-q^2)}{2\sqrt{q^2}m_{D^*}}V_5(q^2)
		+\frac{|\vec{p}|(m_{B^*_{c}}^2-m_{D^*}^2+q^2)}{2\sqrt{q^2}m_{D^*}}V_6(q^2)\,,\\
H_{t00}(q^2) &=&\frac{(m_{B^*_{c}}^2-m_{D^*}^2)(m_{B^*_{c}}^2+m_{D^*}^2-q^2)}{2\sqrt{q^2}m_{B^*_{c}}m_{D^*}}V_1(q^2)
		-\frac{\sqrt{q^2}(m_{B^*_{c}}^2+m_{D^*}^2-q^2)}{2m_{B^*_{c}}m_{D^*}}V_2(q^2)\nonumber\\
		&&+\frac{m_{B^*_{c}}|\vec{p}|^2}{\sqrt{q^2}m_{D^*}}V_3(q^2)
		-\frac{m_{B^*_{c}}|\vec{p}|^2\sqrt{q^2}}{m_{D^*}(m_{B^*_{c}}^2-m_{D^*}^2)}V_4(q^2)
		-\frac{m_{B^*_{c}}|\vec{p}|^2}{\sqrt{q^2}m_{D^*}}V_5(q^2)\nonumber\\
		&&+\frac{m_{B^*_{c}}|\vec{p}|^2}{\sqrt{q^2}m_{D^*}}V_6(q^2)\label{eq:Ht00}\,.
\end{eqnarray}
It should be noted that $H_{t0}(q^2)$, $H_{00}(q^2)$ and $H_{\mp \pm}(q^2)$ are the four helicity amplitudes $H_{\lambda_{W^*}\lambda_{B^*_c}\lambda_{D}}$ for $\bar{B}^*_{c} \to \bar{D}$ transition with $\lambda_{D}=0$ omitted. In addition, it is obvious that only those amplitudes with $\lambda_{B^*_c}=\lambda_{D^{(*)}}-\lambda_{W^*}$ can survive due to helicity conservation.

{In addition, these hadronic helicity amplitudes $H_{\lambda_{W^*}\lambda_{B^*_c}\lambda_{D^{(*)}}}$ are also commonly referred to as helicity-basis form factors in the literature~\cite{Cheng:2022mvd,Cheng:2018ouz,BESIII:2015hty}.
At the kinematic endpoint $q^2_{\max} = (m_{B_c^*} - m_{D^{(*)}})^2$, where the three-momentum of the final-state meson vanishes, $|\vec{p}| \to 0$, the helicity-basis form factors satisfy the following model-independent relations.}

{For the $\bar B_c^* \to \bar D$ transition, one finds
\begin{align}
H_{t0}(q^2_{\max}) &= 0, \label{eq:Ht0_endpoint}\\
H_{-+}(q^2_{\max}) &= H_{+-}(q^2_{\max})=- H_{00}(q^2_{\max})\,. \label{eq:Hpm_endpoint}
\end{align}
These relations follow directly from the explicit $|\vec{p}|$ factor in Eqs.~\eqref{eq:Ht0}--\eqref{eq:Hpm}.
}

{For the $\bar B_c^* \to \bar  D^*$ transition, the endpoint behavior is more involved. The helicity-basis form factors satisfy
\begin{align}
H_{000}(q^2_{\max}) &= 0, \label{eq:H000_endpoint}\\
H_{t++}(q^2_{\max}) &=H_{t--}(q^2_{\max})=H_{t00}(q^2_{\max}), \label{eq:H000_endpoint}\\
H_{0++}(q^2_{\max}) &= H_{-+0}(q^2_{\max})
=H_{-0-}(q^2_{\max}) = -H_{0--}(q^2_{\max})
=-H_{+-0}(q^2_{\max}) = -H_{+0+}(q^2_{\max})\,. \label{eq:Hmp0_endpoint}
\end{align}
These remaining relations are direct consequences of the helicity amplitude definitions in Eqs.~\eqref{eq:H0pp}--\eqref{eq:Ht00}. These relations are independent of the specific dynamical model used
for the form factors, and therefore provide useful consistency checks for the CLFQM numerical results. In particular, the computed helicity-basis form factors are expected to approach these kinematic limits regardless of the detailed $q^2$ dependence of the individual form factors.}

\subsection{Leptonic Helicity Amplitudes and Observables}
For the leptonic part, the leptonic tensor could be expanded in terms of a complete set of Wigner's $d^J$-functions~\cite{Korner:1987kd,Kadeer:2005aq,Fajfer:2012vx}. $L_{\mu\nu}H^{\mu\nu}$ is rewritten to a very compact form
\begin{eqnarray}\label{eq:ampd}
L_{\mu\nu}H^{\mu\nu}&=&\frac{1}{8} \sum_{\lambda_{\ell},\lambda_{\bar{\nu}_{\ell}}, \lambda_{W^*},\lambda_{W^*}^{\prime},\, J,\,J^{\prime}}\,(-1)^{J+J^{\prime}}\,|h_{\lambda_{\ell},\lambda_{\bar{\nu}_{\ell}}}|^2\,\delta_{\lambda_{B^*_c}\,,\lambda_{D^{(*)}}-\lambda_{W^*}}\,\delta_{\lambda_{B^*_c}\,,\lambda_{D^{(*)}}-\lambda_{W^*}^{\prime}}\nonumber\\
&&\times\, d^{J}_{\lambda_{W^*},\lambda_{\ell}-\frac{1}{2}}\,d^{J^{\prime}}_{\lambda_{W^*}^{\prime},\lambda_{\ell}-\frac{1}{2}}\,H_{\lambda_{W^*}\lambda_{B^*_c}\lambda_{D^{(*)}}}\,H_{\lambda_{W^*}^{\prime}\lambda_{B^*_c}\lambda_{D^{(*)}}}\,,
\end{eqnarray}
where $J$ and $J^{\prime}$ run over $1$ and $0$, $\lambda_{W^*}^{(\prime)}$ and $\lambda_{\ell}$ run over their components, and massless right-handed antineutrinos with $\lambda_{\bar{\nu}_{\ell}}=1/2$.

In the $\ell^--\bar{\nu}_\ell$ center-of-mass frame,  $h_{\lambda_{\ell},\lambda_{\bar{\nu}_{\ell}}}$ in Eq.~\eqref{eq:ampd} are the leptonic helicity amplitudes defined as
\begin{eqnarray}
h_{\lambda_{\ell},\lambda_{\bar{\nu}_{\ell}}}=\bar{u}_{\ell}(\lambda_{\ell})\gamma^{\mu}(1-\gamma_5)v_{\bar{\nu}}(\frac{1}{2})\bar{\epsilon}_{\mu}(\lambda_{W^*})\,.
\end{eqnarray}

Taking the exact forms of the spinors and $W^*$ polarization vectors, we finally obtain the following non-vanishing results
\begin{eqnarray}
|h_{-\frac{1}{2},\frac{1}{2}}|^2&=&8(q^2-m_\ell^2)\,,\\
|h_{\frac{1}{2},\frac{1}{2}}|^2&=&8\frac{m_\ell^2}{2q^2}(q^2-m_\ell^2)\,,
\end{eqnarray}
with the cases $\lambda_{\ell}=-1/2$ and $1/2$ are referred to as the non-flip and flip transitions, respectively.

Taking the derived basic building blocks of the amplitudes, we focus on the explicit forms of the observables. The double differential decay rate of $\bar{B}^*_{c} \to \bar{D}^{(*)} \ell^- \bar{\nu}_\ell$ decay is written as
\begin{eqnarray}
\frac{{\rm d}\Gamma[\bar{B}^*_{c} \to \bar{D}^{(*)} \ell^- \bar{\nu}_\ell]}{{\rm d}q^2{\rm d}\cos\theta}=\frac{G_F^2|V_{ub}|^2}{(2\pi)^3}\,\frac{|\vec{p}|}{8m_{B^*_c}^2}\,\frac{1}{3}(1-\frac{m_\ell^2}{q^2})L_{\mu\nu}H^{\mu\nu}\,,
\end{eqnarray}
where the factor $1/3$ is caused by averaging over the spins of initial state $B^*_c$. Using the standard convention for $d^J$-function~\cite{ParticleDataGroup:2024cfk}, we can write the double differential decay rates for different leptonic helicity states as
\begin{eqnarray}\label{eq:DdGDml}
	\frac{{\rm d}^2\Gamma[\bar{B}^*_{c} \to \bar{D} \ell^- \bar{\nu}_\ell]_{\lambda_\ell=-1/2}}{{\rm d}q^2{\rm d}\cos\theta}&=&\frac{G_F^2|V_{ub}|^2|\vec{p}|}{256\pi^3m_{B^*_{c}}^2}\,\frac{1}{3}\,q^2\,(1-\frac{m_\ell^2}{q^2})^2\,\nonumber\\
	&&\times\left[(1-\cos\theta)^2H_{+-}^2+(1+\cos\theta)^2H_{-+}^2+2\sin^2\theta H_{00}^2\right]\,,\\
	\label{eq:DdGDpl}
	\frac{{\rm d}^2\Gamma[\bar{B}^*_{c} \to \bar{D} \ell^- \bar{\nu}_\ell]_{\lambda_\ell=1/2}}{{\rm d}q^2{\rm d}\cos\theta}&=&\frac{G_F^2|V_{ub}|^2|\vec{p}|}{256\pi^3m_{B^*_{c}}^2}\,\frac{1}{3}\,q^2\,(1-\frac{m_\ell^2}{q^2})^2\,\frac{m_\ell^2}{q^2}\nonumber\\
	&&\times\left[\sin^2\theta(H_{+-}^2+H_{-+}^2)+2(H_{t0}-\cos\theta H_{00})^2\right]\,,\\
    \label{eq:DdGDstarml}
	\frac{{\rm d}^2\Gamma\left[\bar{B}^*_{c} \to \bar{D}^* \ell^- \bar{\nu}_\ell\right]_{\lambda_{\ell}=-1/2}}{{\rm d}q^2{\rm d}\cos\theta}&=&\frac{G_F^2|V_{ub}|^2|\vec{p}|}{256\pi^3m_{B^*_{c}}^2}\,
	\frac{1}{3}q^2(1-\frac{m_\ell^2}{q^2})^2\,\nonumber\\
	&&\times\left[(1-\text{cos}\,\theta)^2(H_{+0+}^2+H_{+-0}^2)
	+(1+\text{cos}\,\theta)^2(H_{-0-}^2+H_{-+0}^2)\right.\nonumber\\
	&&\left.+2\text{sin}^2\,\theta(H_{0++}^2+H_{0--}^2+H_{000}^2)\right]\,,\\
	\label{eq:DdGDstarpl}
	\frac{{\rm d}^2\Gamma\left[\bar{B}^*_{c} \to \bar{D}^* \ell^- \bar{\nu}_\ell\right]_{\lambda_{\ell}=1/2}}{{\rm d}q^2{\rm d}\cos\theta}&=&\frac{G_F^2|V_{ub}|^2|\vec{p}|}{256\pi^3m_{B^*_{c}}^2}\,
	\frac{1}{3}q^2(1-\frac{m_\ell^2}{q^2})^2\,\frac{m_{\ell}^2}{q^2}\,\nonumber\\
	&&\times\left[\text{sin}^2\,\theta(H_{+0+}^2+H_{+-0}^2+H_{-0-}^2+H_{-+0}^2)\right.\nonumber\\
	&&+2(H_{t++}-\text{cos}\,\theta\,H_{0++})^2+2(H_{t--}-\text{cos}\,\theta\,H_{0--})^2\nonumber\\
	&&\left.+2(H_{t00}-\text{cos}\,\theta\,H_{000})^2\right]\,.
\end{eqnarray}
Using Eqs.~\eqref{eq:DdGDml}, \eqref{eq:DdGDpl}, \eqref{eq:DdGDstarml} and \eqref{eq:DdGDstarpl}, one can get the explicit forms of observables of $\bar{B}^*_{c} \to \bar{D}^{(*)} \ell^- \bar{\nu}_\ell$ decays.

Performing the integration over $\cos\theta$ and summing over the lepton helicity, we obtain the singly differential decay rates
\begin{eqnarray}\label{eq:dSdGD}
\frac{{\rm d}\Gamma[\bar{B}^*_{c} \to \bar{D} \ell^- \bar{\nu}_\ell]}{{\rm d}q^2}&=&\frac{G_F^2|V_{ub}|^2|\vec{p}|}{96\pi^3m_{B^*_c}^2}\,\frac{1}{3}\,q^2\,(1-\frac{m_\ell^2}{q^2})^2\nonumber\\\,
&&\times\left[(1+\frac{m_\ell^2}{2\,q^2})(H_{+-}^2+H_{-+}^2+H_{00}^2)+\frac{3m_\ell^2}{2q^2}H_{t0}^2\right]\,,\\
\frac{{\rm d}\Gamma[\bar{B}^*_{c} \to \bar{D}^* \ell^- \bar{\nu}_\ell]}{{\rm d}q^2}&=&\frac{G_F^2 |V_{ub}|^2 |\vec{p}|}{96\pi^3 m_{B^*_c}^2}\,\frac{1}{3}\,q^2\,(1 -\frac{m_\ell^2}{q^2})^2 \nonumber\\\label{eq:dSdGDstar}
&& \times\Bigg[\left(1 + \frac{m_\ell^2}{2q^2}\right) (H_{+0+}^2 +H_{+-0}^2 + H_{-0-}^2 + H_{-+0}^2 + H_{000}^2 + H_{0--}^2\nonumber\\
&&+ H_{0++}^2)+\frac{3m_\ell^2}{2q^2} \left(H_{t++}^2 + H_{t--}^2 + H_{t00}^2\right) \Bigg]\,,
\end{eqnarray}
from which the branching fractions can be easily derived. Furthermore, paying attention to the polarization states of daughter $D^*$ meson, one can obtain the longitudinal differential decay width $ {\rm d}\Gamma_L[\bar{B}^*_{c} \to \bar{D}^* \ell^- \bar{\nu}_\ell]/{\rm d}q^2$ by selecting $H_{\lambda_{W^*}\lambda_{B^*_c}\lambda_{D^{*}}}$ terms with $\lambda_{D^*}=0$ in Eq.~\eqref{eq:dSdGDstar}.

Besides above differential decay widths, there are several important observables, the ratios of branching fractions, the longitudinal polarization fraction of the daughter $D^*$ meson, the $\tau$ lepton spin asymmetry and the forward-backward asymmetry, which are defined as
\begin{eqnarray}\label{eq:dsRL}
R_{(L)}^{D^{(*)}}(q^2)&=&\frac{{{\rm d}\Gamma_{(L)}[\bar{B}^*_{c} \to \bar{D}^{(*)} \tau^- \bar{\nu}_\tau]}/{{\rm d}q^2}}{{{\rm d}\Gamma_{(L)}[\bar{B}^*_{c} \to \bar{D}^{(*)} \ell'^- \bar{\nu}_{\ell'}]}/{{\rm d}q^2}}\,,~ \text{for}~\ell'=e,\mu\,,\\\label{eq:dsFL}
F^{D^{*}}_{L}(q^2)&=&\frac{{\rm d}\Gamma_L [\bar{B}^*_{c} \to \bar{D}^* \tau^- \bar{\nu}_\tau]/{\rm d}q^2}{{\rm d}\Gamma/{\rm d}q^2}\,,\\\label{eq:dsAlambda}
A^{D^{(*)}}_{\lambda}(q^2)&=&\frac{{\rm d}\Gamma[\bar{B}^*_{c} \to \bar{D}^{(*)} \tau^- \bar{\nu}_\tau]_{\lambda_\tau=-1/2}/{\rm d}q^2-{\rm d}\Gamma[\bar{B}^*_{c} \to \bar{D}^{(*)} \tau^- \bar{\nu}_\tau]_{\lambda_\tau=1/2}/{\rm d}q^2}{{\rm d}\Gamma[\bar{B}^*_{c} \to \bar{D}^{(*)} \tau^- \bar{\nu}_\tau]_{\lambda_\tau=-1/2}/{\rm d}q^2+{\rm d}\Gamma[\bar{B}^*_{c} \to \bar{D}^{(*)} \tau^- \bar{\nu}_\tau]_{\lambda_\tau=1/2}/{\rm d}q^2}\,,\\\label{eq:dsAtheta}
A^{D^{(*)}}_{\theta}(q^2)&=&\frac{\int_{-1}^0{\rm d}\cos\theta\,( {\rm d}^2\Gamma/{\rm d}q^2{\rm d}\cos\theta)-\int_{0}^1{\rm d}\cos\theta\,( {\rm d}^2\Gamma/{\rm d}q^2{\rm d}\cos\theta)}{ {\rm d}\Gamma/{\rm d}q^2}\,,
\end{eqnarray}
respectively.

{These observables probe complementary aspects of the decay dynamics. The ratios $R_{P,V}$ are sensitive to the relative strength of the transverse, longitudinal and time-like helicity contributions. Expanding the differential ratio in powers of \(m_\tau^2/q^2\), one obtains
\begin{equation}
R_{P,V}(q^2) \approx 1 - \frac{3}{2}\frac{m_\tau^2}{q^2}
+ \frac{3}{2}\frac{m_\tau^2}{q^2} \frac{H_{t\lambda_{B_c^*}\lambda_{D^{(*)}}}^2}{\sum_i H_{i\lambda_{B_c^*}\lambda_{D^{(*)}}}^2}
+ \mathcal{O}\left(\frac{m_\tau^4}{q^4}\right),
\end{equation}
where $i$ runs over $0,\pm$. The first correction term reflects the phase-space suppression of the $\tau$ mode, which is gradually lifted as $q^2$ increases, while the second term arises from the time-like helicity contribution weighted by $m_\tau^2/q^2$. Thus, $R_{P,V}$ generally rises with $q^2$, although the detailed behavior is channel-dependent since the magnitude of the time-like amplitude varies among different transitions.}

{In contrast, the forward-backward asymmetry $A_\theta$ probes the interference between helicity amplitudes. For a pseudoscalar final state,
\begin{equation}
A_{\theta}^P(q^2) = \frac{3}{4} \frac{H_{+-}^2 - H_{-+}^2 + 2x_{\ell} \, H_{t0}H_{00}}{(1 + x_{\ell}/2)(H_{+-}^2 + H_{-+}^2 + H_{00}^2) + (3x_{\ell}/2)H_{t0}^2},
\label{eq:Atheta_P}
\end{equation}
where $x_{\ell} \equiv m_{\ell}^2/q^2$. The numerator contains the interference between transverse helicity amplitudes, as seen from the term $H_{+-}^2 - H_{-+}^2 = -8M_{B_c^*}|\vec{p}|\,A_1V$ (see Eq.~\eqref{eq:Hpm}). Thus $A_\theta$ probes the interference structure of the hadronic current, complementing $R_{P,V}$ and $F_L$, which are mainly sensitive to the modulus of the helicity amplitudes. The expression for a vector final state is analogous.}

{The longitudinal polarization fraction $F_L^{D^*}$ measures the probability that the daughter vector meson is produced with zero helicity, i.e., its longitudinal polarization fraction. In contrast to the total ratios $R_{P,V}$, $F_L$ can be extracted from the angular distribution of the daughter vector meson decay products without relying on an overall normalization, and thus provides complementary information on the polarization composition of the final state.}

Finally, the branching fractions in Eqs.~\eqref{eq:dSdGD} and \eqref{eq:dSdGDstar} are strongly affected by the uncertainties from the CKM matrix element and hadronic form factors, but the situation differs for the observables in Eqs.~\eqref{eq:dsRL}, \eqref{eq:dsFL}, \eqref{eq:dsAlambda}, and \eqref{eq:dsAtheta}. Because these observables are ratios of the branching fractions, their dependence on the CKM matrix element cancels out exactly, and the uncertainties arising from form factors are also largely reduced in these ratios, thus resulting in the predictions with high accuracy.

\section{Numerical Results and Discussions}
\subsection{Input Parameters}
Before presenting our numerical results and analyses for $\bar{B}^*_{c} \to (P,V) \ell^- \bar{\nu}_\ell$ decays, we would like to clarify the input parameters used in our numerical evaluations. For the CKM elements, we take $|V_{ub}|=36.67^{+0.88}_{-0.73}\times 10^{-4}$, $|V_{cb}|=41.45^{+0.35}_{-0.61} \times 10^{-3}$, $|V_{cs}|=97.35^{+0.01}_{-0.01} \times 10^{-2}$ and $|V_{cd}|=22.49^{+0.02}_{-0.02}\times 10^{-2}$ fitted by CKMFitter Group~\cite{CKMfitterGroup}, the central value of the well-known Fermi coupling constant and the masses of mesons and leptons given by PDG~\cite{ParticleDataGroup:2024cfk}.

To evaluate the branching fractions of semileptonic $\bar{B}^*_{c} \to (P, V) \ell^- \bar{\nu}_{\ell}$ decays, the total decay width (or lifetime) $\Gamma_{B^*_{c}}$ is essential. Unfortunately, there is no available experimental and theoretical information for $\Gamma_{B^*_{c}}$ to date. Due to the fact that the radiative process $B^*_{c} \to B_c\gamma$ dominates the decays of $B^*_{c}$ meson, we take the approximation $\Gamma_{\rm{tot}}(B^*_{c}) \simeq \Gamma (B^*_{c} \to B_c \gamma)$ in our following evaluation of branching fractions. Since the photon from the $B^*_{c} \to B_c\gamma$ process is too soft to be easily identified experimentally, the information on $\Gamma (B^*_{c} \to B_c \gamma) $ is derived primarily from theoretical estimations. The predictions for $\Gamma(B^*_{c} \to B_c\gamma)$ have been obtained in various theoretical models~\cite{Barik:1994vd,Eichten:1994gt,Gershtein:1994dxw,Fulcher:1998ka,Ciftci:2001kt,Ebert:2002xz,Jena:2002is,Lahde:2002wj,Ebert:2002pp,Godfrey:2004ya}. In this paper, the CLFQM is employed to evaluate $\Gamma(B^*_{c} \to B_c\gamma)$. The relevant theoretical formulas have been obtained in Ref.~\cite{Choi:2007se}. Using the constituent quark masses and the Gaussian parameters $\beta$ given in Appendix B \cite{Chang:2020wvs},
{which are determined by fitting to the decay constants of pseudoscalar and vector mesons following the procedure of Ref.~\cite{Chang:2018zjq},} we obtain
\begin{eqnarray}\label{eq:GammaBcstar}
 \Gamma_{\rm{tot}}(B^*_{c}) \simeq \Gamma (B^*_{c} \to B_c \gamma)=(39^{+23}_{-17})~\text{eV}\,.
\end{eqnarray}

{The corresponding decay constant of the $B_c^*$ meson is obtained within the same CLFQM framework, defined via
\begin{eqnarray}
\langle 0|\bar{b}\gamma_\mu c|B_c^*(p)\rangle = f_{B_c^*} M_{B_c^*} \epsilon_\mu\,.
\end{eqnarray}
The numerical result is $f_{B_c^*} = 422 \pm 25$~MeV, where the uncertainty is propagated from the Gaussian parameters $\beta$ and the constituent quark masses. Since the Gaussian parameter $\beta_{B_c^*}$ is determined with reference to this decay constant, this value is a model output rather than an independent prediction. For reference, other theoretical determinations include $f_{B_c^*} = 437.9 \pm 4.3$~MeV from LQCD~\cite{Cai:2026xja}, $453 \pm 20$~MeV from NRQCD~\cite{Tao:2022qxa}, and $415^{+22}_{-39}$~MeV from QCD SR~\cite{Narison:2015nxh}.}

Besides the input parameters given above, the form factors of $\bar{B}^*_{c} \to (P,V)$ transitions serve as the fundamental and critical inputs for evaluating observables. Regrettably, there is no experimental data and readily available theoretical results for use at present. In our previous works, we calculate the vector and axial-vector form factors of $V^{\prime} \to V^{\prime\prime}$~\cite{Chang:2019obq} and $P \to V$~\cite{Chang:2019mmh} transitions within the standard LFQM (SLFQM) and CLFQM, investigate the self-consistency and Lorentz covariance of the CLFQM with two types of correspondence schemes, and analyse the zero-mode and valence contributions to the form factors of the relevant transitions in the CLFQM and their relation to the SLFQM results. {The explicit theoretical formulas are given in Appendix~\ref{sec:AppendixA}.

In this work, we adopt the strictly covariant and self-consistent CLFQM to evaluate the form factors of $\bar{B}^*_{c} \to (P,V)$ transitions, and take the Gaussian type wave functions for the initial $B^*_c$-meson and daughter ground and excited mesons as~\cite{Ke:2010vn}
\begin{eqnarray}\label{eq:TWF}
\psi_{1S}(x,\mathbf{k}_{\bot})&=&4\left(\frac{\pi}{\beta^2}\right)^{3/4}\sqrt{\frac{\partial{k_z}}{\partial{x}}}\exp\left(-\frac{k_z^2+\mathbf{k}_{\bot}^2}{2\beta^2}\right)\,,\\
\label{eq:MWF2S}
\psi_{2S}(x,\mathbf{k}_{\bot})&=&4\Big(\frac{\pi}{\beta^2}\Big)^{3/4}\sqrt{\frac{\partial k_z}{\partial x}} \exp \Big(-\frac{{2}^\delta}{2} \frac{k^2_z+\mathbf{k}_{\bot}^2}{\beta^2}\Big)\Big(a_2 -b_2\frac{k^2_z+\mathbf{k}_{\bot}^2}{\beta^2}\Big)\,,\\
\label{eq:MWF3S}
\psi_{3S}(x,\mathbf{k}_{\bot})&=&4\Big(\frac{\pi}{\beta^2}\Big)^{3/4}\sqrt{\frac{\partial k_z}{\partial x}} {\exp} \Big(-\frac{{3}^\delta}{2} \frac{k^2_z+\mathbf{k}_{\bot}^2}{\beta^2}\Big)\Big(a_3-b_3\frac{k^2_z+\mathbf{k}_{\bot}^2}{\beta^2}+c_3\frac{(k^2_z+\mathbf{k}_{\bot}^2)^2}{\beta^4}\Big)\,,
\end{eqnarray}
{with $\delta=1/1.82$, $a_2=1.89$, $b_2=1.55$, $a_3=2.54$, $b_3=5.67$, and $c_3=1.86$ are taken from Ref.~\cite{Ke:2010vn}, where they were obtained by fitting the decay constants of heavy quarkonium states while preserving orthogonality among the wave functions.} It should be noted that the transitions form factors for semileptonic $B^*_c$ decays are derived in the $q^+=0$ frame within the CLFQM, and they are known only for space-like momentum transfer, as $q^2=-q^2_{\perp} \leq 0$. Consequently, the form factors in the physical time-like region require an additional $q^2$ extrapolation. For the phenomenological applications, we adopt the BCL version of $z$-series expansion~\cite{Bourrely:2008za} in the form adopted in Refs.~\cite{Khodjamirian:2017fxg,Gao:2019lta,Gubernari:2018wyi},
\begin{eqnarray}
{F(q^2)=\frac{1}{1-q^2 / M_{\text{pole}}^2}\sum_{k=0}^{N} b_{k} \left[ z(q^{2}, t_{0}) - z(0, t_{0}) \right]^{k}\,,\label{eq:BCL}}
\end{eqnarray}
with$z(q^2,t_0)=\frac{\sqrt{t_+-q^2}-\sqrt{t_+-t_0}}{\sqrt{t_+-q^2}+\sqrt{t_+-t_0}}$,
{$t_{-} = (m_{B^*_c}- m_{P/V})^2$, $t_0\equiv t_{+}(1-\sqrt{1-t_-/t_+})$, and $t_+$ is the squared invariant mass of the lightest multi-hadron state that can be excited from the vacuum by the mediating weak transition current. In Eq.~\eqref{eq:BCL},  $M_{\text{pole}}$ denotes the mass of the excited resonance state of the quarks involved in the weak current of the form factor and shares the quantum numbers of the transition for a particular form factor,  with the corresponding masses listed in Table \ref{tab:PoleMass}. }For our subsequent numerical calculations, we truncate the expansion at $N=1$ in practice, with the parameter $b_k$ determined by fitting to the results computed directly by CLFQM.

\begin{table}[htbp]
\centering
\caption{{The summary of the resonance masses (in units of GeV) with different quantum numbers entering the parametrization of the $B_c^* \to (P, V)$ form factors~\cite{Dowdall:2012ab,ParticleDataGroup:2024cfk}.}}
\vspace{0.5em}
\begin{tabular*}{\linewidth}{@{\extracolsep{\fill}} ccccccc @{}}
\hline\hline \addlinespace[1.5pt]
\multicolumn{2}{c}{$\mathcal{F}(q^2)$} & & & & & \\
\cline{1-2} \addlinespace[1.5pt]
$B_c^* \to P$ & $B_c^* \to V$ & \raisebox{2.8ex}[0pt]{$J^P$} & \raisebox{2.8ex}[0pt]{$c \to d$} & \raisebox{2.8ex}[0pt]{$c \to s$} & \raisebox{2.8ex}[0pt]{$b \to u$} & \raisebox{2.8ex}[0pt]{$b \to c$} \\
\hline \addlinespace[4.5pt]
$A_0(q^2)$ & $A_1(q^2)$ & $0^-$ & 1.870 & 1.968 & 5.279 & 6.274 \\ \addlinespace[7pt]
$V(q^2)$ & $V_{1-6}(q^2)$ & $1^-$ & 2.010 & 2.112 & 5.325 & 6.332 \\ \addlinespace[7pt]
$A_{1,2}(q^2)$ & $A_{2-4}(q^2)$ & $1^+$ & 2.422 & 2.460 & 5.726 & 6.742 \\ \addlinespace[3pt]
\hline\hline
\end{tabular*}
\label{tab:PoleMass}
\end{table}

\begin{table}[t]
	\caption{Parameters of  the $z$-series expansion for $\bar{B}^*_c \to (P,V)$ transitions form factors within CLFQM. The uncertainties are from the Gaussian parameters $\beta$ and the constituent quark masses, respectively.}
	\begin{center}
		\scalebox{0.65}{
			\begin{tabular}{lrrlrrlrr}
				\hline\hline
		$F$ & $F(0)$        &$b_1$    &$F$  & $F(0)$      &$b_1$    &$F$ & $F(0)$       &$b_1$
\\\hline
$V^{\bar{B}_c^* \to \bar{B}}$ & $3.09^{+0.05+0.32}_{-0.06-0.27}$ & $-10.06^{+0.40+1.09}_{-0.40-0.79}$&
$V^{\bar{B}_c^* \to \bar{B}_s}$ & $3.40^{+0.04+0.32}_{-0.05-0.25}$ & $-13.36^{+0.59+0.99}_{-0.58-1.27}$&
$V^{\bar{B}_c^* \to \bar{D}}$ & $0.23^{+0.02+0.02}_{-0.02-0.02}$ & $-1.28^{+0.06+0.14}_{-0.06-0.13}$\\

$A_0^{\bar{B}_c^* \to \bar{B}}$ & $0.60^{+0.01+0.06}_{-0.01-0.06}$ & $-1.99^{+0.07+0.17}_{-0.07-0.17}$&
$A_0^{\bar{B}_c^* \to \bar{B}_s}$ & $0.69^{+0.01+0.05}_{-0.01-0.06}$ & $-2.74^{+0.12+0.23}_{-0.12-0.21}$&
$A_0^{\bar{B}_c^* \to \bar{D}}$ & $0.18^{+0.01+0.01}_{-0.01-0.02}$ & $-0.13^{+0.02+0.02}_{-0.02-0.02}$\\

$A_1^{\bar{B}_c^* \to \bar{B}}$ & $0.65^{+0.01+0.06}_{-0.01-0.06}$ & $-2.56^{+0.06+0.25}_{-0.06-0.27}$&
$A_1^{\bar{B}_c^* \to \bar{B}_s}$ & $0.75^{+0.01+0.08}_{-0.01-0.07}$ & $-3.54^{+0.11+0.40}_{-0.11-0.42}$&
$A_1^{\bar{B}_c^* \to \bar{D}}$ & $0.19^{+0.01+0.02}_{-0.01-0.02}$ & $-0.67^{+0.01+0.06}_{-0.01-0.06}$\\

$A_2^{\bar{B}_c^* \to \bar{B}}$ & $0.91^{+0.07+0.05}_{-0.07-0.08}$ & $-4.70^{+0.35+0.49}_{-0.37-0.53}$&
$A_2^{\bar{B}_c^* \to \bar{B}_s}$ & $0.96^{+0.08+0.06}_{-0.08-0.09}$ & $-5.61^{+0.48+0.68}_{-0.50-0.71}$&
$A_2^{\bar{B}_c^* \to \bar{D}}$ & $0.12^{+0.01+0.01}_{-0.01-0.01}$ & $-0.68^{+0.05+0.07}_{-0.05-0.07}$\\
\hline
$V^{\bar{B}_c^* \to \eta_c(1S)}$ & $0.91^{+0.01+0.11}_{-0.01-0.12}$ & $-5.36^{+0.10+0.62}_{-0.09-0.47}$ &
$V^{\bar{B}_c^* \to \eta_c(2S)}$ & $0.41^{+0.01+0.04}_{-0.01-0.04}$ & $2.32^{+0.21+0.31}_{-0.25-0.31}$ &
$V^{\bar{B}_c^* \to \eta_c(3S)}$ & $0.21^{+0.02+0.02}_{-0.02-0.02}$ & $2.19^{+0.20+0.19}_{-0.19-0.18}$ \\

$A_0^{\bar{B}_c^* \to \eta_c(1S)}$ & $0.66^{+0.01+0.08}_{-0.01-0.08}$ & $2.08^{+0.20+0.27}_{-0.20-0.24}$ &
$A_0^{\bar{B}_c^* \to \eta_c(2S)}$ & $0.27^{+0.01+0.03}_{-0.01-0.04}$ & $5.33^{+0.17+0.48}_{-0.17-0.45}$ &
$A_0^{\bar{B}_c^* \to \eta_c(3S)}$ & $0.11^{+0.01+0.01}_{-0.01-0.01}$ & $3.02^{+0.07+0.24}_{-0.07-0.15}$ \\

$A_1^{\bar{B}_c^* \to \eta_c(1S)}$ & $0.70^{+0.01+0.08}_{-0.01-0.09}$ & $-1.26^{+0.14+0.15}_{-0.14-0.13}$ &
$A_1^{\bar{B}_c^* \to \eta_c(2S)}$ & $0.21^{+0.01+0.02}_{-0.01-0.03}$ & $4.37^{+0.19+0.43}_{-0.20-0.56}$ &
$A_1^{\bar{B}_c^* \to \eta_c(3S)}$ & $0.09^{+0.01+0.01}_{-0.01-0.01}$ & $2.34^{+0.07+0.24}_{-0.07-0.12}$ \\

$A_2^{\bar{B}_c^* \to \eta_c(1S)}$ & $0.59^{+0.01+0.07}_{-0.01-0.07}$ & $-3.35^{+0.11+0.28}_{-0.11-0.27}$ &
$A_2^{\bar{B}_c^* \to \eta_c(2S)}$ & $0.49^{+0.00+0.02}_{-0.00-0.02}$ & $-1.00^{+0.13+0.13}_{-0.13-0.13}$ &
$A_2^{\bar{B}_c^* \to \eta_c(3S)}$ & $0.25^{+0.01+0.02}_{-0.01-0.02}$ & $1.50^{+0.22+0.12}_{-0.15-0.17}$ \\
\hline
$A_1^{\bar{B}_c^* \to \bar{B}^*}$ & $0.43^{+0.01+0.04}_{-0.01-0.04}$ & $-1.45^{+0.07+0.14}_{-0.07-0.11}$ &
$A_1^{\bar{B}_c^* \to \bar{B}_s^*}$ & $0.53^{+0.01+0.05}_{-0.01-0.05}$ & $-2.15^{+0.10+0.17}_{-0.10-0.18}$ &
$A_1^{\bar{B}_c^* \to \bar{D}^*}$ & $0.09^{+0.01+0.01}_{-0.01-0.01}$ & $-0.59^{+0.04+0.06}_{-0.04-0.05}$ \\

$A_2^{\bar{B}_c^* \to \bar{B}^*}$ & $1.17^{+0.09+0.10}_{-0.09-0.11}$ & $-7.72^{+0.55+0.80}_{-0.55-0.90}$ &
$A_2^{\bar{B}_c^* \to \bar{B}_s^*}$ & $1.05^{+0.08+0.09}_{-0.08-0.14}$ & $-8.79^{+0.70+0.96}_{-0.71-1.25}$ &
$A_2^{\bar{B}_c^* \to \bar{D}^*}$ & $0.10^{+0.01+0.01}_{-0.01-0.01}$ & $-0.66^{+0.05+0.07}_{-0.05-0.05}$ \\

$A_3^{\bar{B}_c^* \to \bar{B}^*}$ & $0.81^{+0.02+0.07}_{-0.02-0.09}$ & $-4.85^{+0.19+0.51}_{-0.18-0.54}$ &
$A_3^{\bar{B}_c^* \to \bar{B}_s^*}$ & $0.73^{+0.01+0.05}_{-0.02-0.08}$ & $-5.42^{+0.27+0.66}_{-0.26-0.69}$ &
$A_3^{\bar{B}_c^* \to \bar{D}^*}$ & $0.06^{+0.01+0.01}_{-0.01-0.00}$ & $-0.39^{+0.03+0.03}_{-0.03-0.03}$ \\

$A_4^{\bar{B}_c^* \to \bar{B}^*}$ & $0.89^{+0.02+0.10}_{-0.02-0.07}$ & $-5.12^{+0.13+0.57}_{-0.12-0.50}$ &
$A_4^{\bar{B}_c^* \to \bar{B}_s^*}$ & $0.85^{+0.01+0.09}_{-0.01-0.08}$ & $-6.11^{+0.19+0.70}_{-0.18-0.70}$ &
$A_4^{\bar{B}_c^* \to \bar{D}^*}$ & $0.06^{+0.00+0.00}_{-0.00-0.00}$ & $-0.38^{+0.02+0.03}_{-0.02-0.03}$ \\

$V_1^{\bar{B}_c^* \to \bar{B}^*}$ & $0.53^{+0.02+0.05}_{-0.02-0.05}$ & $-2.04^{+0.07+0.21}_{-0.07-0.23}$ &
$V_1^{\bar{B}_c^* \to \bar{B}_s^*}$ & $0.64^{+0.01+0.05}_{-0.01-0.05}$ & $-2.99^{+0.10+0.28}_{-0.10-0.29}$ &
$V_1^{\bar{B}_c^* \to \bar{D}^*}$ & $0.10^{+0.01+0.01}_{-0.01-0.01}$ & $-0.62^{+0.04+0.06}_{-0.04-0.07}$ \\

$V_2^{\bar{B}_c^* \to \bar{B}^*}$ & $1.18^{+0.09+0.11}_{-0.09-0.11}$ & $-6.75^{+0.48+0.82}_{-0.47-0.73}$ &
$V_2^{\bar{B}_c^* \to \bar{B}_s^*}$ & $1.06^{+0.08+0.08}_{-0.08-0.13}$ & $-7.79^{+0.62+1.14}_{-0.63-0.96}$ &
$V_2^{\bar{B}_c^* \to \bar{D}^*}$ & $0.10^{+0.01+0.01}_{-0.01-0.01}$ & $-0.65^{+0.05+0.05}_{-0.05-0.06}$ \\

$V_3^{\bar{B}_c^* \to \bar{B}^*}$ & $0.41^{+0.01+0.02}_{-0.01-0.04}$ & $-1.56^{+0.08+0.12}_{-0.08-0.13}$ &
$V_3^{\bar{B}_c^* \to \bar{B}_s^*}$ & $0.40^{+0.00+0.01}_{-0.00-0.03}$ & $-2.01^{+0.10+0.22}_{-0.10-0.21}$ &
$V_3^{\bar{B}_c^* \to \bar{D}^*}$ & $0.07^{+0.01+0.01}_{-0.01-0.01}$ & $-0.48^{+0.03+0.04}_{-0.03-0.05}$ \\

$V_4^{\bar{B}_c^* \to \bar{B}^*}$ & $0.02^{+0.00+0.00}_{-0.00-0.00}$ & $0.02^{+0.00+0.00}_{-0.00-0.00}$ &
$V_4^{\bar{B}_c^* \to \bar{B}_s^*}$ & $0.09^{+0.01+0.01}_{-0.01-0.01}$ & $-0.52^{+0.06+0.06}_{-0.06-0.07}$ &
$V_4^{\bar{B}_c^* \to \bar{D}^*}$ & $-0.01^{+0.00+0.00}_{-0.00-0.00}$ & $0.10^{+0.01+0.01}_{-0.01-0.01}$ \\

$V_5^{\bar{B}_c^* \to \bar{B}^*}$ & $3.16^{+0.09+0.26}_{-0.10-0.34}$ & $-11.68^{+0.48+1.02}_{-0.47-1.25}$ &
$V_5^{\bar{B}_c^* \to \bar{B}_s^*}$ & $3.53^{+0.05+0.34}_{-0.05-0.27}$ & $-15.83^{+0.63+1.38}_{-0.64-1.47}$ &
$V_5^{\bar{B}_c^* \to \bar{D}^*}$ & $0.26^{+0.03+0.02}_{-0.02-0.02}$ & $-1.60^{+0.11+0.13}_{-0.11-0.17}$ \\

$V_6^{\bar{B}_c^* \to \bar{B}^*}$ & $2.68^{+0.08+0.19}_{-0.09-0.23}$ & $-9.82^{+0.40+0.91}_{-0.38-0.96}$ &
$V_6^{\bar{B}_c^* \to \bar{B}_s^*}$ & $3.04^{+0.04+0.23}_{-0.05-0.32}$ & $-13.40^{+0.52+1.44}_{-0.51-1.41}$ &
$V_6^{\bar{B}_c^* \to \bar{D}^*}$ & $0.10^{+0.01+0.01}_{-0.01-0.01}$ & $-0.63^{+0.05+0.06}_{-0.05-0.05}$ \\
\hline
$A_1^{\bar{B}_c^* \to \psi(1S)}$ & $0.54^{+0.00+0.07}_{-0.01-0.07}$ & $-3.67^{+0.05+0.52}_{-0.05-0.37}$ &
$A_1^{\bar{B}_c^* \to \psi(2S)}$ & $0.26^{+0.00+0.02}_{-0.00-0.03}$ & $0.97^{+0.10+0.12}_{-0.10-0.13}$ &
$A_1^{\bar{B}_c^* \to \psi(3S)}$ & $0.14^{+0.01+0.01}_{-0.01-0.01}$ & $1.12^{+0.11+0.13}_{-0.10-0.10}$ \\

$A_2^{\bar{B}_c^* \to \psi(1S)}$ & $0.32^{+0.01+0.02}_{-0.01-0.03}$ & $-2.43^{+0.09+0.28}_{-0.09-0.29}$ &
$A_2^{\bar{B}_c^* \to \psi(2S)}$ & $0.32^{+0.00+0.03}_{-0.00-0.02}$ & $-1.71^{+0.02+0.12}_{-0.02-0.15}$ &
$A_2^{\bar{B}_c^* \to \psi(3S)}$ & $0.22^{+0.00+0.02}_{-0.00-0.01}$ & $-0.39^{+0.04+0.05}_{-0.04-0.05}$ \\

$A_3^{\bar{B}_c^* \to \psi(1S)}$ & $0.13^{+0.00+0.01}_{-0.00-0.01}$ & $-1.14^{+0.03+0.13}_{-0.03-0.16}$ &
$A_3^{\bar{B}_c^* \to \psi(2S)}$ & $0.12^{+0.00+0.01}_{-0.00-0.01}$ & $-0.74^{+0.02+0.10}_{-0.01-0.09}$ &
$A_3^{\bar{B}_c^* \to \psi(3S)}$ & $0.06^{+0.00+0.01}_{-0.00-0.01}$ & $-0.09^{+0.01+0.01}_{-0.01-0.01}$ \\

$A_4^{\bar{B}_c^* \to \psi(1S)}$ & $0.14^{+0.00+0.01}_{-0.00-0.01}$ & $-1.23^{+0.02+0.15}_{-0.01-0.13}$ &
$A_4^{\bar{B}_c^* \to \psi(2S)}$ & $0.02^{+0.00+0.00}_{-0.00-0.00}$ & $0.47^{+0.05+0.04}_{-0.05-0.05}$ &
$A_4^{\bar{B}_c^* \to \psi(3S)}$ & $0.00^{+0.00+0.00}_{-0.00-0.00}$ & $0.26^{+0.03+0.03}_{-0.03-0.02}$ \\

$V_1^{\bar{B}_c^* \to \psi(1S)}$ & $0.56^{+0.01+0.08}_{-0.01-0.07}$ & $-3.83^{+0.05+0.47}_{-0.05-0.48}$ &
$V_1^{\bar{B}_c^* \to \psi(2S)}$ & $0.28^{+0.00+0.02}_{-0.00-0.03}$ & $0.86^{+0.10+0.10}_{-0.10-0.09}$ &
$V_1^{\bar{B}_c^* \to \psi(3S)}$ & $0.15^{+0.01+0.01}_{-0.01-0.01}$ & $1.11^{+0.11+0.14}_{-0.11-0.08}$ \\

$V_2^{\bar{B}_c^* \to \psi(1S)}$ & $0.33^{+0.01+0.02}_{-0.01-0.03}$ & $-2.35^{+0.09+0.27}_{-0.09-0.28}$ &
$V_2^{\bar{B}_c^* \to \psi(2S)}$ & $0.32^{+0.00+0.03}_{-0.00-0.02}$ & $-1.58^{+0.03+0.15}_{-0.03-0.16}$ &
$V_2^{\bar{B}_c^* \to \psi(3S)}$ & $0.22^{+0.00+0.02}_{-0.00-0.01}$ & $-0.27^{+0.04+0.03}_{-0.03-0.03}$ \\

$V_3^{\bar{B}_c^* \to \psi(1S)}$ & $0.20^{+0.00+0.01}_{-0.00-0.02}$ & $-1.55^{+0.02+0.19}_{-0.02-0.16}$ &
$V_3^{\bar{B}_c^* \to \psi(2S)}$ & $0.08^{+0.00+0.01}_{-0.00-0.01}$ & $0.18^{+0.02+0.02}_{-0.02-0.02}$ &
$V_3^{\bar{B}_c^* \to \psi(3S)}$ & $0.04^{+0.00+0.01}_{-0.00-0.00}$ & $0.11^{+0.01+0.01}_{-0.01-0.01}$ \\

$V_4^{\bar{B}_c^* \to \psi(1S)}$ & $-0.01^{+0.00+0.00}_{-0.00-0.00}$ & $0.12^{+0.01+0.01}_{-0.02-0.01}$ &
$V_4^{\bar{B}_c^* \to \psi(2S)}$ & $-0.11^{+0.00+0.01}_{-0.00-0.01}$ & $1.34^{+0.03+0.13}_{-0.03-0.18}$ &
$V_4^{\bar{B}_c^* \to \psi(3S)}$ & $-0.07^{+0.00+0.01}_{-0.00-0.01}$ & $0.41^{+0.03+0.05}_{-0.03-0.03}$ \\

$V_5^{\bar{B}_c^* \to \psi(1S)}$ & $1.16^{+0.01+0.15}_{-0.01-0.15}$ & $-7.73^{+0.12+0.71}_{-0.11-0.81}$ &
$V_5^{\bar{B}_c^* \to \psi(2S)}$ & $0.56^{+0.00+0.05}_{-0.00-0.05}$ & $1.82^{+0.20+0.22}_{-0.20-0.23}$ &
$V_5^{\bar{B}_c^* \to \psi(3S)}$ & $0.31^{+0.01+0.02}_{-0.01-0.03}$ & $2.01^{+0.22+0.23}_{-0.20-0.19}$ \\

$V_6^{\bar{B}_c^* \to \psi(1S)}$ & $0.64^{+0.01+0.08}_{-0.01-0.08}$ & $-4.08^{+0.08+0.51}_{-0.05-0.54}$ &
$V_6^{\bar{B}_c^* \to \psi(2S)}$ & $0.38^{+0.00+0.02}_{-0.00-0.03}$ & $0.86^{+0.12+0.11}_{-0.10-0.11}$ &
$V_6^{\bar{B}_c^* \to \psi(3S)}$ & $0.23^{+0.01+0.01}_{-0.01-0.01}$ & $1.50^{+0.15+0.21}_{-0.14-0.16}$ \\
\hline\hline
\end{tabular}}
\vspace{-0.8em}
\end{center}
\label{tab:formfactor}
\end{table}

\begin{figure*}[htbp]
\begin{center}
\vspace{-0.8cm}
\includegraphics[scale=0.36]{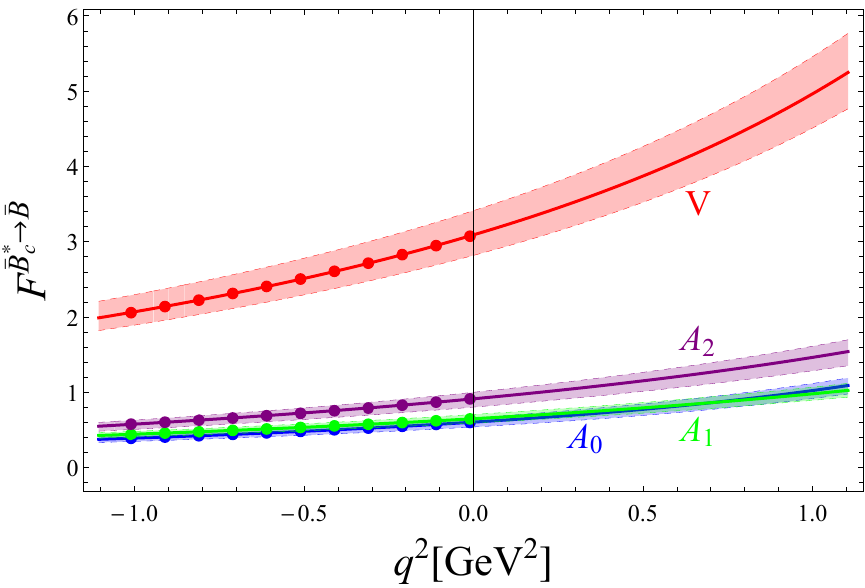}
\includegraphics[scale=0.36]{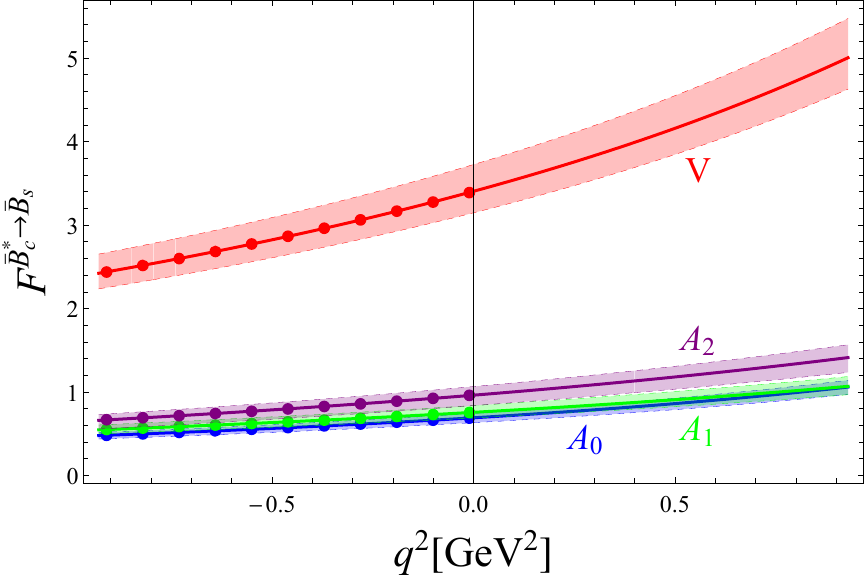}
\includegraphics[scale=0.36]{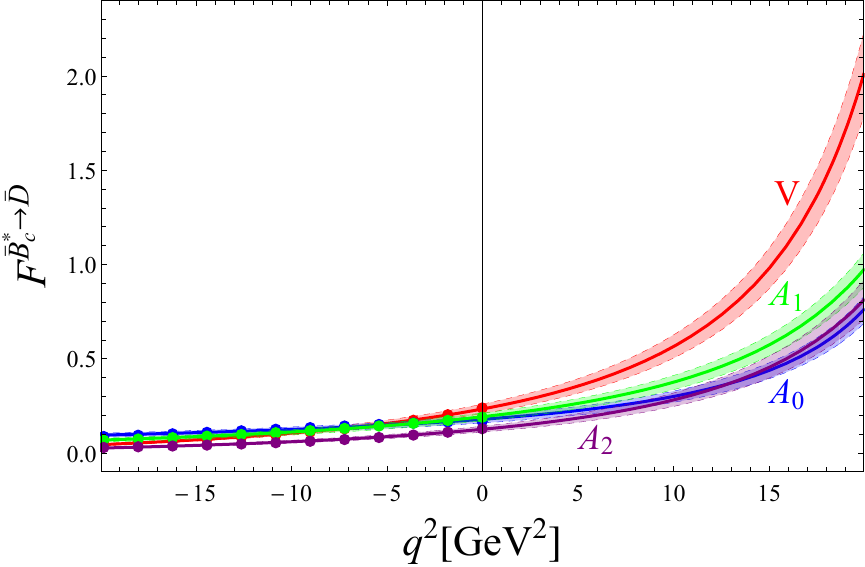}\\
\includegraphics[scale=0.36]{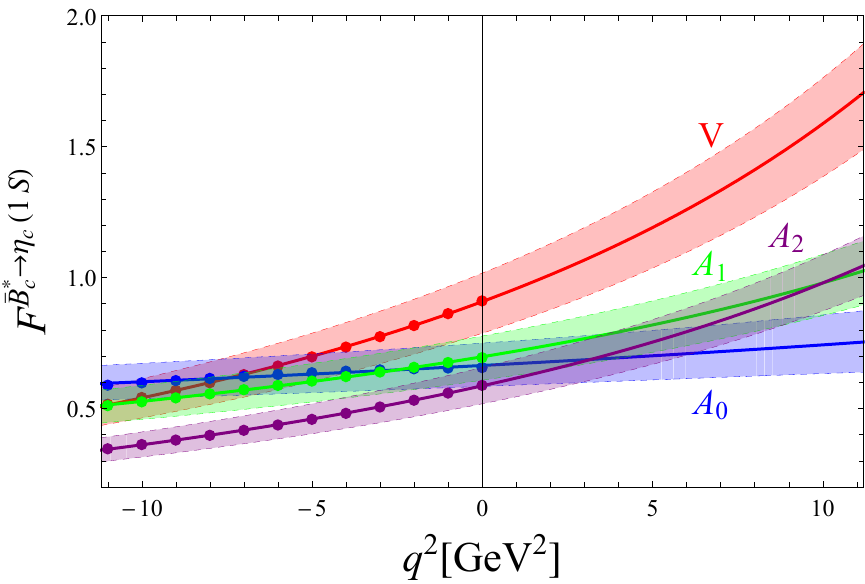}
\includegraphics[scale=0.36]{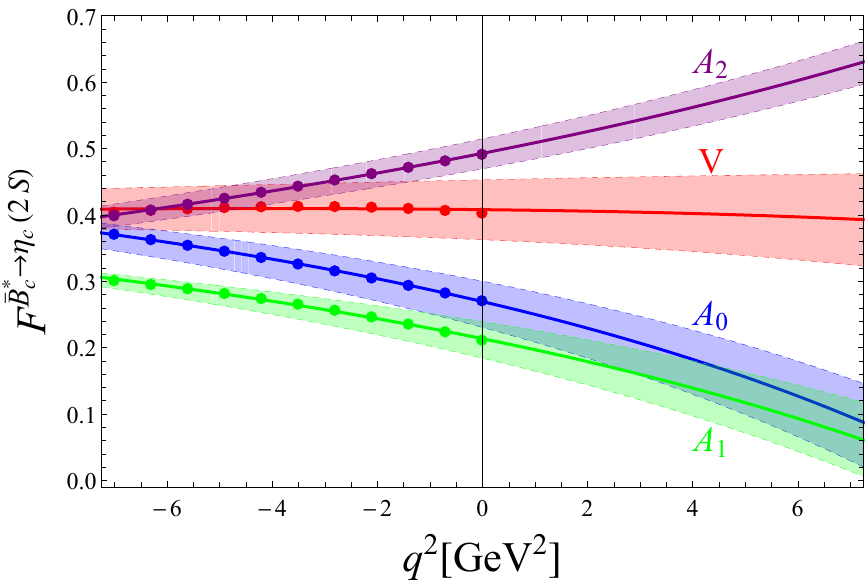}
\includegraphics[scale=0.36]{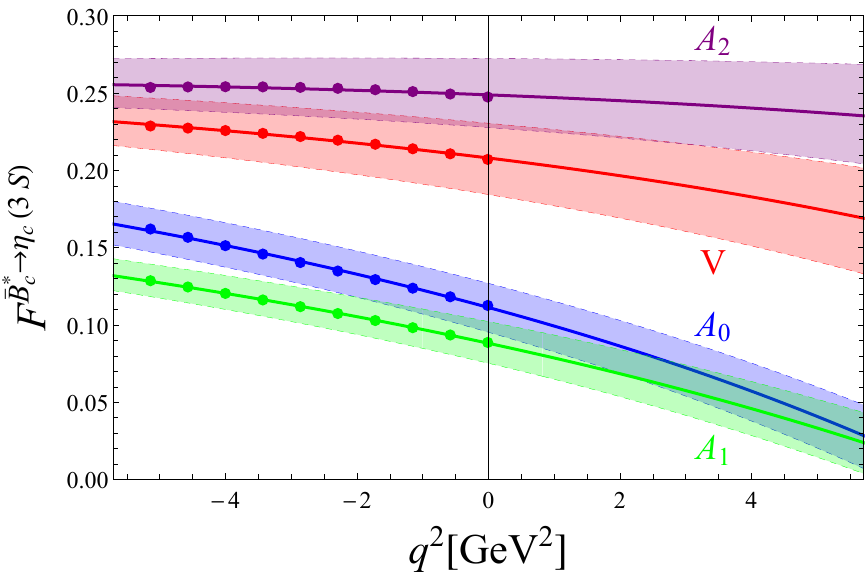}\\
\includegraphics[scale=0.36]{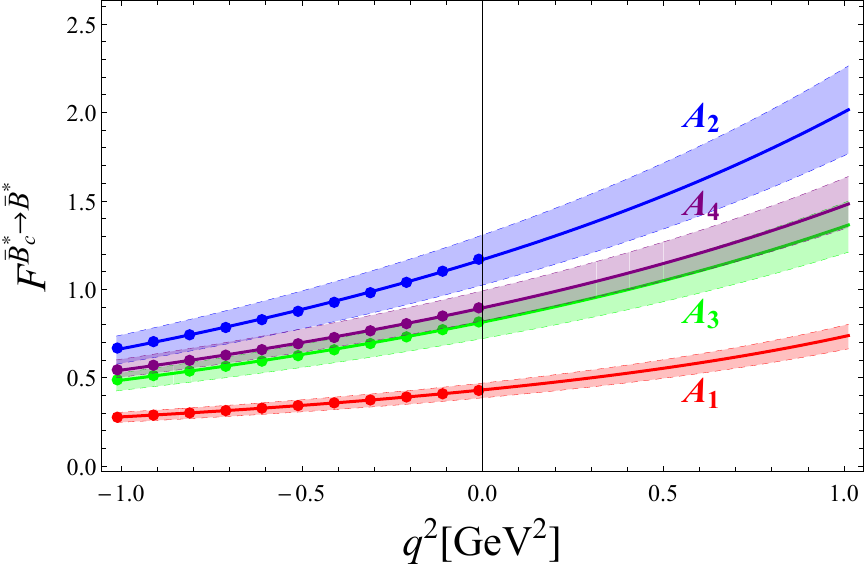}
\includegraphics[scale=0.36]{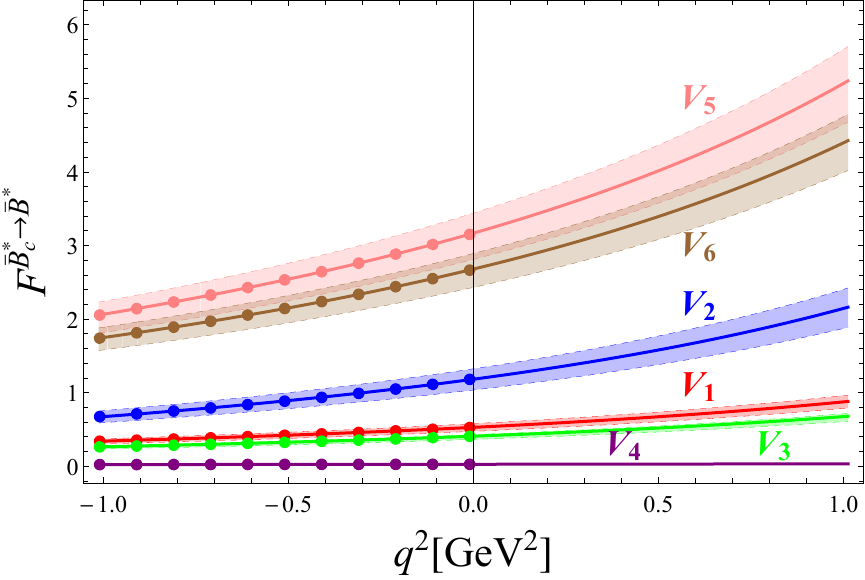}
\includegraphics[scale=0.36]{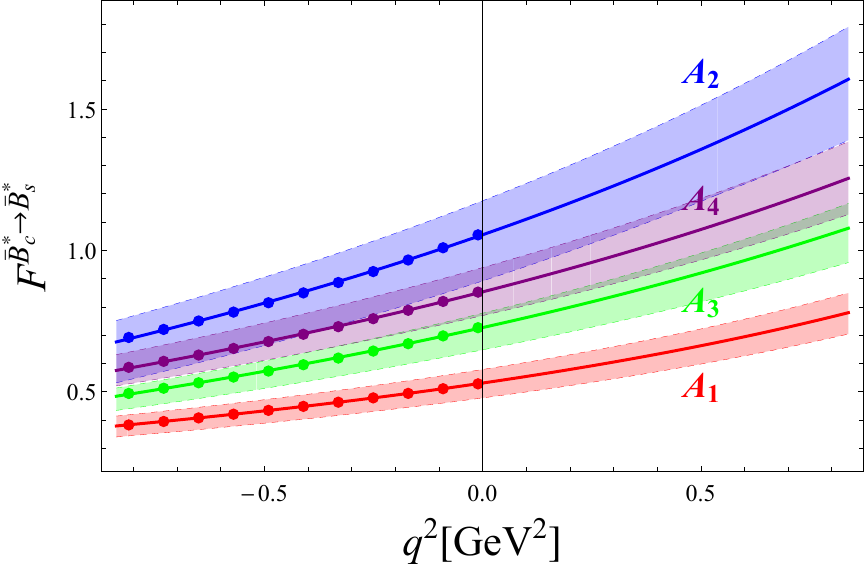}\\
\includegraphics[scale=0.36]{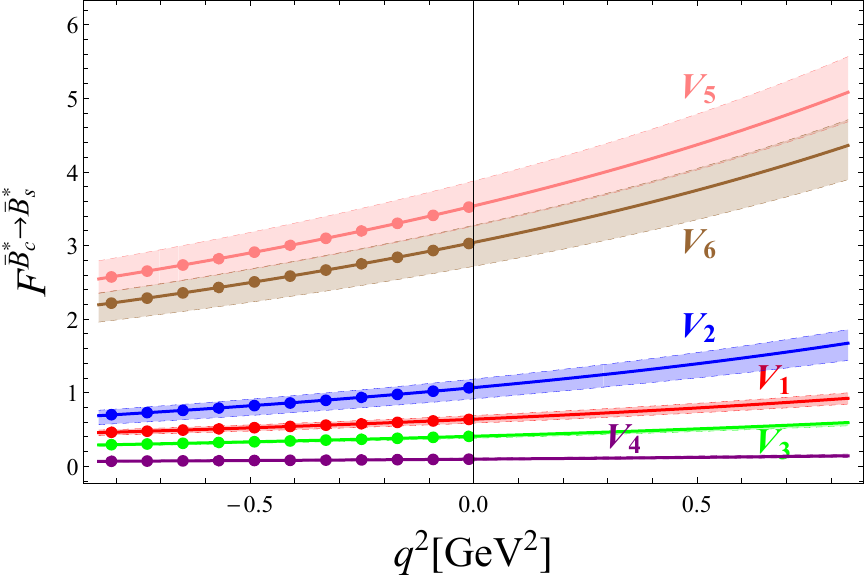}
\includegraphics[scale=0.36]{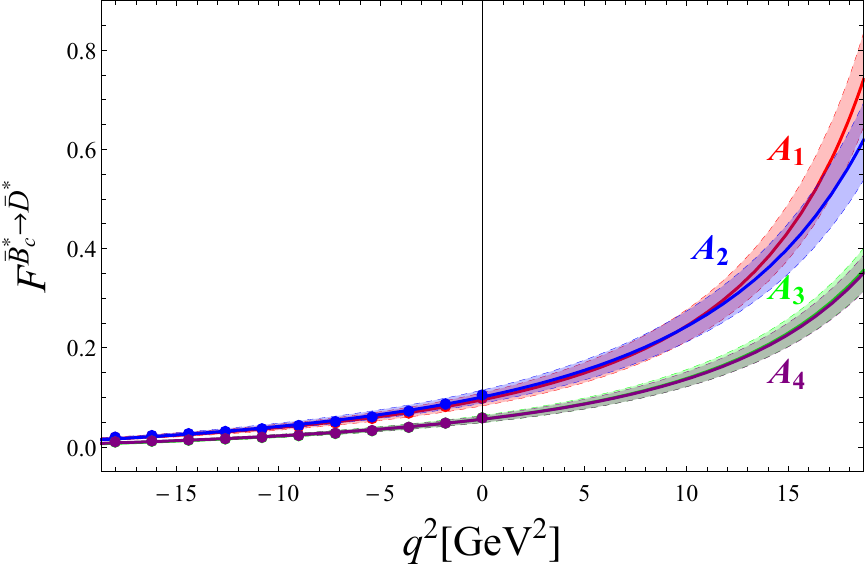}
\includegraphics[scale=0.36]{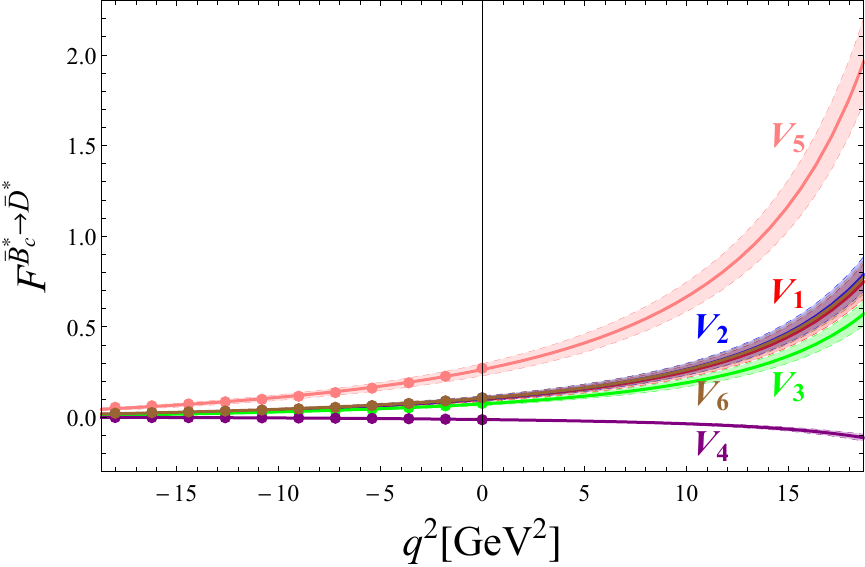}\\
\includegraphics[scale=0.36]{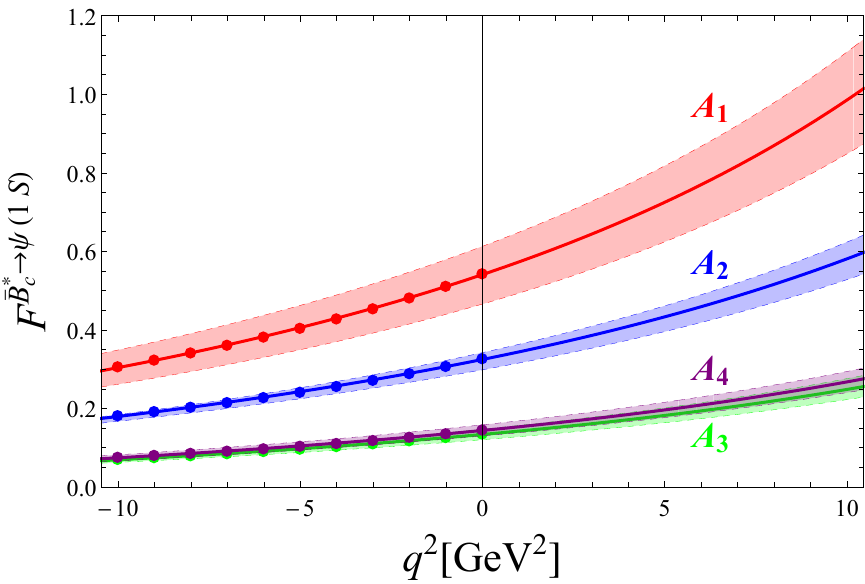}
\includegraphics[scale=0.36]{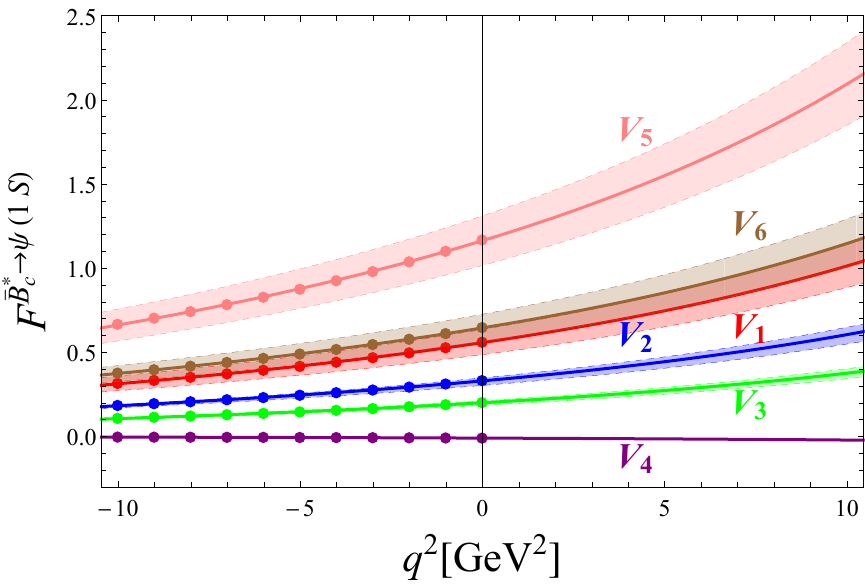}
\includegraphics[scale=0.36]{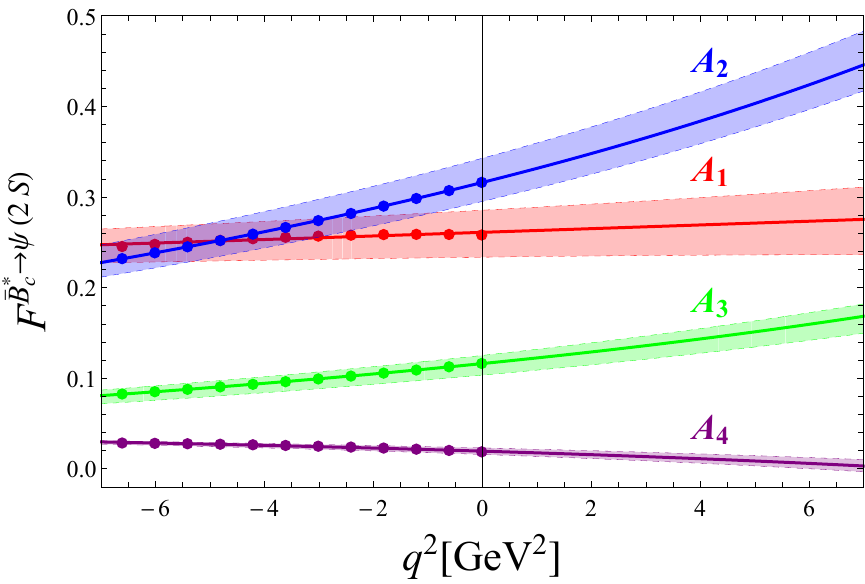}\\
\includegraphics[scale=0.36]{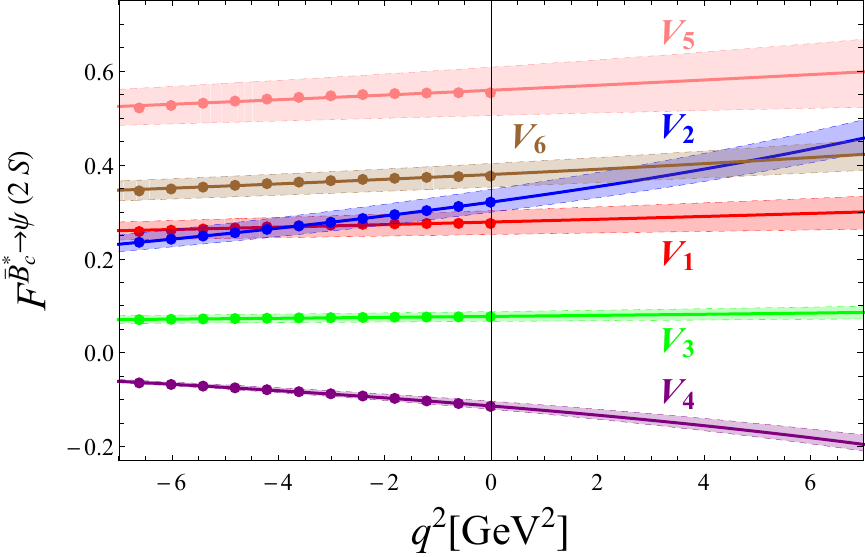}
\includegraphics[scale=0.36]{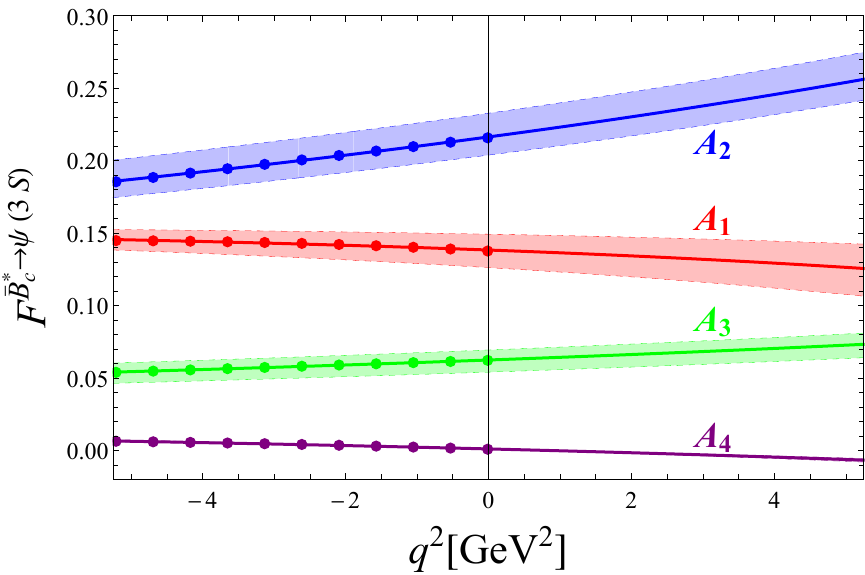}
\includegraphics[scale=0.36]{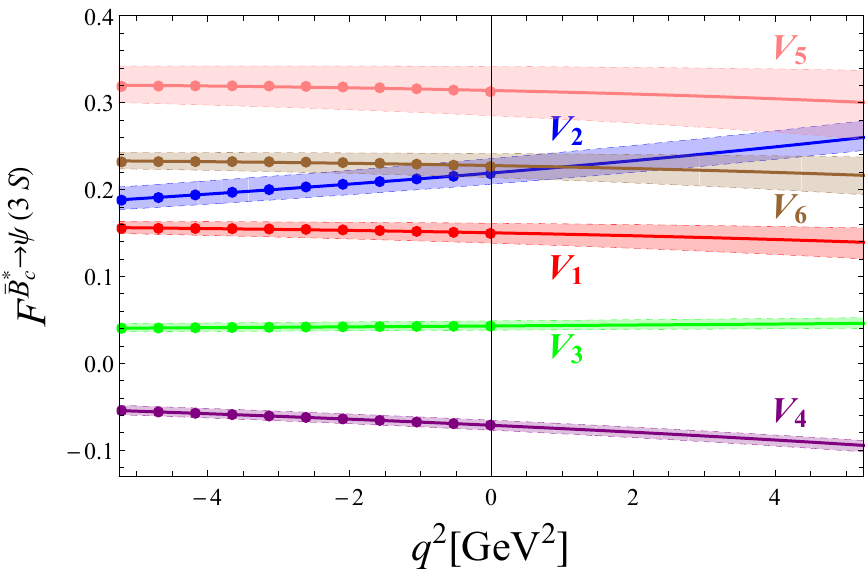}
\end{center}
\caption{The $q^2$ dependences of form factors of $\bar{B}^*_c \to (P,V)$ transitions. {The error bands reflect the uncertainties from the Gaussian parameters $\beta$ and the constituent quark masses.}}
\label{fig:FFBstar2V}
\end{figure*}

Using the values of constituent quark masses and Gaussian parameters fitted by $f_{P,V}$ in our previous studies\cite{Chang:2019obq,Chang:2020wvs}, we present our theoretical predictions for the form factors of $c \to (d,\,s)$ induced $\bar{B}^*_{c} \to \bar{B}^{(*)}_d$, $\bar{B}^{(*)}_s$ transitions and $b \to (u,\,c)$ induced $\bar{B}^*_{c} \to \bar{D}^{(*)}$, $\eta_c(1S,\,2S,\,3S)$, $\psi(1S,\,2S,\,3S)$ transitions in Table  \ref{tab:formfactor}. The $q^2$ dependences of form factors in whole region, $-(m_{B^*_c}-m_f)^2 \leq q^2 \leq (m_{B^*_c}-m_f)^2$, are shown in Fig.~\ref{fig:FFBstar2V}.

\begin{table}[!htb]
\vspace{-0.8cm}
\caption{{The multiplicative helicity-basis form factors $\sqrt{q^2}\,H_{\lambda_{W^*}\lambda_{B_c^*}}$  (in units of GeV$^2$) for the $\bar{B}^*_c \to \bar B_s$ transition at ten equally spaced interior points of the physical $q^2$ region. The uncertainties are from the Gaussian parameters $\beta$ and the constituent quark masses.}}
\label{tab:MHFF_Bcstar2Bs_errorbars}
\centering
\renewcommand{\arraystretch}{0.87}  
\setlength{\tabcolsep}{8pt}  
{\begin{tabular}{ccccc}
\hline\hline
$q^2\,(\mathrm{GeV}^2)$ & $\sqrt{q^2}H_{t0}(q^2)$ & $\sqrt{q^2}H_{00}(q^2)$ & $\sqrt{q^2}H_{-+}(q^2)$ & $\sqrt{q^2}H_{+-}(q^2)$ \\
\hline
0.08 & $7.68^{+0.59}_{-0.65}$ & $8.94^{+0.82}_{-0.77}$ & $-1.71^{+0.17}_{-0.20}$ & $-3.59^{+0.31}_{-0.37}$ \\
0.17 & $7.55^{+0.58}_{-0.64}$ & $9.20^{+0.87}_{-0.80}$ & $-2.56^{+0.26}_{-0.30}$ & $-5.16^{+0.45}_{-0.54}$ \\
0.25 & $7.39^{+0.57}_{-0.62}$ & $9.47^{+0.91}_{-0.83}$ & $-3.32^{+0.34}_{-0.39}$ & $-6.43^{+0.56}_{-0.67}$ \\
0.34 & $7.17^{+0.55}_{-0.60}$ & $9.76^{+0.95}_{-0.87}$ & $-4.07^{+0.41}_{-0.48}$ & $-7.54^{+0.66}_{-0.79}$ \\
0.42 & $6.89^{+0.53}_{-0.58}$ & $10.05^{+1.00}_{-0.90}$ & $-4.83^{+0.49}_{-0.56}$ & $-8.55^{+0.75}_{-0.90}$ \\
0.51 & $6.54^{+0.50}_{-0.55}$ & $10.35^{+1.05}_{-0.93}$ & $-5.64^{+0.56}_{-0.65}$ & $-9.49^{+0.84}_{-1.00}$ \\
0.59 & $6.08^{+0.47}_{-0.51}$ & $10.66^{+1.10}_{-0.97}$ & $-6.51^{+0.65}_{-0.75}$ & $-10.36^{+0.93}_{-1.10}$ \\
0.68 & $5.48^{+0.42}_{-0.46}$ & $10.98^{+1.15}_{-1.01}$ & $-7.46^{+0.74}_{-0.85}$ & $-11.15^{+1.00}_{-1.19}$ \\
0.76 & $4.66^{+0.36}_{-0.39}$ & $11.31^{+1.20}_{-1.05}$ & $-8.53^{+0.84}_{-0.97}$ & $-11.85^{+1.08}_{-1.27}$ \\
0.84 & $3.43^{+0.26}_{-0.29}$ & $11.66^{+1.26}_{-1.09}$ & $-9.81^{+0.95}_{-1.10}$ & $-12.38^{+1.14}_{-1.33}$ \\
\hline\hline
\end{tabular}}
\end{table}

\begin{figure*}[!htb]
\vspace{0.2cm}
\begin{center}
\includegraphics[scale=0.47]{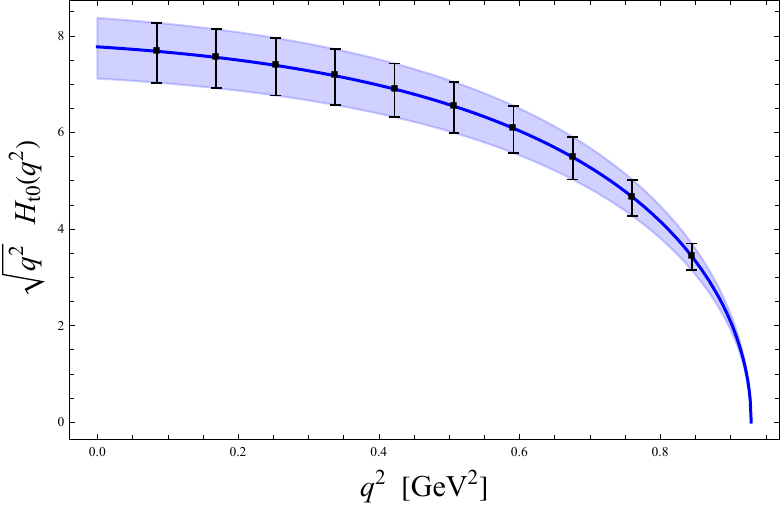}\qquad
\includegraphics[scale=0.47]{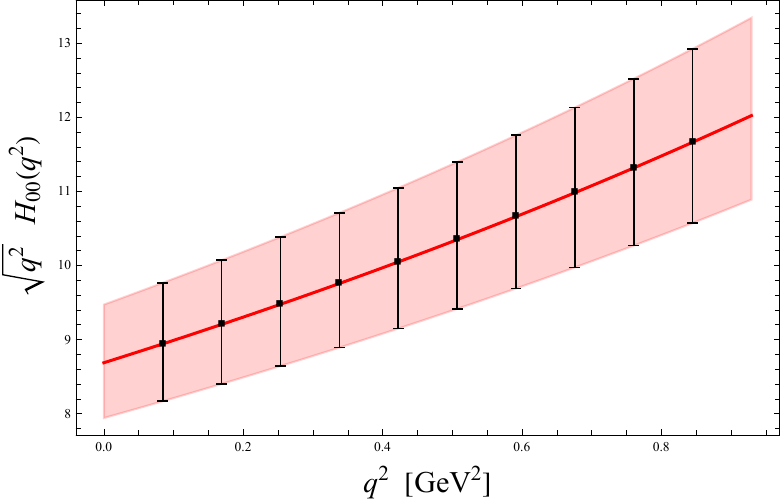}\\
\includegraphics[scale=0.47]{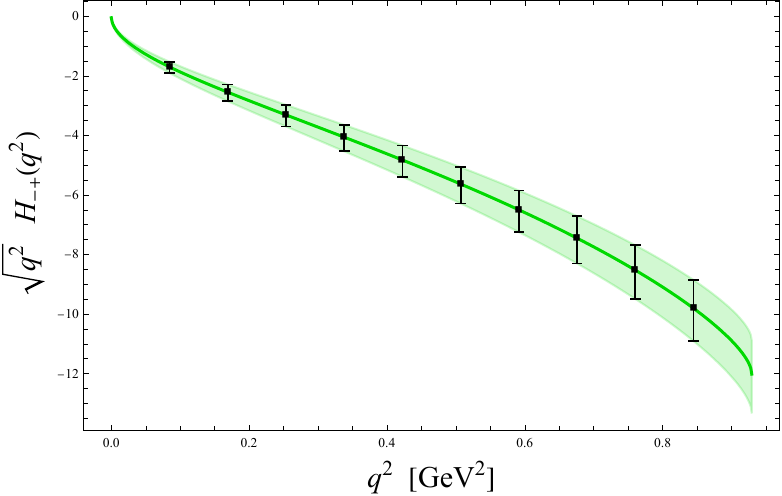}\qquad
\includegraphics[scale=0.47]{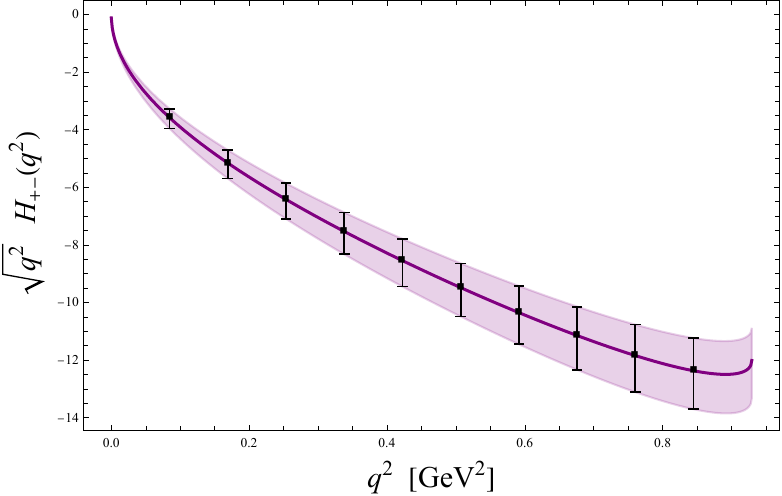}\\
\end{center}
\caption{{The $q^2$ dependences of the helicity-basis form factors $\sqrt{q^2} H_{\lambda_{W^*}\lambda_{B_c^*}\lambda_{B_s^{(*)}}}$ for $\bar{B}^*_c \to \bar B_s$ decays. The multiplicative factor $\sqrt{q^2}$ removes the kinematic singularities near $q^2=0$. The data points with black error bars correspond to the numerical values listed in Table~\ref{tab:MHFF_Bcstar2Bs_errorbars}, and the error bands reflect the uncertainties from the Gaussian parameters $\beta$ and the constituent quark masses.}}
\label{fig:HTFFPs}
\end{figure*}

\begin{table}[htbp]
\caption{{The multiplicative helicity-basis form factors $\sqrt{q^2}\,H_{\lambda_{W^*}\lambda_{B_c^*}\lambda_{B_s^*}}$ (in units of GeV$^2$) for the $\bar{B}^*_c \to \bar B_s^*$ transition at ten equally spaced interior points of the physical $q^2$ region. The uncertainties are from the Gaussian parameters $\beta$ and the constituent quark masses.}}
\label{tab:MHFF_Bcstar2Bsstar_errorbars}
\centering
{\begin{tabular}{cccccc}
\hline\hline
$q^2\,(\mathrm{GeV}^2)$ & $\sqrt{q^2}H_{000}(q^2)$ & $\sqrt{q^2}H_{t++}(q^2)$ & $\sqrt{q^2}H_{t--}(q^2)$ & $\sqrt{q^2}H_{t00}(q^2)$ & $\sqrt{q^2}H_{0++}(q^2)$ \\
\hline
0.08 & $6.73{}^{+0.46}_{-0.63}$ & $1.36{}^{+0.07}_{-0.05}$ & $12.59{}^{+1.07}_{-1.15}$ & $6.98{}^{+0.50}_{-0.65}$ & $0.92{}^{+0.05}_{-0.05}$ \\
0.15 & $6.60{}^{+0.46}_{-0.62}$ & $1.61{}^{+0.11}_{-0.07}$ & $12.61{}^{+1.09}_{-1.15}$ & $7.12{}^{+0.53}_{-0.66}$ & $0.68{}^{+0.07}_{-0.07}$ \\
0.23 & $6.42{}^{+0.45}_{-0.61}$ & $1.89{}^{+0.15}_{-0.10}$ & $12.62{}^{+1.11}_{-1.15}$ & $7.26{}^{+0.57}_{-0.68}$ & $0.39{}^{+0.10}_{-0.10}$ \\
0.30 & $6.21{}^{+0.44}_{-0.59}$ & $2.21{}^{+0.19}_{-0.14}$ & $12.59{}^{+1.12}_{-1.15}$ & $7.40{}^{+0.60}_{-0.69}$ & $0.07{}^{+0.13}_{-0.15}$ \\
0.38 & $5.94{}^{+0.43}_{-0.56}$ & $2.58{}^{+0.25}_{-0.18}$ & $12.52{}^{+1.13}_{-1.15}$ & $7.55{}^{+0.64}_{-0.70}$ & $-0.32{}^{+0.17}_{-0.19}$ \\
0.46 & $5.61{}^{+0.41}_{-0.53}$ & $3.00{}^{+0.30}_{-0.23}$ & $12.40{}^{+1.14}_{-1.14}$ & $7.70{}^{+0.68}_{-0.71}$ & $-0.77{}^{+0.22}_{-0.25}$ \\
0.53 & $5.19{}^{+0.38}_{-0.49}$ & $3.50{}^{+0.37}_{-0.28}$ & $12.20{}^{+1.14}_{-1.12}$ & $7.85{}^{+0.72}_{-0.73}$ & $-1.31{}^{+0.27}_{-0.31}$ \\
0.61 & $4.65{}^{+0.34}_{-0.44}$ & $4.10{}^{+0.45}_{-0.35}$ & $11.91{}^{+1.14}_{-1.10}$ & $8.01{}^{+0.77}_{-0.74}$ & $-1.98{}^{+0.33}_{-0.39}$ \\
0.69 & $3.93{}^{+0.29}_{-0.38}$ & $4.86{}^{+0.54}_{-0.43}$ & $11.47{}^{+1.12}_{-1.06}$ & $8.17{}^{+0.81}_{-0.76}$ & $-2.83{}^{+0.42}_{-0.48}$ \\
0.76 & $2.87{}^{+0.22}_{-0.28}$ & $5.90{}^{+0.65}_{-0.54}$ & $10.75{}^{+1.08}_{-1.00}$ & $8.33{}^{+0.86}_{-0.77}$ & $-4.02{}^{+0.52}_{-0.59}$ \\
\hline
$q^2\,(\mathrm{GeV}^2)$ & $\sqrt{q^2}H_{-+0}(q^2)$ & $\sqrt{q^2}H_{-0-}(q^2)$ & $\sqrt{q^2}H_{0--}(q^2)$ & $\sqrt{q^2}H_{+-0}(q^2)$ & $\sqrt{q^2}H_{+0+}(q^2)$ \\
\hline
0.08 & $-2.50{}^{+0.26}_{-0.25}$ & $-2.53{}^{+0.24}_{-0.28}$ & $12.53{}^{+1.07}_{-1.15}$ & $0.86{}^{+0.13}_{-0.10}$ & $0.90{}^{+0.13}_{-0.13}$ \\
0.15 & $-3.55{}^{+0.37}_{-0.37}$ & $-3.60{}^{+0.35}_{-0.40}$ & $12.50{}^{+1.09}_{-1.15}$ & $1.29{}^{+0.20}_{-0.15}$ & $1.34{}^{+0.20}_{-0.19}$ \\
0.23 & $-4.38{}^{+0.46}_{-0.45}$ & $-4.43{}^{+0.43}_{-0.49}$ & $12.44{}^{+1.11}_{-1.15}$ & $1.68{}^{+0.26}_{-0.20}$ & $1.74{}^{+0.25}_{-0.24}$ \\
0.30 & $-5.09{}^{+0.54}_{-0.53}$ & $-5.14{}^{+0.50}_{-0.57}$ & $12.34{}^{+1.12}_{-1.15}$ & $2.07{}^{+0.31}_{-0.24}$ & $2.14{}^{+0.31}_{-0.29}$ \\
0.38 & $-5.70{}^{+0.60}_{-0.60}$ & $-5.76{}^{+0.57}_{-0.64}$ & $12.18{}^{+1.13}_{-1.15}$ & $2.48{}^{+0.37}_{-0.29}$ & $2.55{}^{+0.36}_{-0.34}$ \\
0.46 & $-6.25{}^{+0.66}_{-0.67}$ & $-6.30{}^{+0.63}_{-0.71}$ & $11.97{}^{+1.13}_{-1.14}$ & $2.93{}^{+0.42}_{-0.34}$ & $2.99{}^{+0.42}_{-0.39}$ \\
0.53 & $-6.74{}^{+0.72}_{-0.73}$ & $-6.78{}^{+0.68}_{-0.77}$ & $11.67{}^{+1.13}_{-1.12}$ & $3.42{}^{+0.48}_{-0.40}$ & $3.47{}^{+0.48}_{-0.45}$ \\
0.61 & $-7.15{}^{+0.76}_{-0.79}$ & $-7.18{}^{+0.73}_{-0.82}$ & $11.26{}^{+1.12}_{-1.10}$ & $3.97{}^{+0.55}_{-0.46}$ & $4.02{}^{+0.54}_{-0.51}$ \\
0.69 & $-7.46{}^{+0.80}_{-0.84}$ & $-7.49{}^{+0.77}_{-0.86}$ & $10.67{}^{+1.10}_{-1.06}$ & $4.62{}^{+0.62}_{-0.53}$ & $4.66{}^{+0.62}_{-0.57}$ \\
0.76 & $-7.63{}^{+0.83}_{-0.87}$ & $-7.64{}^{+0.80}_{-0.89}$ & $9.76{}^{+1.05}_{-1.00}$ & $5.44{}^{+0.71}_{-0.62}$ & $5.46{}^{+0.70}_{-0.65}$ \\
\hline\hline
\end{tabular}}

\end{table}

\begin{figure*}[htbp]
\vspace{-0.9cm}
\begin{center}
\includegraphics[scale=0.47]{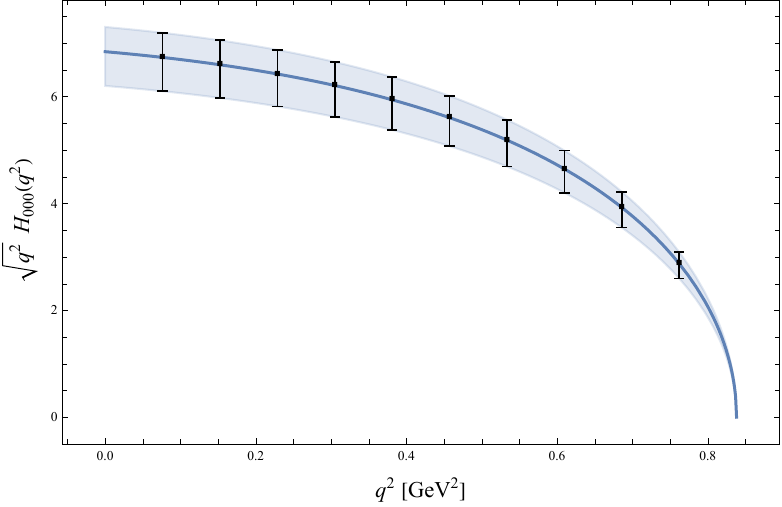}\qquad
\includegraphics[scale=0.47]{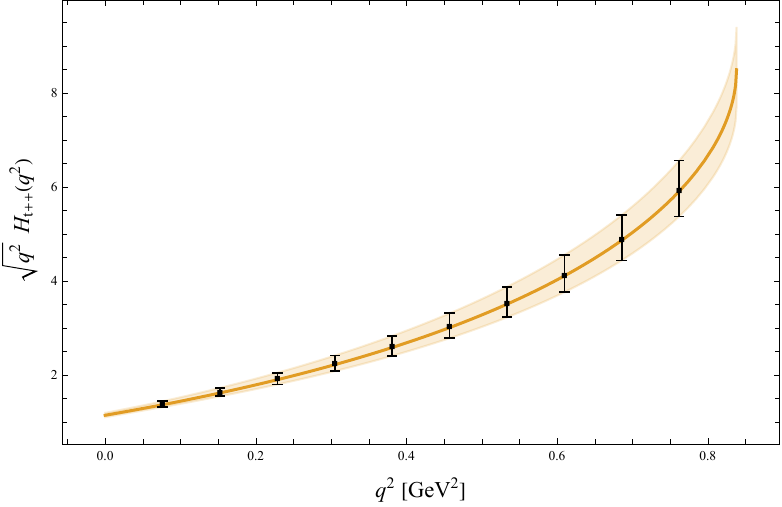}\\
\includegraphics[scale=0.47]{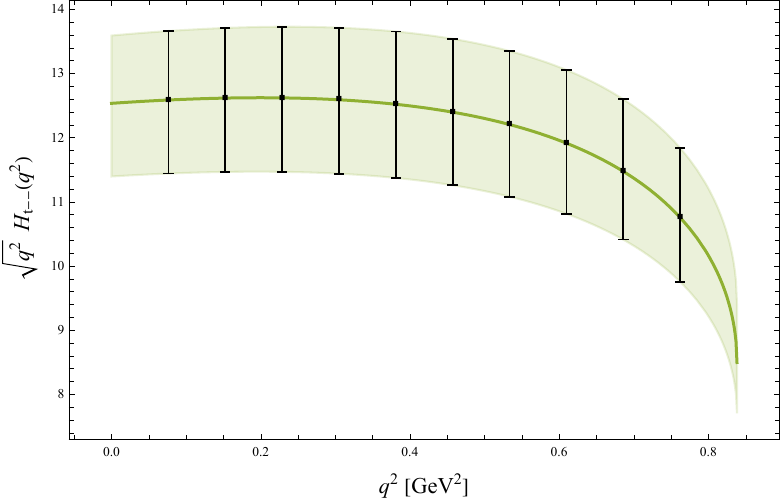}\qquad
\includegraphics[scale=0.47]{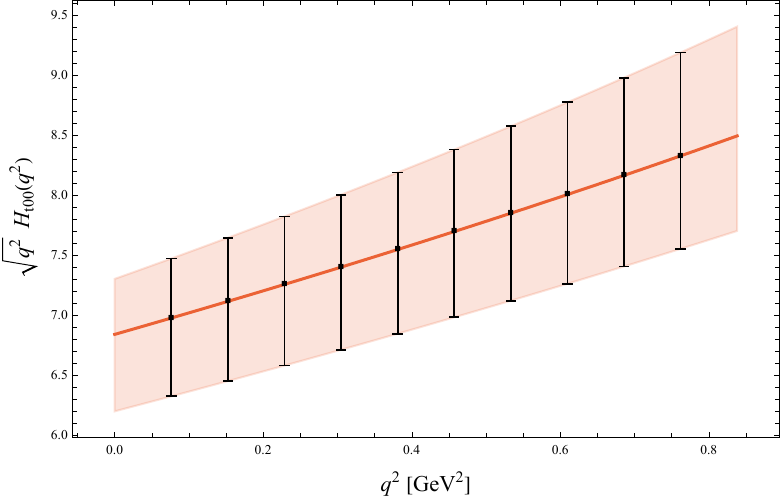}\\
\includegraphics[scale=0.47]{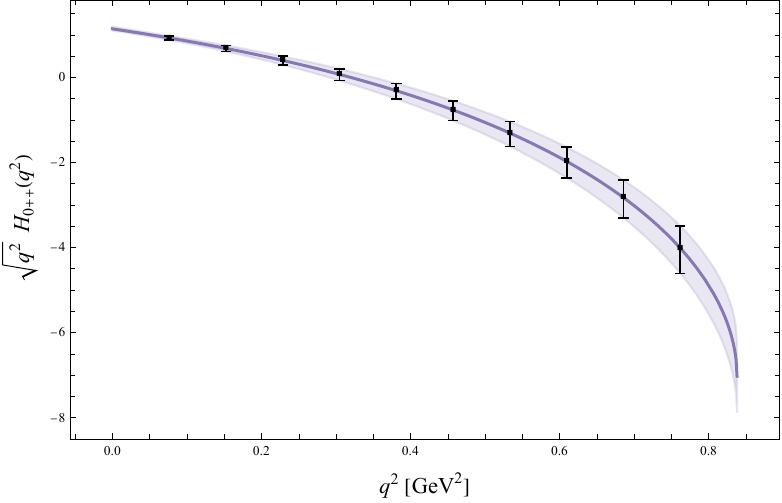}\qquad
\includegraphics[scale=0.47]{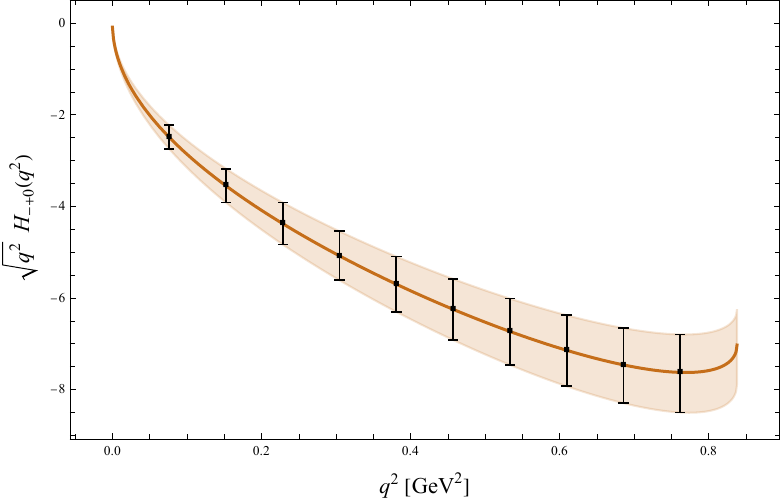}\\
\includegraphics[scale=0.47]{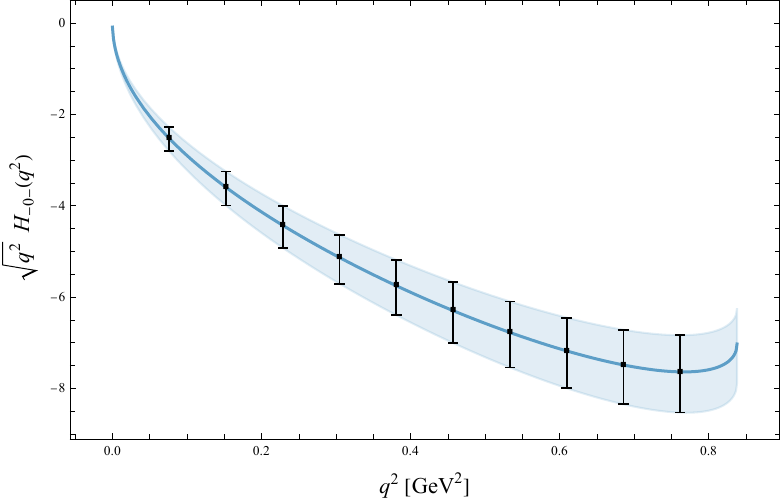}\qquad
\includegraphics[scale=0.47]{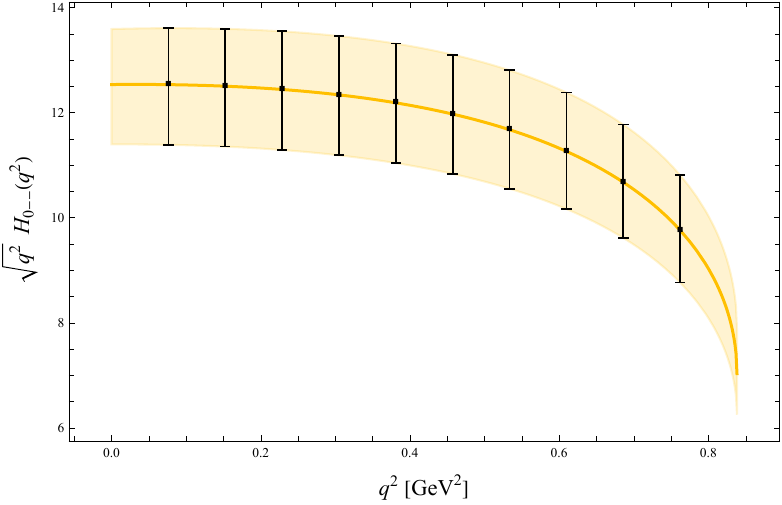}\\
\includegraphics[scale=0.47]{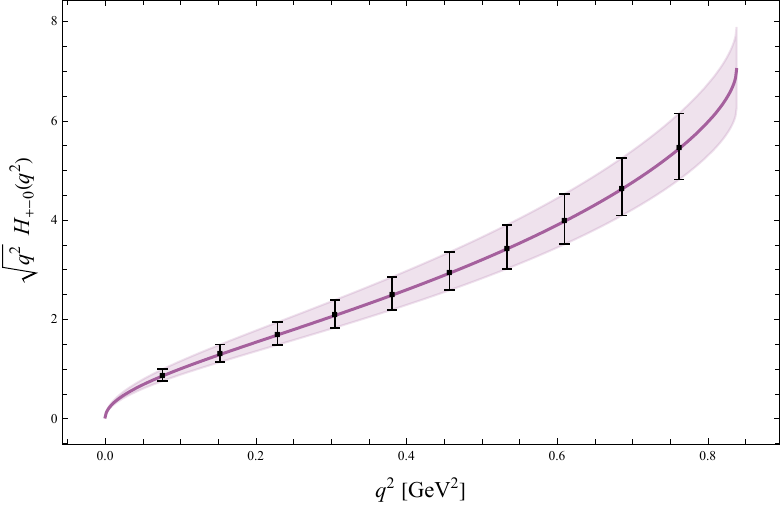}\qquad
\includegraphics[scale=0.47]{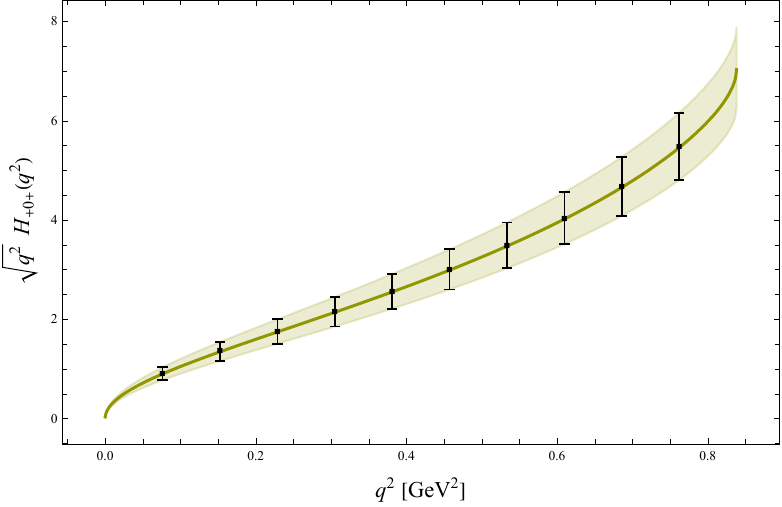}\\
\end{center}
\vspace{-0.5cm}
\caption{{The $q^2$ dependences of the helicity-basis form factors $\sqrt{q^2} H_{\lambda_{W^*}\lambda_{B_c^*}\lambda_{B_s^{(*)}}}$ for $\bar{B}^*_c \to \bar B_s^*$ decays. The multiplicative factor $\sqrt{q^2}$ removes the kinematic singularities near $q^2=0$. The data points with black error bars correspond to the numerical values listed in Table~\ref{tab:MHFF_Bcstar2Bsstar_errorbars}, and the error bands reflect the uncertainties from the Gaussian parameters $\beta$ and the constituent quark masses.}}
\label{fig:HTFFVs}
\end{figure*}

{Besides the invariant form factors listed in Table~3 and shown in Fig.~1, the helicity-basis form factors are also useful for analyzing the helicity structure of these semileptonic decays. Using the relations given in
Eqs.~\eqref{eq:Ht0}--\eqref{eq:Ht00}, we obtain them from the BCL-parametrized invariant form factors. For conciseness, we present only the results for the Cabibbo-favored
\(\bar B_c^*\to\bar B_s\ell^{\prime-}\bar\nu_{\ell^{\prime}}\) and
\(\bar B_c^*\to\bar B_s^*\ell^{\prime-}\bar\nu_{\ell^{\prime}}\) modes, which have relatively
large branching fractions among the channels considered in this work (see below). The $q^2$ dependences of the helicity-basis form factors $\sqrt{q^2}\,H_{\lambda_{W^*}\lambda_{B_c^*}\lambda_{B_s^{(*)}}}$ are shown in Figs.~\ref{fig:HTFFPs} and~\ref{fig:HTFFVs}, with the error bars corresponding to the numerical values in Tables~\ref{tab:MHFF_Bcstar2Bs_errorbars} and~\ref{tab:MHFF_Bcstar2Bsstar_errorbars}, respectively.
The factor \(\sqrt{q^2}\) is introduced to remove the kinematic singularities at $q^2=0$. The uncertainties are obtained by propagating the uncertainties in the Gaussian parameters $\beta$ and the constituent quark masses.}

{A closer inspection of these helicity-basis form factors reveals the following features. For the $\bar B_c^* \to \bar B_s$ transition, $H_{00}$ and $H_{t0}$ dominate at low $q^2$, while $H_{\mp\pm}$ become significant only near the endpoint, where $H_{t0}$ vanishes as required by kinematics. For the $\bar B_c^* \to\bar  B_s^*$ transition, the $V-A$ structure creates a clear hierarchy, $H_{0--}$ and $H_{t--}$ are strongly favored over $H_{0++}$ and $H_{t++}$, reflecting the chiral selectivity of the charged current, with $H_{t00}$ also sizable. Moreover, $H_{000}$ in the vector case exhibits a similar $q^2$ dependence as $H_{t0}$ in the pseudoscalar case,  as both amplitudes are large at low $q^2$ and decrease toward the endpoint; this similarity is largely kinematic, since both amplitudes are subject to the same phase-space suppression as $q^2$ approaches $q^2_{\max}$. These observations are consistent with the endpoint relations derived in Sec.~2.3 (see Eqs.~\eqref{eq:Ht0_endpoint}--\eqref{eq:Hmp0_endpoint}) and provide a consistency check of our numerical calculation. The observed hierarchy also indicates which angular contributions are expected to be most accessible in future experimental analyses.}

{In addition, the uncertainties vary among helicity amplitudes. The largest relative errors occur in amplitudes that become zero at the endpoint, such as \(H_{t0}\) and \(H_{000}\), where small values amplify the absolute errors. In contrast, \(H_{00}\) and \(H_{t00}\) have larger uncertainties over the entire \(q^2\) region due to multiple form-factor combinations.}

\subsection{Theoretical Prediction and Discussion}
\begin{table}[!htb]
	\caption{The theoretical predictions for the branching fractions of the semileptonic $\bar{B}^*_c \to (P,V)\ell^- \bar{\nu}_\ell$ decays. The theoretical errors are caused by uncertainties of the Gaussian parameters $\beta$, the constituent quark masses, CKM factors and total decay width, respectively.}
	\begin{center}
	\scalebox{0.9}{
			\begin{tabular}{lccc}
				\hline\hline
			Decay mode          &This work      &BS method ~\cite{Wang:2018ryc}              &CLFQM ~\cite{Wang:2024cyi}  \\\hline
$\bar{B}^*_c \to \bar{B}e^-\bar{\nu}_{e}       $&$6.61^{+0.12+1.20}_{-0.12-1.11}$$^{+0.01}_{-0.01}$$^{+5.03}_{-2.50}\times10^{-8}$                        &$5.78^{+0.81}_{-0.71}\times10^{-8}            $&$4.37^{+1.11+0.14+0.27}_{-0.74-0.12-0.27}\times10^{-8}$\\
$\bar{B}^*_c \to \bar{B}\mu^-\bar{\nu}_{\mu}   $&$6.33^{+0.11+1.15}_{-0.12-1.06}$$^{+0.01}_{-0.01}$$^{+4.82}_{-2.39}\times10^{-8}$                        &$5.57^{+0.78}_{-0.68}\times10^{-8}            $&$4.15^{+1.06+0.11+0.24}_{-0.70-0.10-0.24}\times10^{-8}$\\
$\bar{B}^*_c \to \bar{B}_s e^-\bar{\nu}_{e}    $&$9.80^{+0.12+2.12}_{-0.13-1.70}$$^{+0.00}_{-0.00}$$^{+7.46}_{-3.71}\times10^{-7}$                        &$9.43^{+1.25}_{-1.11}\times10^{-7}            $&$7.51^{+1.92+0.17+0.10}_{-1.27-0.08-0.37}\times10^{-7}$\\
$\bar{B}^*_c \to \bar{B}_s\mu^-\bar{\nu}_{\mu} $&$9.28^{+0.11+2.01}_{-0.12-1.61}$$^{+0.00}_{-0.00}$$^{+7.07}_{-3.51}\times10^{-7}$                        &$8.96^{+1.19}_{-1.05}\times10^{-7}            $&$7.06^{+1.80+0.14+0.06}_{-0.19-0.10-0.32}\times10^{-7}$\\
$\bar{B}^*_c \to \bar{D}\ell'^-\bar{\nu}_{\ell'}     $&$2.27^{+0.22+0.39}_{-0.21-0.40}$$^{+0.11}_{-0.09}$$^{+1.73}_{-0.86}\times10^{-9}$                        &$1.60^{+0.31}_{-0.26}\times10^{-9}            $&$1.36^{+0.35+0.13+0.20}_{-0.23-0.13-0.15}\times10^{-9}$\\
$\bar{B}^*_c \to \bar{D}\tau^-\bar{\nu}_{\tau}       $&$1.41^{+0.13+0.24}_{-0.13-0.25}$$^{+0.07}_{-0.06}$$^{+1.08}_{-0.53}\times10^{-9}$                        &$1.08^{+0.20}_{-0.17}\times10^{-9}            $&$0.65^{+0.17+0.07+0.08}_{-0.11-0.06-0.07}\times10^{-9}$\\
$\bar{B}^*_c \to \eta_c(1S)\ell'^-\bar{\nu}_{\ell'}   $&$3.80^{+0.04+0.89}_{-0.05-0.92}$$^{+0.06}_{-0.11}$$^{+2.89}_{-1.44}\times10^{-7}$             &$4.19^{+0.41}_{-0.37}\times10^{-7}        $&$4.45^{+1.14+0.27+0.39}_{-0.75-0.10-0.49}\times10^{-7}$\\
$\bar{B}^*_c \to \eta_c(1S) \tau^- \bar{\nu}_{\tau}    $&$1.06^{+0.01+0.25}_{-0.01-0.26}$$^{+0.02}_{-0.03}$$^{+0.81}_{-0.40}\times10^{-7}$            &$1.26^{+0.13}_{-0.11}\times10^{-7}        $&$1.03^{+0.26+0.07+0.07}_{-0.17-0.04-0.10}\times10^{-7}$\\
$\bar{B}^*_c \to \eta_c(2S) \ell'^- \bar{\nu}_{\ell'}  $&$7.79^{+1.01+2.62}_{-0.97-2.37}$$^{+0.13}_{-0.23}$$^{+5.94}_{-2.95}\times10^{-9}$            &$-                                        $&$5.60^{+1.43+0.02+0.60}_{-0.95-0.03-0.29}\times10^{-8}$\\
$\bar{B}^*_c \to \eta_c(2S) \tau^- \bar{\nu}_{\tau}    $&$3.12^{+0.87+2.44}_{-0.77-1.85}$$^{+0.05}_{-0.09}$$^{+2.38}_{-1.18}\times10^{-10}$           &$-                                        $&$4.92^{+1.26+0.11+0.12}_{-0.83-0.11-0.23}\times10^{-9}$\\
$\bar{B}^*_c \to \eta_c(3S) \ell'^- \bar{\nu}_{\ell'}  $&$8.04^{+2.52+2.18}_{-2.24-1.73}$$^{+0.14}_{-0.23}$$^{+6.12}_{-3.04}\times10^{-10}$           &$-$   &   $-$\\
$\bar{B}^*_c \to \eta_c(3S) \tau^- \bar{\nu}_{\tau}    $&$9.52^{+6.42+6.29}_{-5.07-4.60}$$^{+0.16}_{-0.28}$$^{+7.25}_{-3.60}\times10^{-12}$           &$-$   &   $-$\\
$\bar{B}^*_c \to \bar{B}^* e^- \bar{\nu}_{e}                 $&$6.73^{+0.30+1.25}_{-0.31-1.30}$$^{+0.01}_{-0.01}$$^{+5.13}_{-2.55}\times10^{-8}
$&$1.59^{+0.21}_{-0.21}\times10^{-7} $&$-$\\
$\bar{B}^*_c \to \bar{B}^* \mu^- \bar{\nu}_{\mu}             $&$6.44^{+0.28+1.20}_{-0.30-1.25}$$^{+0.01}_{-0.01}$$^{+4.91}_{-2.44}\times10^{-8}             $&$1.53^{+0.22}_{-0.19}\times10^{-7} $&$-$\\
$\bar{B}^*_c \to \bar{B}^*_s e^- \bar{\nu}_{e}               $&$1.12^{+0.02+0.22}_{-0.02-0.21}$$^{+0.00}_{-0.00}$$^{+0.85}_{-0.42}\times10^{-6}
$&$2.47^{+0.33}_{-0.29}\times10^{-6} $&$-$\\
$\bar{B}^*_c \to \bar{B}^*_s \mu^- \bar{\nu}_{\mu}           $&$1.06^{+0.02+0.21}_{-0.02-0.20}$$^{+0.00}_{-0.00}$$^{+0.81}_{-0.40}\times10^{-6}
$&$2.34^{+0.31}_{-0.28}\times10^{-6} $&$-$\\
$\bar{B}^*_c \to \bar{D}^* \ell'^- \bar{\nu}_{\ell'}         $&$1.35^{+0.24+0.30}_{-0.22-0.25}$$^{+0.07}_{-0.05}$$^{+1.03}_{-0.51}\times10^{-9}
$&$4.61^{+0.92}_{-0.76}\times10^{-9} $&$-$\\
$\bar{B}^*_c \to \bar{D}^*\tau^- \bar{\nu}_{\tau}            $&$8.17^{+1.41+1.84}_{-1.33-1.52}$$^{+0.40}_{-0.32}$$^{+6.23}_{-3.09}\times10^{-10}
$&$3.22^{+0.61}_{-0.51}\times10^{-9} $&$-$\\
$\bar{B}^*_c \to \psi(1S) \ell'^- \bar{\nu}_{\ell'}    $&$6.42^{+0.09+1.95}_{-0.10-1.79}$$^{+0.11}_{-0.19}$$^{+4.89}_{-2.43}\times10^{-7}
$&$1.13^{+0.11}_{-0.10}\times10^{-6} $&$-$\\
$\bar{B}^*_c \to \psi(1S)\tau^- \bar{\nu}_{\tau}       $&$1.69^{+0.02+0.54}_{-0.02-0.50}$$^{+0.03}_{-0.05}$$^{+1.29}_{-0.64}\times10^{-7}
$&$3.13^{+0.32}_{-0.28}\times10^{-7} $&$-$\\
$\bar{B}^*_c \to \psi(2S) \ell'^- \bar{\nu}_{\ell'}    $&$3.63^{+0.12+0.82}_{-0.12-0.86}$$^{+0.06}_{-0.11}$$^{+2.76}_{-1.37}\times10^{-8}$
&$-$  &  $-$\\
$\bar{B}^*_c \to \psi(2S)\tau^- \bar{\nu}_{\tau}       $&$2.24^{+0.13+0.69}_{-0.13-0.72}$$^{+0.04}_{-0.07}$$^{+1.70}_{-0.85}\times10^{-9}$
&$-$  &  $-$\\
$\bar{B}^*_c \to \psi(3S) \ell'^- \bar{\nu}_{\ell'}    $&$4.96^{+0.60+0.85}_{-0.60-0.92}$$^{+0.08}_{-0.14}$$^{+3.78}_{-1.88}\times10^{-9}$
&$-$  &  $-$\\
$\bar{B}^*_c \to \psi(3S)\tau^- \bar{\nu}_{\tau}       $&$7.09^{+1.44+2.11}_{-1.39-2.23}$$^{+0.12}_{-0.21}$$^{+5.40}_{-2.68}\times10^{-11}$
&$-$  &  $-$\\
				\hline\hline
		\end{tabular}}
	\end{center}
	\label{tab:BranchingRatios1}
\end{table}	
\begin{figure}[!htb]
    \vspace{-0.6cm}
\begin{center}
\includegraphics[scale=0.55]{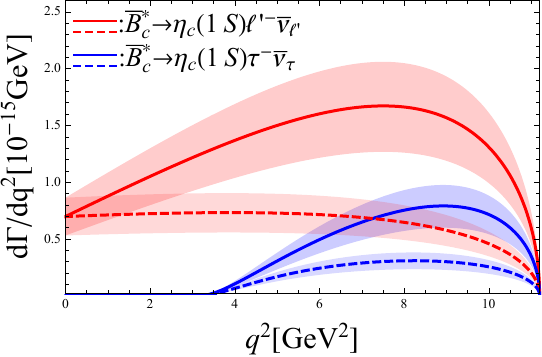}\quad
\includegraphics[scale=0.55]{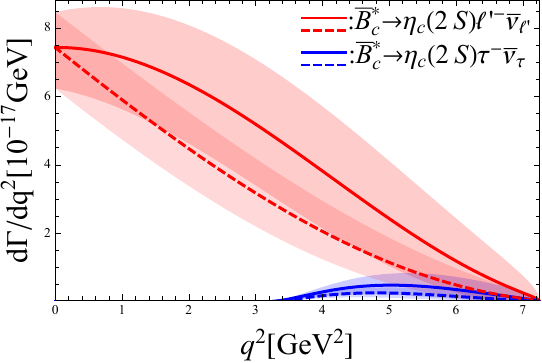}\quad
\includegraphics[scale=0.55]{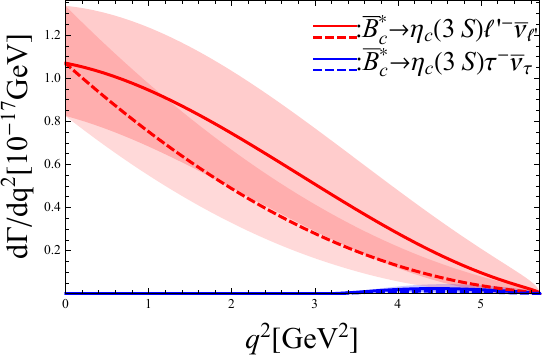}\\
\includegraphics[scale=0.55]{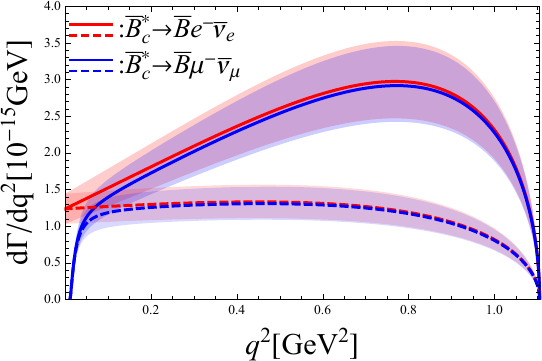}\quad
\includegraphics[scale=0.55]{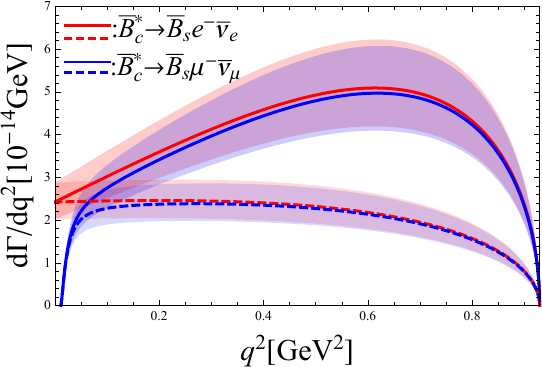}\quad
\includegraphics[scale=0.55]{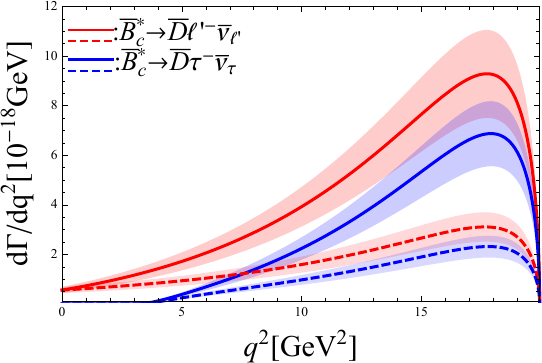}\\
\includegraphics[scale=0.55]{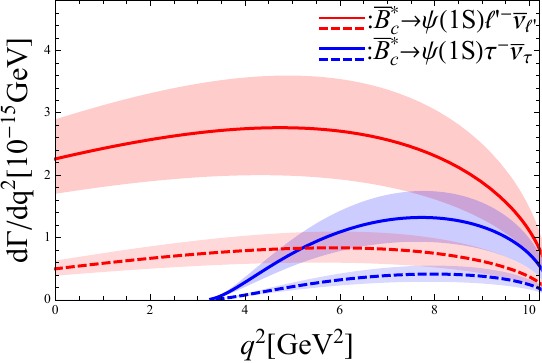}\quad
\includegraphics[scale=0.55]{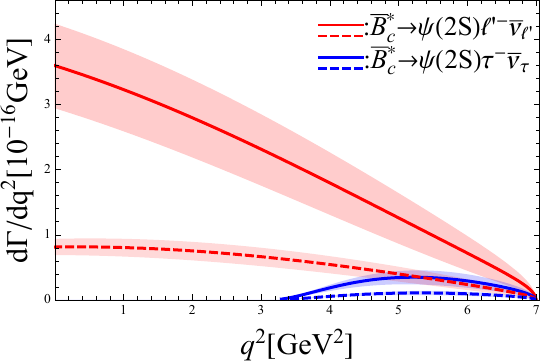}\quad
\includegraphics[scale=0.55]{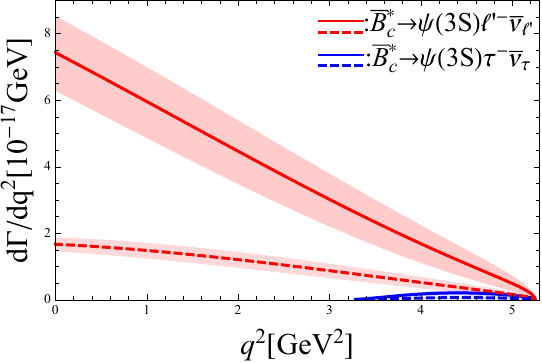}\\
\includegraphics[scale=0.55]{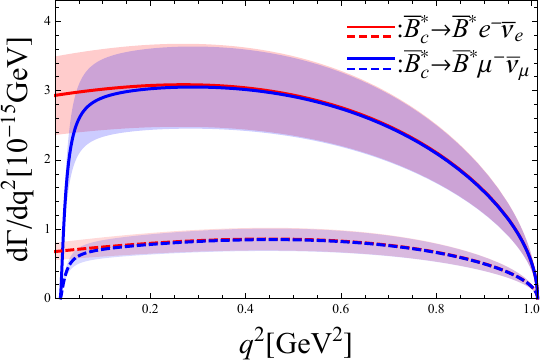}\quad
\includegraphics[scale=0.55]{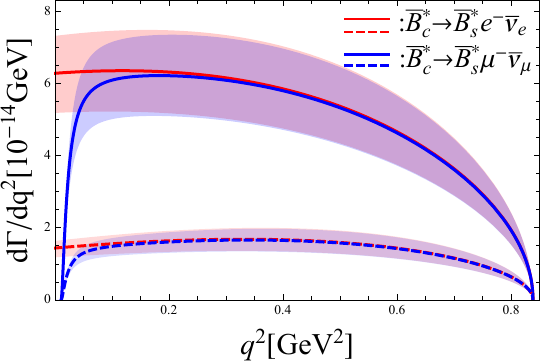}\quad
\includegraphics[scale=0.55]{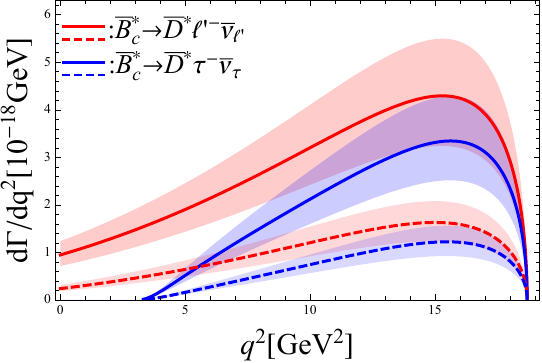}
\end{center}
\vspace{-10pt}
\caption{The $q^2$ dependences of differential decay rates ${\rm d}\Gamma/{\rm d}q^2$ (solid lines) and ${\rm d}\Gamma_L/{\rm d}q^2$ (dashed lines) for semileptonic $\bar{B}^*_c \to (P,V)\ell^- \bar{\nu}_{\ell}$ decays.
The error bands represent the uncertainties propagated from the Gaussian parameters $\beta$ and the constituent quark masses.}
\label{fig:dGBstar}
\end{figure}
\begin{table}[!htb]
		\caption{Predictions for $q^2$-integrated observables within the SM, The theoretical errors are caused by uncertainties of the Gaussian parameters $\beta$ and the constituent quark masses, respectively.}
    \vspace{-0.6cm}
		\begin{center}
			\scalebox{0.85}{
				\begin{tabular}{lclrlrlr}
					\hline\hline
				Obs.     &Prediction     &Obs.       &Prediction     &Obs.    &Prediction     &Obs.   &Prediction  \\\hline
$R^{D^*}                  $&$ 0.606^{+0.001+0.003}_{-0.003-0.003} $&$R^{D^*}_{L}          $&$ 0.592^{+0.001+0.003}_{-0.003-0.004}
$&$A_\lambda^{D^*}        $&$ 0.514^{+0.001+0.003}_{-0.001-0.003} $&$A_\theta^{D^*}       $&$-0.126^{+0.001+0.004}_{-0.000-0.004}$\\
$R^{\psi(1S)}             $&$ 0.263^{+0.001+0.004}_{-0.001-0.005} $&$R^{\psi(1S)}_{L}     $&$ 0.278^{+0.001+0.005}_{-0.001-0.006}
$&$A_\lambda^{\psi(1S)}   $&$ 0.221^{+0.000+0.000}_{-0.000-0.011} $&$A_\theta^{\psi(1S)}  $&$ 0.074^{+0.000+0.004}_{-0.000-0.000}$\\
$R^{\psi(2S)}             $&$ 0.062^{+0.002+0.005}_{-0.002-0.007} $&$R^{\psi(2S)}_{L}     $&$ 0.068^{+0.002+0.006}_{-0.002-0.008}
$&$A_\lambda^{\psi(2S)}   $&$ 0.024^{+0.003+0.015}_{-0.003-0.022} $&$A_\theta^{\psi(2S)}  $&$ 0.169^{+0.003+0.016}_{-0.003-0.009}$\\
$R^{\psi(3S)}             $&$ 0.014^{+0.001+0.002}_{-0.001-0.002} $&$R^{\psi(3S)}_{L}     $&$ 0.017^{+0.001+0.002}_{-0.001-0.003}
$&$A_\lambda^{\psi(3S)}   $&$-0.077^{+0.005+0.021}_{-0.006-0.025} $&$A_\theta^{\psi(3S)}  $&$ 0.188^{+0.007+0.019}_{-0.006-0.012}$\\
$R^{D}                    $&$0.621^{+0.002+0.001}_{-0.002-0.000} $&$F^{D^*}_{L}           $&$0.353^{+0.002+0.002}_{-0.001-0.002}
$&$A_\lambda^{D}          $&$0.757^{+0.001+0.000}_{-0.001-0.000} $&$A_\theta^{D}          $&$-0.224^{+0.003+0.001}_{-0.003-0.002}$\\
$R^{\eta_c(1S)}           $&$0.279^{+0.001+0.001}_{-0.001-0.001} $&$F^{\psi(1S)}_{L}      $&$0.306^{+0.000+0.002}_{-0.000-0.002}
$&$A_\lambda^{\eta_c(1S)} $&$0.591^{+0.001+0.003}_{-0.001-0.004} $&$A_\theta^{\eta_c(1S)} $&$-0.086^{+0.001+0.004}_{-0.001-0.001}$\\
$R^{\eta_c(2S)}           $&$0.040^{+0.005+0.014}_{-0.005-0.015} $&$F^{\psi(2S)}_{L}      $&$0.303^{+0.001+0.005}_{-0.001-0.004}
$&$A_\lambda^{\eta_c(2S)} $&$0.381^{+0.016+0.024}_{-0.019-0.019} $&$A_\theta^{\eta_c(2S)} $&$-0.049^{+0.002+0.016}_{-0.002-0.023}$\\
$R^{\eta_c(3S)}           $&$0.012^{+0.004+0.004}_{-0.004-0.004} $&$F^{\psi(3S)}_{L}      $&$0.304^{+0.001+0.001}_{-0.001-0.001}
$&$A_\lambda^{\eta_c(3S)} $&$0.373^{+0.012+0.018}_{-0.017-0.040} $&$A_\theta^{\eta_c(3S)} $&$-0.074^{+0.020+0.020}_{-0.026-0.028}$\\
					\hline\hline
			\end{tabular}}
		\end{center}
		\label{tab:Observables}
	\end{table}
\begin{figure}[!htb]
\begin{center}
\includegraphics[scale=0.55]{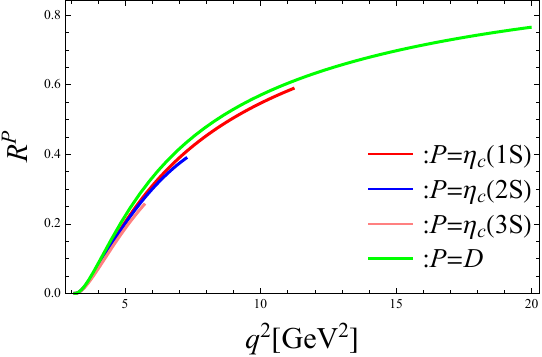}\quad
\includegraphics[scale=0.55]{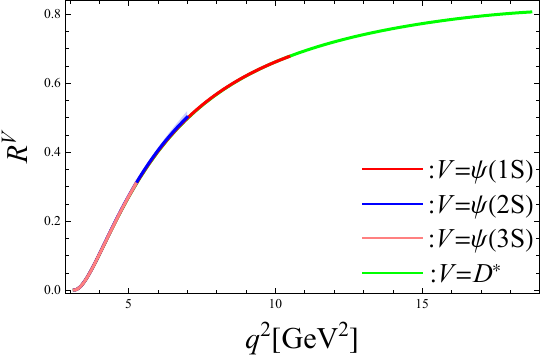}\quad
\includegraphics[scale=0.55]{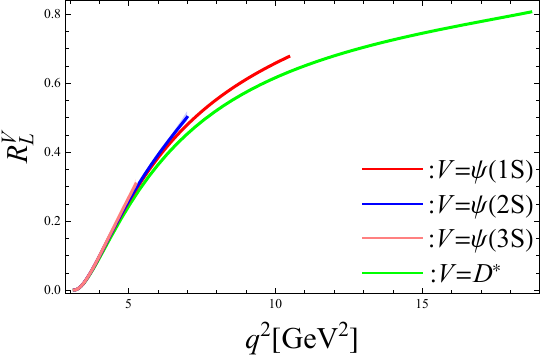}\\
\includegraphics[scale=0.55]{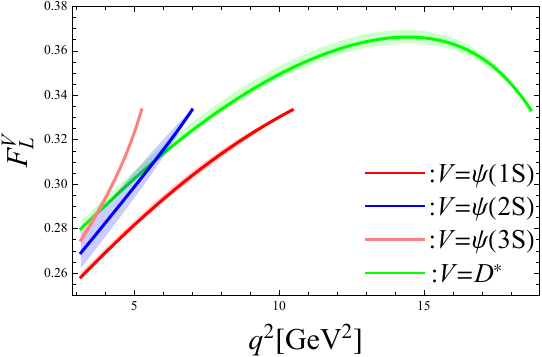}\quad
\includegraphics[scale=0.55]{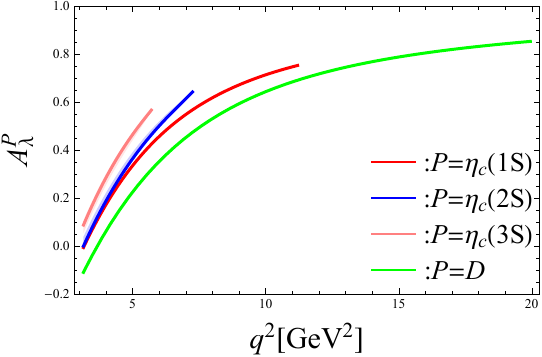}\quad
\includegraphics[scale=0.55]{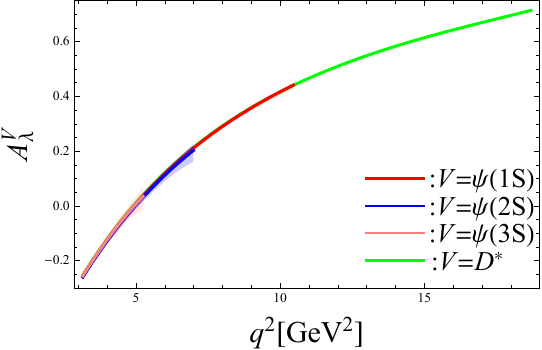}\\
\includegraphics[scale=0.55]{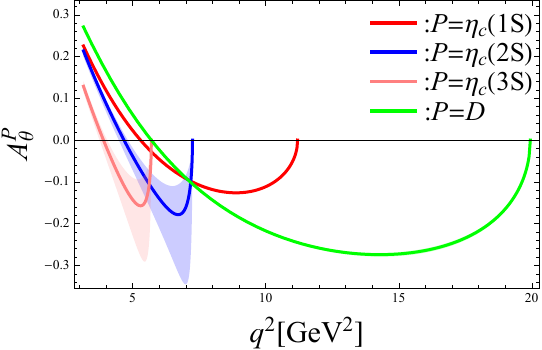}\qquad
\includegraphics[scale=0.55]{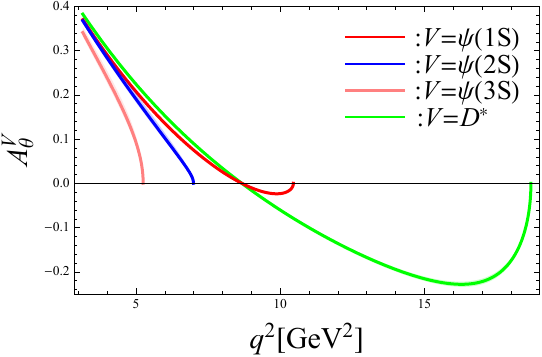}\\
\end{center}
\vspace{-10pt}
\caption{The $q^2$ dependences of $R_{(L)}(q^2)$, $F_L(q^2)$, $A_{\lambda}(q^2)$ and $A_{\theta}(q^2)$. The error bands represent the uncertainties propagated from the Gaussian parameters $\beta$ and the constituent quark masses.}
\label{fig:dsdobserv}
\end{figure}
Using the theoretical formulas given in the last section and inputs given above, we then present our numerical results for the branching fractions of $\bar{B}^*_{c} \to (P,V) \ell^- \bar{\nu}_\ell$ decays in Table~\ref{tab:BranchingRatios1}, in which the four errors are caused by the Gaussian parameters $\beta$, the constituent quark masses, CKM factors and $\Gamma_{B^*_{c}}$, respectively. For the other $q^2$-integrated observables listed in Table~\ref{tab:Observables}, the theoretical uncertainties are caused only by the Gaussian parameters $\beta$ and the constituent quark masses. Besides, the $q^2$ dependences of differential decay rates $ {\rm d}\Gamma_{(L)}/{\rm d}q^2$, $R_{(L)}$, $F_L^V$ and $A_{\lambda,\theta}$, are shown in Figs.~\ref{fig:dGBstar} and \ref{fig:dsdobserv}, with the corresponding error bands. The follows are some explanations and discussions:
\begin{enumerate}
\item For branching fractions, it could be found from Table~\ref{tab:BranchingRatios1} that our results are consistent with the theoretical predictions of $\bar{B}^*_{c} \to (P,V) \ell^- \bar{\nu}_\ell$ $(P=B$, $B_s$, $D$, $\eta_c(1S)$; $V=B^*$, $B_s^*$, $D^*$, $\psi(1S))$ decays within the BS method~\cite{Wang:2018ryc} and $\bar{B}^*_{c} \to P \ell^- \bar{\nu}_\ell$ $(P=B$, $B_s$, $D$, $\eta_c(1S,\,2S))$ decays within the CLFQM~\cite{Wang:2024cyi}. It is important to highlight that the CLFQM adopted in this paper yields the self-consistent and covariant results for the form factors of $V \to P$ and $V' \to V''$ transitions, as discussed in Refs.~\cite{Chang:2019obq,Chang:2019mmh}. The input parameters adopted in our calculation for form factors differ from those used in the CLFQM analysis of Ref.~\cite{Wang:2024cyi}, which leads to slight discrepancies between our predictions and the results reported therein.

    In Table~\ref{tab:BranchingRatios1}, our theoretical predictions for branching fractions have larger uncertainties than some previous estimates~\cite{Wang:2018ryc,Wang:2024cyi}.
    {This is because we propagate the full uncertainties from all input parameters (quark masses, Gaussian parameters $\beta$, CKM elements, and the total width $\Gamma_{B_c^*}$), whereas earlier works often considered only partial errors.}
    The largest uncertainty from $\Gamma_{B^*_{c}}$, which is induced by the input parameters used in the calculation of $\Gamma (B^*_{c} \to B_c \gamma)$.
    {Our value $\Gamma(B_c^*\to B_c\gamma) = (39^{+23}_{-17})$~eV differs from the result $(59\pm12)$~eV in Ref.~\cite{Wang:2024cyi},} and this uncertainty will be reduced by the precise measurement of $\Gamma_{B^*_{c}}$ in future high-luminosity experiments. The smallest uncertainty arises from CKM factors, particularly for the $\bar{B}^*_c \to \bar{B}^{(*)}_s \ell'^- \bar{\nu}_{\ell'}$ decays induced by $\bar{c} \to \bar{s} \ell'^- \bar{\nu}_{\ell'}$ transition at quark level, which is attributed to the current precise experimental measurements of CKM matrix elements. In addition, the results of $\bar{B}^*_{c} \to \eta_c(3S) \ell^- \bar{\nu}_\ell$, $\psi(2S,\,3S) \ell^- \bar{\nu}_\ell$ decays in this paper are first presented.

    Owing to the hierarchical relations between the CKM matrix elements $|V_{ub}| < |V_{cb}| < |V_{cd}| < |V_{cs}|$, there are some clear hierarchical relationships among branching ratios for the $B^*_c$ decays induced either by the bottom quark decay or by the charm quark decay, i.e.,
    \begin{eqnarray}
    \mathcal{B}(\bar{B}^*_{c} \to \bar{D} \ell^- \bar{\nu}_\ell) < \mathcal{B}(\bar{B}^*_{c} \to \bar{B}\ell'^- \bar{\nu}_{\ell'}) < \mathcal{B}(\bar{B}^*_{c} \to \eta_c(1S)\ell^-\bar{\nu}_\ell) < \mathcal{B}(\bar{B}^*_{c} \to \bar{B}_s \ell'^-\bar{\nu}_{\ell'}),\\
     \mathcal{B}(\bar{B}^*_{c} \to \bar{D}^* \ell^- \bar{\nu}_\ell) < \mathcal{B}(\bar{B}^*_{c} \to \bar{B}^* \ell'^-\bar{\nu}_{\ell'}) < \mathcal{B}(\bar{B}^*_{c} \to \psi(1S) \ell^-\bar{\nu}_\ell) < \mathcal{B}(\bar{B}^*_{c} \to \bar{B}^*_s \ell'^-\bar{\nu}_{\ell'}).
    \end{eqnarray}
    It should be pointed out that the phase spaces in $b \to c$ induced $\bar{B}^*_{c} \to \eta_c(1S)$, $\psi(1S)$ transitions are much larger than $c \to d$ induced $\bar{B}^*_{c} \to \bar{B}^{(*)}$ transitions and thus can compensate for the small CKM matrix element in $\bar{B}^*_{c} \to \eta_c(1S) \ell^- \bar{\nu}_\ell$, $\psi(1S) \ell^- \bar{\nu}_\ell$ decays.

    Due to the hierarchical relations between masses $m_{\eta_c(1S)}$ $<m_{\eta_c(2S)}$ $<m_{\eta_c(3S)}$ and $m_{\psi(1S)}$ $<m_{\psi(2S)}$ $<m_{\psi(3S)}$, the phase spaces of final states decrease with the increase in the radial quantum number n. There are some clear hierarchical relations, i.e.,
    \begin{eqnarray}
    \mathcal{B}(\bar{B}^*_{c} \to \eta_c(3S) \ell^- \bar{\nu}_\ell)<\mathcal{B}(\bar{B}^*_{c} \to \eta_c(2S) \ell^- \bar{\nu}_\ell)<\mathcal{B}(\bar{B}^*_{c} \to \eta_c(1S) \ell^- \bar{\nu}_\ell),\\
    \mathcal{B}(\bar{B}^*_{c} \to \psi(3S) \ell^- \bar{\nu}_\ell)<\mathcal{B}(\bar{B}^*_{c} \to \psi(2S) \ell^- \bar{\nu}_\ell)<\mathcal{B}(\bar{B}^*_{c} \to \psi(1S) \ell^- \bar{\nu}_\ell).
    \end{eqnarray}
     In addition, {from the numerical results in Table~\ref{tab:BranchingRatios1}, the branching fractions of $\bar B_c^* \to V \ell^- \bar{\nu}_\ell$ decays are generally larger than those of the corresponding $P$ modes, except for the $\bar D^{(*)}$ case where the larger phase space of the $\bar D$ mode compensates for the fewer helicity amplitudes. This leads to the following hierarchical relations, }i.e.,
     \begin{eqnarray}\label{eq:relation38}
    \mathcal{B}(\bar{B}^*_{c} \to \eta_c(nS) \ell^- \bar{\nu}_\ell)<\mathcal{B}(\bar{B}^*_{c} \to \psi(nS) \ell^- \bar{\nu}_\ell),\\\label{eq:relation39}
    \mathcal{B}(\bar{B}^*_{c} \to \bar{B}_{(s)} \ell'^- \bar{\nu}_{\ell'})<\mathcal{B}(\bar{B}^*_{c} \to \bar{B}^*_{(s)} \ell'^- \bar{\nu}_{\ell'}).
    \end{eqnarray}
     As for the hierarchical relations between the branching ratios of $\bar{B}^*_{c} \to \bar{D}^{(*)} \ell^- \bar{\nu}_\ell$ decays,
       \begin{eqnarray}
    \mathcal{B}(\bar{B}^*_{c} \to \bar{D} \ell^- \bar{\nu}_\ell)>\mathcal{B}(\bar{B}^*_{c} \to \bar{D}^* \ell^- \bar{\nu}_\ell),
    \end{eqnarray}
     which differ from the above predictions in Eqs.~\eqref{eq:relation38} and \eqref{eq:relation39}. {This can be attributed to two factors. One is the larger phase space available in the $\bar B_c^* \to \bar D$ mode; the other is that the form factors at $q^2=0$ are also larger for the $\bar D$ mode than for the $\bar D^*$ mode, as listed in Table~\ref{tab:formfactor}.}

    In Table~\ref{tab:BranchingRatios1}, one may find that the CKM-favored $\bar{B}^*_{c} \to \bar{B}^{(*)}_s \ell'^{-} \bar{\nu}_{\ell'}$, $\bar{B}^{(*)} \ell'^{-} \bar{\nu}_{\ell'}$, $\psi(1S) \ell^{-} \bar{\nu}_{\ell}$, $\psi(2S) \ell'^{-} \bar{\nu}_{\ell'}$ and  $\eta_c(1S) \ell^{-} \bar{\nu}_{\ell}$ decays have the large branching fractions, $\gtrsim$ ${\cal O}(10^{-8})$. In particular,  the branching fractions of the  $\bar{B}^*_{c} \to \bar{B}^*_s e^- \bar{\nu}_{e}$ and $\bar{B}^*_s \mu^- \bar{\nu}_{\mu}$  decays can reach up to ${\cal O}(10^{-6})$ and therefore should be prioritized for searches and are expected to be first observed in future LHCb experiments. Owing to the relatively small CKM factors and form factors, the $\bar{B}^*_{c} \to \bar{D}^{(*)} \ell^- \bar{\nu}_\ell$ and other excited charmonium modes possess much smaller branching fractions of ${\cal O}(10^{-12}) \sim {\cal O}(10^{-9})$, which implies that these decays are hardly observable for a long time.

    \item At present, no direct evidence of NP has been reported, there still exists some discrepancies with the SM predictions in semileptonic $B$ decays. As mentioned above, the BaBar, Belle, Belle II and LHCb collaborations reported a combined $3.8\sigma$ deviation of their measured ratios $R_{D^{(*)}}$ from the SM predictions. Besides, the ratio $R_{J/\psi}$ in semileptonic $B_c$ decays measured by the LHCb collaboration lies $1.8\sigma$ above the SM prediction. The above mentioned large deviations in $R_{D^{(*)}}$ and possible anomaly in $B_c$ decay channels  imply possible hints of NP. Therefore, it would be interesting to investigate whether such similar anomalies also exist independently in semileptonic $B^*_c$ decay modes.

        Our numerical results for the ratios $R^{P(V)}$ defined in $\bar{B}^*_c \to (P,\,V) \ell^- \bar{\nu}_{\ell}$ decays are summarized in Table \ref{tab:Observables}, and the $q^2$ dependences of these observables are shown in Fig. \ref{fig:dsdobserv}. Using the results collected in Table \ref{tab:BranchingRatios1}, we obtain the predictions within the CLFQM~\cite{Wang:2024cyi} and BS method~\cite{Wang:2018ryc},
        \begin{eqnarray}
        &R^{D}= 0.478,~R^{\eta_c(1S)}= 0.229,~R^{\eta_c(2S)}= 0.087,~&\text{CLFQM}\\
        &R^{D}= 0.675,~R^{\eta_c(1S)}= 0.300,~R^{D^*}= 0.698,~R^{\psi(1S)}= 0.277.~& \text{BS method}
        \end{eqnarray}
        One can see that these results are slightly smaller (larger) than ours, due to the adoption of distinct models and parameterizations in form factor evaluation. Notably, these ratios are less model-dependent and can be used to verify the consistency between theoretical predictions and experimental observations in future experiments.

        From Table \ref{tab:Observables}, it could be found that
          \begin{eqnarray}
            R^{\psi(3S)}_{(L)}<R^{\psi(2S)}_{(L)}<R^{\psi(1S)}_{(L)}, \\
            R^{\eta_c(3S)}<R^{\eta_c(2S)}<R^{\eta_c(1S)}\,.
          \end{eqnarray}
        {This hierarchy is mainly correlated with the available phase space, which is more restricted for the excited states, and also reflects the differences in the helicity compositions among these states.}

         In contrast to $R^V_{L}$, $F_L^V$ has identical decay modes and the ranges of $q^2$ integration in both its numerator and denominator. one can find that $F^{\psi(1S)}_{L} \simeq F^{\psi(2S)}_{L} \simeq F^{\psi(3S)}_{L} \simeq 30\% $ and $F^{D^*_u}_{L} \simeq 35\%$, which implies that $\bar{B}^*_{c} \to V \tau^- \bar{\nu}_\tau$ decay dominated by the transverse polarization.

         Moreover, we also consider the ratios
          \begin{eqnarray}
         R_{e}^{B_s/B}&=&\frac{\mathcal{B}(\bar{B}^*_{c} \to \bar{B}_s e^- \bar{\nu}_e)}{\mathcal{B}(\bar{B}^*_{c} \to \bar{B} e^- \bar{\nu}_{e})}={14.8^{+0.1+0.5}_{-0.1-0.4}},\\
         R_{\mu}^{B_s/B}&=&\frac{\mathcal{B}(\bar{B}^*_{c} \to \bar{B}_s \mu^- \bar{\nu}_{\mu})}{\mathcal{B}(\bar{B}^*_{c} \to \bar{B} {\mu}^- \bar{\nu}_{\mu})}={14.7^{+0.1+0.5}_{-0.1-0.4}},\\
         R_{e}^{B^*_s/B^*}&=&\frac{\mathcal{B}(\bar{B}^*_{c} \to \bar{B}^*_s e^- \bar{\nu}_e)}{\mathcal{B}(\bar{B}^*_{c} \to \bar{B}^* e^- \bar{\nu}_{e})}={16.6^{+0.5+0.6}_{-0.4-0.5}},\\
         R_{\mu}^{B^*_s/B^*}&=&\frac{\mathcal{B}(\bar{B}^*_{c} \to \bar{B}^*_s \mu^- \bar{\nu}_{\mu})}{\mathcal{B}(\bar{B}^*_{c} \to \bar{B}^* \mu^- \bar{\nu}_{\mu})}={16.5^{+0.5+0.6}_{-0.4-0.5}},
          \end{eqnarray}
         { where the first uncertainty comes from the Gaussian parameter $\beta$ and the second from the constituent quark masses. The larger $\beta$ induced uncertainty for $R^{B^*_s/B^*}$ arises because vector meson wave functions are more sensitive to $\beta$ than pseudoscalar ones, while the quark mass uncertainties affect both ratios similarly.}
         These results suggest a relatively large SU(3) symmetry breaking effects in $\bar{B}^*_{c} \to \bar{B}_{(s)} \ell'^- \bar{\nu}_{\ell'}$ decays.

\item  Furthermore, we also take into account another two physical observables, the lepton spin asymmetry $A^{P(V)}_{\lambda}$ and the forward-backward asymmetry $A^{P(V)}_{\theta}$, which are very sensitive to the NP effects and widely studied in $B$ system within various NP scenarios~\cite{Li:2016vvp,Hu:2018lmk,Hu:2020yvs,Celis:2012dk}. Our numerical results for $q^2$-integrated $A^{P(V)}_{\lambda}$ and $A^{P(V)}_{\theta}$ are collected in Table \ref{tab:Observables}, and the $q^2$ dependences of $A^{P(V)}_{\lambda}$ and $A^{P(V)}_{\theta}$ are shown in Fig.~\ref{fig:dsdobserv}.

     One can notice that the theoretical uncertainties of these two observables listed in Table \ref{tab:Observables} are significantly small compared with the branching fractions. Therefore, {the precise SM predictions for \(A^{P(V)}_{\lambda}\) and \(A^{P(V)}_{\theta}\) obtained in this work may provide useful references for future experimental measurements.}

    A detailed look at the \(q^2\)-dependent distributions and the integrated observables reveals several features. {For $A_\theta$ in Table~\ref{tab:Observables}, the uncertainties are generally smaller for ground-state channels than for radially excited states. This reflects the fact that excited-state form factors are more sensitive to the detailed shape of the wave functions, which introduces additional parametric dependence in the CLFQM. Consequently, ground-state predictions are under better theoretical control, while excited-state results are more affected by the model inputs. This channel dependence is also visible in the error bands at high $q^2$ for $\eta_c(2S)$ and $\eta_c(3S)$, as shown in Fig.~\ref{fig:dsdobserv}.}
   In addition, the differential $q^2$ dependences show the following features: $A^{V}_{\lambda}$ crosses zero at $q^2 \approx 4.9 \text{GeV}^2$, $A^{\eta_c(1S),\eta_c(2S),D}_{\lambda}$ cross zero at $q^2 \approx (3.2,\,3.2,\,3.7) \text{GeV}^2$ respectively, while $A^{\eta_c(3S)}_{\lambda}$ exhibits no zero crossing in the physical $q^2$ region. These distinct patterns offer complementary information for testing theoretical predictions.

\item Interestingly, the LHCb collaboration has measured the ratio $R_{\pi/\mu\bar{\nu}_{\mu}}$ relating the nonleptonic and semileptonic branching fractions of $B_c$ meson\cite{LHCb:2014rck},
    \begin{align}
	R_{\pi/\mu\bar{\nu}_{\mu}}=\frac {\mathcal{B} (B^-_c\to J/\psi \pi^-)}{\mathcal{B} (B^-_c\to J/\psi \mu^- \bar{\nu}_{\mu})} = 0.0469 \pm 0.0028(\text{stat}) \pm 0.0046(\text{syst}).
      \end{align}
    This ratio takes the advantage of being independent of both the $B_c$ production rate and $V_{cb}$. Similarly, using the predictions of nonleptonic $B_c^*$ decays given in our previous work~\cite{Yang:2022jqu} and the results of semileptonic $B_c^*$ decays calculated in this paper, we also consider the theoretical predictions of the following ratios:
\begin{eqnarray}
&&R^B_{\pi/ \mu \bar{\nu}_\mu}=\frac {\mathcal{B}(\bar{B}_c^* \to \bar{B} \pi^-)}{\mathcal{B}(\bar{B}_c^* \to \bar{B} \mu^- \bar{\nu}_\mu)}=0.332\,,\nonumber\\
&&R^{B_s}_{\pi/ \mu \bar{\nu}_\mu}=\frac {\mathcal{B}(\bar{B}_c^* \to \bar{B}_s \pi^-)}{\mathcal{B}(\bar{B}_c^* \to \bar{B}_s\mu^- \bar{\nu}_\mu)}=0.474\,,\nonumber\\
&&R^{\eta_c(1S)}_{\pi/ \ell' \bar{\nu}_{\ell'}}=\frac {\mathcal{B}(\bar{B}_c^* \to \eta_c(1S) \pi^-)}{\mathcal{B}(\bar{B}_c^* \to \eta_c(1S) \ell^{'-} \bar{\nu}_{\ell'})}=0.039\,,\nonumber\\
&&R^{\eta_c(2S)}_{\pi/ \ell' \bar{\nu}_{\ell'}}=\frac {\mathcal{B}(\bar{B}_c^* \to \eta_c(2S) \pi^-)}{\mathcal{B}(\bar{B}_c^* \to \eta_c(2S) \ell^{'-} \bar{\nu}_{\ell'})}=0.526\,,\nonumber\\
&&R^{B^*}_{\pi/ \mu \bar{\nu}_\mu}=\frac {\mathcal{B}(\bar{B}_c^* \to \bar{B}^* \pi^-)}{\mathcal{B}(\bar{B}_c^* \to \bar{B}^* \mu^- \bar{\nu}_\mu)}=0.978\,,\nonumber\\
&&R^{B^*_s}_{\pi/ \mu \bar{\nu}_\mu}=\frac {\mathcal{B}(\bar{B}_c^* \to \bar{B}^*_s \pi^-)}{\mathcal{B}(\bar{B}_c^*  \to \bar{B}^*_s \mu^- \bar{\nu}_\mu)}=1.321\,,\nonumber\\
&&R^{\psi(1S)}_{\pi/ \ell' \bar{\nu}_{\ell'}}=\frac {\mathcal{B}(\bar{B}_c^* \to \psi(1S) \pi^-)}{\mathcal{B}(\bar{B}_c^* \to \psi(1S) \ell^{'-} \bar{\nu}_{\ell'})}=0.079\,,\nonumber\\
&&R^{\psi(2S)}_{\pi/ \ell' \bar{\nu}_{\ell'}}=\frac {\mathcal{B}(\bar{B}_c^* \to \psi(2S) \pi^-)}{\mathcal{B}(\bar{B}_c^* \to \psi(2S) \ell^{'-} \bar{\nu}_{\ell'})}=0.441\,.
\end{eqnarray}
Theoretically, these ratios can be predicted more precisely within the SM and will be more suitable for probing NP beyond the SM.
\end{enumerate}

\section{Summary}
With the high luminosities of the running LHC and the future Super Electron Positron Collider, it is expected that a huge amount of the $B^*_{c}$ data samples would be accumulated, offering a valuable opportunity to investigate the $B^*_{c}$ weak decays. In this paper, we have studied the tree-dominated semileptonic $\bar{B}^*_{c}\to (P,V) \ell^- \bar{\nu}_{\ell}$ ($P=D$, $B_{d,s}$, $\eta_c(1S,\,2S,\,3S)$; $V=D^*$, $B^*_{d,s}$, $\psi(1S,\,2S,\,3S)$) decays, induced by the charm quark and bottom quark weak decay. The form factors for the $\bar{B}^*_{c}\to (P,V)$ transitions are calculated using the CLFQM. The helicity amplitudes are calculated in detail, and the predictions of observables including branching fraction (decay rate), the longitudinal polarization fraction of the daughter vector meson, lepton spin asymmetry, forward-backward asymmetry and ratios $R_{(L)}$ are presented in Tables \ref{tab:BranchingRatios1}, \ref{tab:Observables} and Figs.~\ref{fig:dGBstar} and \ref{fig:dsdobserv}. It is found that the CKM-favored $\bar{B}^*_{c}$ ${\to}$ $\bar{B}^*_{s} e^- \bar{\nu}_e$ and $\bar{B}^*_{s} \mu^- \bar{\nu}_\mu$ decay has large branching fraction of ${\cal O}(10^{-6})$. Therefore, it is likely to be observed in future LHCb experiments or can serve as golden channel to search for the $B^*_{c}$ meson in experiments. In addition, the ratios $R_{\pi/\ell'\bar{\nu}_{\ell'}}$ relating the nonleptonic and semileptonic branching fractions of $B^*_{c}$ meson are also predicted in discussion. This study provides a ready and helpful reference for experimental discovery and investigation of $B^*_{c}$ mesons in the future.

\begin{appendices}
\section{Details of the CLFQM calculation}
\label{sec:AppendixA}

{In this appendix, we provide the explicit expressions for the form factors of $V\to (P,V)$ transitions in the CLFQM. {The notation used in this Appendix follows the CLFQM conventions of Ref.~\cite{Chang:2019mmh}. }

{Following Ref.~\cite{Chang:2019mmh}, the $V \to P$ transition matrix elements are parameterized in terms of the auxiliary functions $g(q^2)$, $f(q^2)$, $a_+(q^2)$, and $a_-(q^2)$ as
\begin{align}
\langle P(p'',\lambda)|\bar{q}_1'' \gamma_\mu q_1' |V(p')\rangle &= i \epsilon_{\mu\nu\alpha\beta} \epsilon^{\nu} P^\alpha q^\beta g(q^2), \\
\langle P(p'',\lambda)|\bar{q}_1'' \gamma_\mu \gamma_5 q_1' |V(p')\rangle
&= -f(q^2) \epsilon_\mu - \epsilon\cdot P \left[ a_+(q^2) P_\mu + a_-(q^2) q_\mu \right],
\end{align}
where $P = p' + p''$ and $q = p' - p''$. These auxiliary functions are related to the physical form factors $V(q^2)$, $A_0(q^2)$, $A_1(q^2)$, and $A_2(q^2)$ by
\begin{align}
	&V(q^2)= -(M'+M'')g(q^2)\,,\\
	&A_0(q^2)=-\frac{1}{2M'}\left[ -(M'^2-M''^2)a_+(q^2)+f(q^2)+q^2a_-(q^2)\right]\,,\\
	&A_1(q^2)=-\frac{f(q^2)}{M'+M''}\,,\\
	&A_2(q^2)=(M'+M'')a_+(q^2)\,.
\end{align}}
Their explicit expressions in the CLFQM are given by the following integrands:
\begin{equation}\label{eq:FCLF}
[\mathcal{F}(q^2)]_{\mathrm{CLF}} =
N_c \int \frac{\mathrm{d}x \, \mathrm{d}^2 \mathbf{k}'_\perp}{2(2\pi)^3}
\frac{\chi_M' \chi_M''}{\bar x}
\widetilde{\mathcal{F}}^{\,\mathrm{CLF}}(x, \mathbf{k}'_\perp, q^2)\,,
\end{equation}
{where $\mathcal{F}$ denotes the CLFQM form factor obtained within the self-consistent scheme. For $V \to P$ transition, the integrands $\widetilde{\mathcal{F}}$ are given by}
{\small
\begin{align}
\label{eq:VCLF}
\widetilde g^{V\to P}(q^2)=&-2\bigg\{\bar x m'_1+xm_2+(m'_1-m''_1)\frac{\kb\cdot\qb}{q^2}+\frac{2}{D_{V,{\rm con}}}\left[ \kb'^2+ \frac{(\kb'\cdot\qb)^2}{q^2}
	\right]\bigg\}\,,
	\\
\widetilde a^{V\to P}_{+}(q^2)=&2m'_1(1-A^{(1)}_1-A^{(1)}_2)+2m''_1(A^{(1)}_1-A^{(1)}_2-4A^{(2)}_2+4A^{(2)}_3)+4m_2(A^{(1)}_1+2A^{(2)}_2-2A^{(2)}_3)\nonumber\\
	&-\frac{4\kb''\cdot\qb}{\bar x q^2 D_{V,{\rm con}}}\Big[\kb\cdot\kb''+(\bar xm''_1+xm_2)(xm_2-\bar xm'_1)\Big]
	\,,\\
\widetilde f^{V\to P}(q^2)=&-2\bigg\{m'_1\Big[M''^2-(m''_1-m_2)^2-q^2-\hat N'_1-\hat Z_2\Big] +m''_1(M'^2-m'^2_1-m_2^2-\hat N'_1-\hat Z_2 +4A^{(2)}_1)\nonumber\\
	&+2m_2( m'^2_1+\hat N'_1-2A^{(2)}_1)+\frac{2}{D_{V,{\rm con}}}\Big[(2x-1)M'^2+M''^2-2(m'_1-m''_1)(m'_1+m_2)\nonumber\\
	&+2xM'^2_0-q^2-\frac{2\kb\cdot\qb}{q^2}(M'^2-M''^2+q^2)\Big]\Big[ \kb^2+ \frac{(\kb\cdot\qb)^2}{q^2}
	\Big]
	\bigg\}\,,
	\\
\widetilde a_-^{V\to P}(q^2)=&2\bigg\{m'_1(A^{(1)}_1+A^{(1)}_2+1)+m''_1(A^{(1)}_2-A^{(1)}_1+4A^{(2)}_3-4A^{(2)}_4)+2m_2(A^{(1)}_1-2A^{(1)}_2-2A^{(2)}_3
	\nonumber\\
	&+2A^{(2)}_4)-\frac{1}{D_{V,{\rm con}}}\Big[ -2(M'^2+M''^2+2(m'_1+m_2)(m''_1-m_2)-q^2)(A^{(2)}_3+A^{(2)}_4-A^{(1)}_2)\nonumber\\
	&+\Big(2M'^2+(m'_1+m''_1)^2-2(m'_1+m_2)^2-q^2-\hat N'_1+\hat N''_1\Big)(A^{(1)}_1+A^{(1)}_2-1)\nonumber\\
	&+2\hat Z_2(2A^{(2)}_4-3A^{(1)}_2+1)
	+2\frac{M'^2-M''^2}{q^2}(4A^{(1)}_2A^{(2)}_1-3A^{(2)}_1)
	\Big]\bigg\}\,.
\end{align}}
{For $V \to V$ transition, the integrands $\widetilde{\mathcal{F}}$ are given by}
{\small
\begin{align}
\label{eq:V1CLF}
\widetilde{V}_1^{V\to V}(q^2)=&2\bigg\{x(m'_1+m''_1)m_2+\frac{1}{\bar x}(\kb'^2+x^2m_2^2)-{\bf k'_\bot\cdot q_\bot}+\bar xm'_1m''_1+8A_1^{(3)}\nonumber\\
&-\frac{2}{D'_{ V,{\rm con}}}\left[m'_1\left(-A_1^{(2)}+4A_1^{(3)}\right)+m'_1 A_1^{(2)}+4m_2A^{(3)}_1\right]\nonumber\\
&-\frac{2}{D''_{ V,{\rm con}}}\left[m'_1 A_1^{(2)}+m''_1\left(-A_1^{(2)}+4A_1^{(3)}\right)+4m_2 A^{(3)}_1\right]\nonumber\\
&-\frac{4}{D'_{ V,{\rm con}}D''_{ V,{\rm con}}}A_1^{(2)}\Big[\frac{1}{\bar x}({\bf k'_\bot}^{2}+x^2m_2^2)-{\bf k'_\bot\cdot q_\bot}+\bar xm'_1m''_1-x(m'_1+m''_1)m_2\Big]
\bigg\}\,, \\
\widetilde{V}_2^{V\to V}(q^2)=&2{M'}^2-2(m_1'-m_2)^2+(m_1'-m_1'')^2-q^2-\hat N'_1+\hat N''_1-2Z_2-16A_2^{(3)}\nonumber\\
&+2\left[2(m_2-m'_1)(m_2-m''_1)-{M'}^{2}-{M''}^{2}+q^2+2Z_2\right]A_2^{(1)}+4\left(2+\frac{{M'}^{2}-{M''}^{2}}{q^2}\right)A_1^{(2)}\nonumber\\
&+\frac{4}{D'_{ V,{\rm con}}}\left[m'_1\left(4A_2^{(3)}-3A_1^{(2)}\right)+m''_1 A_1^{(2)}+2m_2\left(2A_2^{(3)}-A_1^{(2)}\right)\right]\nonumber\\
&+\frac{4}{D''_{ V,{\rm con}}}\left[-m'_1 A_1^{(2)}+m''_1\left(4A_2^{(3)}-A_1^{(2)}\right)+2m_2\left( 2A_2^{(3)}-A_1^{(2)}\right)\right]\nonumber\\
&+\frac{4}{D'_{ V,{\rm con}}D''_{ V,{\rm con}}}\bigg\{\Big[-2{M'}^{2}-(m'_1-m''_1)^2+2(m'_1-m_2)^2+q^2+\hat N'_1-\hat N''_1+2Z_2\nonumber\\
&-\frac{4({M'}^{2}-{M''}^{2})}{3q^2}A_1^{(2)}\Big]A_1^{(2)}+2\Big[{M'}^{2}+{M''}^{2}-2(m'_1+m_2)(m''_1+m_2)-q^2\nonumber\\
&-2Z_2\Big]A_2^{(3)}\bigg\}\,,\\
\widetilde{V}_3^{V\to V}(q^2)=&4\left({M'}^{2}-{M''}^{2}\right)\bigg\{4\left(A_3^{(2)}-A_2^{(2)}+A_3^{(3)}-A_5^{(3)}\right)\nonumber\\
&+\frac{1}{D'_{ V,{\rm con}}}\Big[(x-\bar x)m'_1\left(A_1^{(1)}-A_2^{(1)}-A_2^{(2)}+A_4^{(2)}\right)
+m''_1\left(A_2^{(2)}-2A_3^{(2)}+A_4^{(2)}\right)\nonumber\\
&+xm_2\left(A_1^{(1)}-A_2^{(1)}-2A_2^{(2)}+2A_4^{(2)}\right)\Big]\nonumber\\
&+\frac{1}{D''_{ V,{\rm con}}}\Big[m'_1\left(\bar x-2A_2^{(1)}+A_2^{(2)}+2A_3^{(2)}+A_4^{(2)}\right)+(x-\bar x)m''_1\big(A_1^{(1)}-A_2^{(1)}-A_2^{(2)}\nonumber\\
&+A_4^{(2)}\big)+2m_2\left(A_1^{(1)}+A_2^{(2)}-3A_3^{(2)}-2A_3^{(3)}+2A_5^{(3)}\right)\Big]\nonumber\\
&+\frac{2}{D'_{ V,{\rm con}}D''_{ V,{\rm con}}}\left(A_1^{(1)}-A_2^{(1)}-A_2^{(2)}+A_4^{(2)}\right)\Big[\frac{1}{\bar x}({\bf k'_\bot}^{2}+x^2m_2^2)-{\bf k'_\bot\cdot q_\bot}\nonumber\\
&+\bar xm'_1m''_1-x(m'_1+m''_1)m_2\Big]
\bigg\}\,,\\
\widetilde{V}_4^{V\to V}(q^2)=&-4q^2\bigg\{2\left(-A_1^{(1)}+A_2^{(1)}+A_2^{(2)}+2A_3^{(2)}-3A_4^{(2)}-2A_4^{(3)}+2A_6^{(3)}\right)\nonumber\\
&+\frac{1}{D'_{ V,{\rm con}}}\Big[m'_1\left(3A_1^{(1)}-3A_2^{(1)}-3A_2^{(2)}-4A_3^{(2)}+7A_4^{(2)}+4A_4^{(3)}-4A_6^{(3)}\right)\nonumber\\
&+m''_1\left(A_2^{(2)}-2A_3^{(2)}+A_4^{(2)}\right)-2m_2\left(A_2^{(2)}+A_3^{(2)}-2A_4^{(2)}-2A_4^{(3)}+2A_6^{(3)}\right)\Big]\nonumber\\
&+\frac{1}{D''_{ V,{\rm con}}}\Big[m'_1\left(-\bar x+2A_2^{(1)}-A_2^{(2)}-2A_3^{(2)}-A_4^{(2)}\right)+m''_1\Big(A_1^{(1)}-A_2^{(1)}-A_2^{(2)}-4A_3^{(2)}\nonumber\\
&+5A_4^{(2)}+4A_4^{(3)}-4A_6^{(3)}\Big)+2m_2\big(A_1^{(1)}-2A_2^{(1)}-A_2^{(2)}-A_3^{(2)}+4A_4^{(2)}+2A_4^{(3)}\nonumber\\
&-2A_6^{(3)}\big)\Big]+\frac{1}{D'_{ V,{\rm con}}D''_{ V,{\rm con}}}\Big[2\left({M'}^{2}+{M''}^{2}-2(m'_1+m_2)(m''_1+m_2)-q^2\right)\Big(A_4^{(2)}-A_3^{(2)}\nonumber\\
&+A_4^{(3)}-A_6^{(3)}\Big)+\Big(2{M'}^{2}+(m'_1-m''_1)^2-2(m'_1+m_2)^2-q^2-\hat N'_1+\hat N''_1\Big)\Big(A_1^{(1)}-A_2^{(1)}\nonumber\\
&-A_2^{(2)}+A_4^{(2)}\Big)+\frac{2({M'}^{2}-{M''}^{2})}{q^2}\Big(A_1^{(2)}-6A_2^{(1)}A_1^{(2)}+6A_2^{(1)}A_2^{(3)}-\frac{2}{q^2}(A_1^{(2)})^2\Big)\nonumber\\
&+Z_2\left(2A_2^{(1)}-6A_4^{(2)}+4A_6^{(3)}\right)\Big]\bigg\}
+\widetilde{V}_3^{\rm CLF}(x,{\bf k'_\bot},q^2)\,,\\
\widetilde{V}_5^{V\to V}(q^2)=&2\bigg\{{M'}^{2}-(m'_1-m_2)^2-\hat N'_1-Z_2-8\left(A_1^{(3)}-A_2^{(3)}\right)\nonumber\\
&+\left[{M'}^{2}-{M''}^{2}+2(m''_1-m'_1)(m''_1-m_2)-q^2
+2\hat N''_1\right]\left(A_1^{(1)}-A_2^{(1)}\right)\nonumber\\
&+\frac{2}{D'_{ V,{\rm con}}}\Big[m'_1\Big(\big({M''}^{2}-{m''_1}^{2}-m_2^{2}-\hat{N}_1''\big)\big(A_1^{(1)}-A_2^{(1)}\big)+4\big(A_1^{(3)}-A_2^{(3)}\big)+Z_2A_2^{(1)}\nonumber\\
&+\frac{{M'}^{2}-{M''}^{2}}{q^2}A_1^{(2)}\Big)-m''_1\Big(\big({M'}^{2}-{m'_1}^{2}-m_2^{2}-\hat N'_1\big)\big(A_1^{(1)}-A_2^{(1)}\big)+Z_2A_2^{(1)}\nonumber\\
&+\frac{{M'}^{2}-{M''}^{2}}{q^2}A_1^{(2)}\Big)-m_2\left(4A_2^{(3)}-4A_1^{(3)}\right)-m_2\left((m'_1-m''_1)^2-q^2+\hat N'_1+\hat N''_1\right)\big(A_1^{(1)}\nonumber\\
&-A_2^{(1)}\big)\Big]-\frac{2}{D''_{ V,{\rm con}}}\Big[2m'_1A_1^{(2)} -4m''_1\left(A_1^{(3)}-A_2^{(3)}\right)
-2m_2\left(A_1^{(2)}+2A_1^{(3)}-2A_2^{(3)}\right)\Big]\nonumber\\
&+\frac{4}{D'_{ V,{\rm con}}D''_{ V,{\rm con}}}\Big[\Big({M'}^{2}+{M''}^{2}-2(m'_1+m_2)(m''_1+m_2)-q^2\Big)\Big(A_1^{(3)}-A_2^{(3)}\Big)\nonumber\\
&+2Z_2A_2^{(3)}+\frac{2({M'}^{2}-{M''}^{2})}{3q^2}\Big(A_1^{(2)}\Big)^2\Big]
\bigg\}\,,\\
\widetilde{V}_6^{V\to V}(q^2)=&2\bigg\{{M'}^{2}-(m'_1-m''_1)^2-(m'_1-m_2)^2+q^2-2\hat N'_1-\hat N''_1-Z_2+8\left(A_1^{(2)}-A_1^{(3)}-A_2^{(3)}\right)\nonumber\\
&+\Big[{M''}^{2}-{M'}^{2}+2(m'_1-m''_1)(m'_1-m_2)-q^2+2\hat N'_1\Big]\Big(A_1^{(1)}+A_2^{(1)}\Big)\nonumber\\
&+\frac{2}{D'_{ V,{\rm con}}}\Big[4m'_1\left(A_1^{(3)}+A_2^{(3)}-A_1^{(2)}\right)-2m''_1 A_1^{(2)}-2m_2\left(A_1^{(2)}-2A_1^{(3)}-2A_2^{(3)}\right)\Big]\nonumber\\
&+\frac{2}{D''_{ V,{\rm con}}}\Big[m'_1\Big({M''}^{2}-{m''_1}^{2}-m_2^{2}-\hat N''_1-Z_2+Z_2A_2^{(1)}+\frac{M^{'2}-M^{''2}}{q^2}A_1^{(2)}\Big)\nonumber\\
&+m_1'({m''_1}^{2}+m_2^{2}-{M''}^{2}+\hat N''_1)\left(A_1^{(1)}+A_2^{(1)}\right)+m''_1\left({M'}^{2}-{m'_1}^{2}-m_2^{2}-\hat N'_1\right)\big(A_1^{(1)}\nonumber\\
&+A_2^{(1)}\big)+m''_1\big({m'_1}^{2}+m_2^{2}-{M'}^{2}+\hat N'_1+Z_2-4A_1^{(2)}+4A_1^{(3)}+4A_2^{(3)}-Z_2A_2^{(1)}\nonumber\\
&-\frac{({M'}^{2}-{M''}^{2})}{q^2}A_1^{(2)}\big)+m_2\big((m'_1-m''_1)^2-q^2+\hat N'_1+\hat N''_1\big)\big(1-A_1^{(1)}-A_2^{(1)}\big)\nonumber\\
&+4m_2\big(-A_1^{(2)}+A_1^{(3)}+A_2^{(3)}\big)\Big]\nonumber\\
&+\frac{4}{D'_{ V,{\rm con}}D''_{ V,{\rm con}}}\Big[\big(-{M'}^{2}-{M''}^{2}+q^2+2(m'_1+m_2)(m''_1+m_2)\big)\big(A_1^{(2)}-A_1^{(3)}-A_2^{(3)}\big)\nonumber\\
&+2Z_2(A_1^{(2)}-A_2^{(3)})-\frac{2({M'}^{2}-{M''}^{2})}{3q^2}\big(A_1^{(2)}\big)^2\Big]\bigg\}\,,
\end{align}
\begin{align}
\widetilde{A}_1^{V\to V}(q^2)=&-(m''_1-m'_1)^2+q^2-\hat N'_1-\hat N''_1+8A_1^{(2)}+2\Big[M'^{2}+M''^{2}+2(m'_1-m_2)(m_2-m''_1)\nonumber\\
&-q^2\Big] A_1^{(1)}-4(m_1'+m_1'')\left(\frac{1}{D_{ V,{\rm con}}'}+\frac{1}{D_{ V,{\rm con}}''}\right)A_1^{(2)}\,,\\
\widetilde{A}_2^{\rm{ V\to V}}(q^2)=&-\frac{q^2}{M'^{2}-{M''}^{2}} \bigg\{(m'_1-m''_1)^2-2(m_2-m'_1)^2+2{M'}^2-q^2-2Z_2-\hat N'_1+\hat N''_1\nonumber\\
&+\frac{4({M'}^2-{M''}^2)}{q^2}A_1^{(2)}+2\left[2(m_2-m'_1)(m_2-m''_1)-{M'}^2-{M''}^2+q^2+2Z_2 \right]A_2^{(1)}\nonumber\\
&-\frac{4}{D'_{ V,{\rm con}}}\left(m'_1-m''_1+2m_2\right)A_1^{(2)}-\frac{4}{D''_{ V,{\rm con}}}\left(m'_1-m''_1-2m_2\right)A_1^{(2)}\bigg\}+\widetilde{A}_1^{\rm{ CLF}}(x,{\bf k'_\bot},q^2)\,,
\\
\widetilde{A}_3^{V\to V}(q^2)=&-4({M'}^{2}-{M''}^{2})\bigg\{A_1^{(1)}-A_2^{(1)}-A_2^{(2)}+A_4^{(2)}
\nonumber\\
&+\frac{1}{D_{ V,{\rm con}}''}\Big[ m'_1\left(-\bar x+2A_2^{(1)}-A_2^{(2)}-2A_3^{(2)}-A_4^{(2)}\right)+m''_1\left(-A_1^{(1)}+A_2^{(1)}+A_2^{(2)}-A_4^{(2)}\right)\nonumber\\
&+ 2m_2\left(-A_1^{(1)}+A_2^{(2)}+A_3^{(2)}\right) \Big]+\frac{2}{D'_{ V,{\rm con}}D''_{ V,{\rm con}}}\left(A_1^{(2)}-A_1^{(3)}-A_2^{(3)}\right)
\bigg\}\,,\\
\widetilde{A}_4^{V\to V}(q^2)=&-4({M'}^{2}-{M''}^{2})\bigg\{A_1^{(1)}-A_2^{(1)}-A_2^{(2)}+A_4^{(2)}\nonumber\\
&+\frac{1}{D'_{ V,{\rm con}}}\Big[m'_1\left(A_2^{(1)}-A_1^{(1)}+A_2^{(2)}-A_4^{(2)}\right)+m''_1\left(-A_2^{(2)}+2A_3^{(2)}-A_4^{(2)}\right)\nonumber\\
&+2m_2\left(A_2^{(2)}-A_3^{(2)}\right)\Big]+\frac{2}{D'_{ V,{\rm con}}D''_{ V,{\rm con}}}\left(-A_1^{(3)}+A_2^{(3)}\right)
\bigg\}\,. \label{eq:A4CLF}
\end{align}}
{In the above expressions, $M'$ and $M''$ are the initial and final meson masses, $m'_1$, $m''_1$, and $m_2$ are the constituent quark masses, and $D^{\prime(\prime\prime)}_{V,\mathrm{con}} = M^{\prime(\prime\prime)} + m^{\prime(\prime\prime)}_1 + m_2$ is the denominator function for the vector vertex.}}
\end{appendices}
\section*{Acknowledgments}
This work is supported by the National Natural Science Foundation of China (Grant nos.~12275067 and~12305101), Natural Science Foundation of Henan Province (Grant no.~22520081\-0030), Science and Technology Innovation Leading Talent Support Program of Henan Province  (Grant no.~254200510039), National Key R$\&$D Program of China (Grant no.~2023YFA1606000), the China Postdoctoral Science Foundation (Grant no.~2025M783381) and Key Research Project Plan for Higher Education Institutions of Henan Province (Grant no.~23 A140012).


\end{document}